\documentclass[pra,twocolumn,showpacs,amsmath,amssymb,superscriptaddress,nofootinbib]{revtex4-1}
\usepackage{braket}
\usepackage{cases}
\usepackage{color}
\usepackage{graphicx,subfigure}
\usepackage{float,physics}
\usepackage{bm}
\usepackage{hyperref}
\hypersetup{
    colorlinks=true, 
    linkcolor=cyan,
    citecolor=magenta, 
    filecolor=magenta, 
    urlcolor=cyan,
    runcolor=cyan
}

\newcommand{\ignore}[1]{}

\usepackage{commath}
\usepackage{color}
\usepackage{amsmath}
\usepackage{adjustbox}
\usepackage[normalem]{ulem}

\def\be{\begin{equation}}
\def\ee{\end{equation}}

\newcommand{\bes} {\begin{subequations}}
\newcommand{\ees} {\end{subequations}}
\newcommand{\beq}{\begin{equation}}
\newcommand{\eeq}{\end{equation}}

\def\>{\rangle}
\def\<{\langle}
\def\Tr{\mathrm{Tr}}

\newcommand{\ketb}[2]{|{#1}\>\!\<#2|}

\newcommand{\1}{\leavevmode{\rm 1\ifmmode\mkern  -4.8mu\else\kern -.3em\fi I}}

\newtheorem{thm}{\protect\theoremname}
\newtheorem{prop}[thm]{\protect\propositionname}

\providecommand{\propositionname}{Proposition}
\providecommand{\theoremname}{Theorem}
\providecommand{\conjecturename}{Conjecture}

\begin{document}
\title{Demonstration of error-suppressed quantum annealing via boundary cancellation}
\author{Humberto Munoz-Bauza}
\affiliation{Department of Physics \& Astronomy, University of Southern California, Los Angeles,
CA 90089, USA}
\affiliation{Center for Quantum Information Science \& Technology, University of
Southern California, Los Angeles, CA 90089, USA}
\author{Lorenzo Campos Venuti}
\affiliation{Department of Mechanical Engineering, Massachusetts Institute of Technology, 77 Massachusetts Av., Cambridge, MA 02139, USA}
\affiliation{Department of Physics \& Astronomy, University of Southern California, Los Angeles,
CA 90089, USA}
\affiliation{Center for Quantum Information Science \& Technology, University of
Southern California, Los Angeles, CA 90089, USA}
\author{Daniel Lidar}
\affiliation{Department of Physics \& Astronomy, University of Southern California, Los Angeles,
CA 90089, USA}
\affiliation{Center for Quantum Information Science \& Technology, University of
Southern California, Los Angeles, CA 90089, USA}
\affiliation{Department of Electrical Engineering, University of Southern California,
Los Angeles, CA 90089, USA}
\affiliation{Department of Chemistry, University of Southern California, Los Angeles,
CA 90089, USA}
\date{\today}
\begin{abstract}
The boundary cancellation theorem for open systems extends the standard quantum adiabatic theorem: 
assuming  the gap of the Liouvillian does not vanish,
the distance between a state prepared by
a boundary cancelling adiabatic protocol and the steady state of the
system shrinks as a power of the number of vanishing time derivatives
of the Hamiltonian at the end of the preparation. Here we generalize the boundary cancellation theorem so that it applies also to the case where the Liouvillian gap vanishes, and consider the effect of dynamical freezing of the evolution. We experimentally test the predictions of the boundary cancellation theorem using quantum annealing hardware, and find qualitative agreement with the predicted error suppression despite using annealing schedules that only approximate the required smooth schedules. Performance is further improved by using quantum annealing correction, and we demonstrate that the boundary cancellation protocol is significantly more robust to parameter variations than protocols which employ pausing to enhance the probability of finding the ground state.
\end{abstract}
\maketitle

\section{Introduction}

Adiabatic quantum computing~\cite{farhiQuantum00} and quantum annealing~\cite{kadowakiQuantum98}
can be used to prepare ground states and thermal states of quantum
systems, a central desideratum of quantum computation. The quality
of ground state preparation and Gibbs sampling is central to the performance
of quantum algorithms for optimization as well as machine learning
problems~\cite{Amin:2016}. The performance of these computational
paradigms is quantified by the adiabatic theorem in its different
forms, whether for closed or open quantum systems. The theorem, or
family of theorems, places a bound on the adiabatic error: the distance
between the state that we set out to prepare (usually the solution
to a computational problem) and the state the system actually ends up in through
the dynamical evolution. In general, if a spectral gap condition is
satisfied, the adiabatic error is bounded by $C/t_{f}$ for some constant
$C$, where $t_{f}$ is the total evolution time~\cite{katoAdiabatic50,jansenBounds07,joye_general_2007}.
The boundary cancellation theorem (BCT) shows that an improvement
is possible if the schedule from the initial to the final generator
(Hamiltonian or Liouvillian) has certain properties. In the closed
system case the adiabatic error is bounded by $C_{k}/t_{f}^{k+1}$
for some constant $C_{k}$ independent of $t_f$, if the time-dependent Hamiltonian $H(t)$
has vanishing derivatives up to order $k$ at the \emph{beginning
and end} of the evolution~\cite{garridoDegree62}, a bound that
can be improved to exponentially small in $t_{f}$ under additional
assumptions~\cite{nenciu_linear_1993,Hagedorn:2002kx,lidarAdiabatic09,wiebeImproved12,Ge:2015wo}.
This theorem has recently been extended to the open system setting
for a general class of time-dependent Liouville operators $\mathcal{L}(t)$,
where it can be used to prepare steady states of Liouvillians instead
of ground states of Hamiltonians. The theorem can be succinctly stated
as follows: For a time-dependent Liouvillian $\mathcal{L}(t)$ with
a unique steady state $\sigma(t)$ separated by a gap at each time
$t$, the adiabatic error is likewise bounded by $C_{k}/t_{f}^{k+1}$,
but it is sufficient for the derivatives to vanish up to order $k$
\emph{only at the end time} $t_{f}$~\cite{camposvenutiError18}.
Boundary cancellation (BC) plays a significant role in the theoretical
analysis of error suppression in Hamiltonian quantum computing~\cite{albashDecoherence15,Lidar:2019ab}.

In this work we set out to demonstrate the BCT predictions in experiments
using the D-Wave 2000Q (DW2KQ) quantum annealer. The DW2KQ implements the transverse field Ising model $H(t) = A(t)H_X +B(t)H_Z$, where $H_X$ and $H_Z$ are transverse and longitudinal (Ising) Hamiltonians, respectively~\cite{johnsonQuantum11,DW2KQ}.
It allows
limited programmability of the control schedules $A(t)$ and $B(t)$
of the system Hamiltonian, which we exploit to implement boundary cancellation
protocols. 

To guide our intuition we model the behavior of the D-Wave
quantum annealer with the adiabatic master equation (AME) derived in \cite{albashQuantum12}.
This is a time-dependent Davies-like master equation~\cite{davies_markovian_1974} which has been successfully used to interpret several D-Wave experiments, e.g.,~\cite{albashConsistency15,albashReexamination15} (though not always~\cite{bando2021breakdown}).
When combining the AME for dephasing with boundary cancellation, we encounter a problem. Namely, as we explain in detail below, the Liouvillian gap vanishes at $t=t_f$, which prevents us  from directly applying the BCT in the form given in Ref.~\cite{camposvenutiError18}. 
To circumvent this problem, in this work we generalize the BCT and identify the conditions on the Liouvillian gap under which BC does or does not remain effective. We find that the generalized BCT 
plays an important role in the D-Wave implementation.

There is another 
significant consideration regarding
the implementation of BC on D-Wave:
the phenomenon of freezing. In essence, freezing refers to
a significant increase in all relaxation timescales well before the
end of the anneal ($t<t_{f}$), in such a manner that the system is
frozen in a state that does not correspond to the steady state of
$\mathcal{L}(t_{f})$~\cite{albashQuantum12}. However, the frozen state is not truly static but quasi-static~\cite{Amin:2015qf}, i.e., relaxation is not fully switched off but instead proceeds on a timescale that may be much longer than practically realizable anneal times $t_f$. Such glassy-like behavior can take place even in very simple systems that do not have a glassy landscape. We give an explanation of freezing in Sec.~\ref{sec:freezing} below, 
based on the AME. Essentially, it arises due to the fact that the system Hamiltonian commutes with the system-bath
coupling when $A(s)=0$, and the fact that the schedule $A(t)$  has nearly vanished long before $t=t_{f}$
(see Fig.~\ref{fig:dw2kq_schedule}). This is true in particular 
of currently available commercial quantum annealers manufactured by D-Wave~\cite{boothby2020nextgeneration}, which can be effectively
described by longitudinal field coupling to a bosonic bath, which
commutes with the system Hamiltonian at sufficiently low transverse
fields. The state of the system is insensitive
to the details of the schedules past the freezing point, so that BC becomes ineffective. This freezing mechanism plays an adverse role in demonstrating the predictions of the BCT, since the experimentally accessible anneal times are significantly shorter than the relaxation time required to reach the true steady state of $\mathcal{L}(t_f)$.

In order to overcome both issues, i.e., the gaplessness of the Liouvillian at
$t=t_{f}$ and freezing at $t\approx0.5t_{f}$, we
design the programmable DW2KQ schedule so that it flattens (and thus implements BC) at a point $t_{BC}$ before freezing, followed by a ramp to the final values of $B(t_{f})$
and $A(t_{f})$, upon which the system is measured in computational basis (eigenbasis of $H_Z$).  

We apply the protocol to two 8-qubit gadgets: 1) a ``ferromagnetic chain gadget''
(FM-gadget), and 2) a ``tunneling gadget'' (T-gadget) with a larger
tunneling barrier between the ground state and the first excited state
around the minimum gap~\cite{AlbashDemonstration18}. We estimate
the scaling of the adiabatic error for the FM-gadget  and compare
the experimental results of the boundary cancellation protocol (BCP)
with those predicted by open system simulations using the AME. Based on these simulations,
we discuss the conditions under which the BCT-predicted scaling may
be observed. We also apply the protocol to an error-corrected version of
the FM-gadget using quantum annealing correction
(QAC)~\cite{pudenzErrorcorrected14}. The QAC encoding mitigates
the effects of Hamiltonian programming errors~\cite{youngAdiabatic13,pearsonAnalog19}
and effectively lowers the temperature of the environment~\cite{Matsuura:2016aa,Vinci:2017ab}; we find that QAC improves the performance of the BCP.

The structure of this paper is as follows. In Sec.~\ref{sec:Theory} we first review the BCT, and the formulate a new version which accounts for the possibility of the Liouvillian gap closing at the same point as where the boundary cancellation conditions are enforced, which is relevant for our experiments. The rest of the section is devoted to a discussion of freezing, the effect of a ramp after the BC point, and anomalous heating, all of which are phenomena affecting the performance of the BCP due to their presence in our experiments. In Sec.~\ref{sec:methods} we describe our methods: the BCP we used in detail, the two $8$-qubit gadgets used in our experiments, and encoding with quantum annealing correction. Our results are presented in Sec.~\ref{sec:results}; we start with a standard linear control schedule and then present the results for the experimental BCP implementation. We include simulation results for a higher precision version of the BCP for reference, and then report on experiments from the QAC-encoded version of one of our gadgets. Finally, we compare the performance of the BCP to that of the pausing protocol. We conclude with a discussion and outlook in Sec.~\ref{sec:discussion}. The appendix includes background on the adiabatic master equation, proofs of various Propositions, supporting numerical results,  additional information needed to reproduce our experiments and analysis, and experimental results from two other D-Wave processors.

For convenience, we include a glossary of the acronyms, terminology, and key symbols we used in Table~\ref{tab:gloss}.

\begin{table}[t]
\begin{center}
\begin{tabular}{|c|c|c|}
$\eta$ & Adiabatic error scaling exponent & Sec.~\ref{sec:everything} \\
$\rho$ & Solution of the master equation & Eq.~\eqref{eq:MEode}\\
$\sigma$ & Gibbs state at end of BCP & Sec.~\ref{sec:ramp} \\
$s_{\text{BC}}$ & Termination point of BCP & Sec.~\ref{sec:bcp}\\
$s_0,t_0$ & Dynamical freezing point & Eq.~\eqref{eq:freezing_point}\\
AME & Adiabatic master equation & Sec.~\ref{sec:Theory-A}, \cite{albashQuantum12}\\
BCP & Boundary cancellation protocol & Sec.~\ref{sec:bcp} \\
BCT  & Boundary cancellation theorem & Sec.~\ref{sec:Theory}, \cite{camposvenutiError18}\\
$D_{\text{GS}}$ & adiabatic error & Eq.~\eqref{eq:D_GS} \\
DW2KQ & D-Wave 2000Q annealer & \cite{dwave-manual,DW-manual}\\
FM-gadget & Ferromagnetic chain gadget & Sec.~\ref{sec:8QG} \\
NTS & Number of tries-to-solution & Eq.~\eqref{eq:NTS} \\
$P_{\text{GS}}$ & Measured ground state prob. & Eq.~\eqref{eq:19a} \\
$P_{\text{GS}}^*$ & Gibbs ground state prob. & Eq.~\eqref{eq:19b} \\
PR & Pause-ramp protocol & Sec.~\ref{sec:tries} \\
QAC & Quantum annealing correction & Sec.~\ref{sec:QAC}, \cite{pudenzErrorcorrected14}\\
T-gadget & Tunneling gadget & Sec.~\ref{sec:8QG}, \cite{AlbashDemonstration18}
\end{tabular}
\end{center}
\label{tab:gloss}
\caption{Glossary of key symbols and acronyms used in this work in alphabetical order, along with the sections, equations, or references in which they are defined.}
\end{table}%

\section{Theory}
\label{sec:Theory}

\subsection{Review of the BCT results of Ref.~\cite{camposvenutiError18}}
\label{sec:Theory-A}

We consider a finite-dimensional system whose density matrix $\rho_{t_{f}}$
satisfies the master equation $d\rho_{t_{f}}/dt=\mathcal{L}_{t_{f}}(t)\rho_{t_{f}}(t)$,
where $\mathcal{L}_{t_{f}}(t)$ is a time-dependent Liouvillian superoperator.
If $\mathcal{L}_{t_{f}}(t)$ depends on the time $t$ only through the ``anneal parameter'' $\tau:=t/t_{f}$,
where $t_{f}$ is the total evolution time, then defining $\rho(\tau):=\rho_{t_{f}}(t)$
and $\mathcal{L}(\tau):=\mathcal{L}_{t_{f}}(t)$, the master equation
takes the form 

%
\begin{equation}
\frac{d\rho}{d\tau}=t_{f}\mathcal{L}(\tau)\rho(\tau)\ .
\label{eq:MEode}
\end{equation}
The main result of Ref.~\cite{camposvenutiError18} is summarized
in the following proposition:

\begin{prop}
\label{prop:BCT} 
Assume that $\mathcal{L}(\tau)$ is such that $\left\Vert \mathcal{L}(\tau)\right\Vert _{1,1}$
is summable in $\tau\in[0,1]$,\footnote{The superoperator norm we use is the induced $1,1$ norm: \unexpanded{$\left\Vert \mathcal{A}\right\Vert_{1,1} :=\sup_{\left\Vert y\right\Vert _{1}=1}\left\Vert \mathcal{A}y\right\Vert _{1}$}
for a superoperator $\mathcal{A}$, where the trace-norm is defined
as \unexpanded{$\|A\|_{1}:=\Tr\sqrt{A^{\dagger}A}$}, for any operator
$A$, i.e., the sum of $A$'s singular values. A function is summable if its integral is finite on every compact subset of its domain.} $\mathcal{L}(\tau)$ is differentiable to order $k+2$ in
a neighborhood of $\tau=1$, and generates a trace-preserving and
hermiticity-preserving contractive semigroup, i.e., $\left\Vert e^{r\mathcal{L}(\tau)}\right\Vert _{1,1}\le1$
for all $r\ge0$, $\tau\in\left[0,1\right]$. Suppose $\mathcal{L}(\tau)$ has a unique steady state $\sigma(\tau)$ with eigenvalue zero
separated from the rest of the spectrum by a nonzero gap $\forall \tau\in[0,1]$. Let $\rho(\tau)$
be the solution of Eq.~\eqref{eq:MEode} with initial condition $\rho(0)=\sigma(0)$.
If $\mathcal{L}(\tau)$ has vanishing derivatives at $\tau=1$ to
order $k$, i.e., $\left.\partial_{\tau}^{(j)}\mathcal{L}\right|_{\tau=1}=0$
$\forall j=1,2,\ldots,k$, then there is a constant $C_{k}$ independent
of $t_{f}$ such that 
\begin{equation}
\left\Vert \rho(1)-\sigma(1)\right\Vert _{1}\leq\frac{C_{k}}{t_{f}^{k+1}}\ .
\label{eq:BCT}
\end{equation}
\end{prop}
We note that the smoothness assumptions of
Prop.~\ref{prop:BCT} are more relaxed than
in Ref.~\cite{camposvenutiError18}.  

From here on we focus in particular on the AME, which we briefly review in Sec.~\ref{sec:Theory-AME}. 
It was shown in \cite{camposvenutiError18} that, under the action
of the AME Liouvillian, it is possible to enforce the BCT as required in
Prop.~\ref{prop:BCT} by controlling just the time-dependent system
Hamiltonian $H_S$:

\begin{prop}
\label{prop:AME_gap}
Assume the system evolves according to the time
dependent Liouvillian in Eq.~\eqref{eq:AME}. Assume further that
$\left.\partial_{t}^{(j)}H_{S}(t)\right|_{t=t_{f}}=0$, for $j=1,2,\ldots,k$.
Then $\left.\partial_{t}^{(j)}\mathcal{L}\right|_{t=t_{f}}=0$ $\forall j=1,2,\ldots,k$. 
\end{prop}

This means that rather than having to directly check that the entire Liouvillian
satisfies the conditions of vanishing derivatives at the end, it suffices
to check that the system Hamiltonian $H_{S}(t)$ satisfies the
same conditions. However, the Prop.~\ref{prop:BCT} requirement that the Liouvillian has a finite
gap above its zero eigenvalue throughout the evolution window
must be checked independently. For the finite dimensional systems that we consider here, this is equivalent
to the requirement that the steady state is unique throughout the
evolution, or alternatively, that there are no ``level crossings''
of $\mathcal{L}(\tau)$'s zero eigenvalue throughout the evolution.%
\footnote{In principle one must also check that %
    $\int_0^1 ds \left\Vert \mathcal{L}(s)\right\Vert _{1,1} <\infty$,
but this follows automatically if the control functions satisfy
$\int_0^1 ds \abs{A(s)}<\infty$ and 
$\int_0^1 ds \abs{B(s)}<\infty$,
which is a natural assumption.} Since the system Hamiltonian is naturally the controllable component
of the total Hamiltonian or Liouvillian, this is also a physically
sensible controllability condition. 

Moreover, it was verified numerically in \cite{camposvenutiError18} that Eq.~\eqref{eq:BCT} remains valid for a more general Liouvillian in time-dependent Redfield form, even though it does not satisfy the hypothesis of Props.~\ref{prop:BCT} and \ref{prop:AME_gap}, i.e.~for which the vanishing of the derivatives of the system Hamiltonian does not imply the vanishing of the derivatives of the Liouvillian. See \cite{camposvenutiError18} for details.

We next review the AME for quantum annealing problems and discuss a modified form of the BCT, which is tailored to the situation we encounter in our experiments using the D-Wave annealer.

\subsection{Adiabatic master equation for quantum annealing}
\label{sec:Theory-AME}

\begin{figure}
    \includegraphics[width=.95\columnwidth]{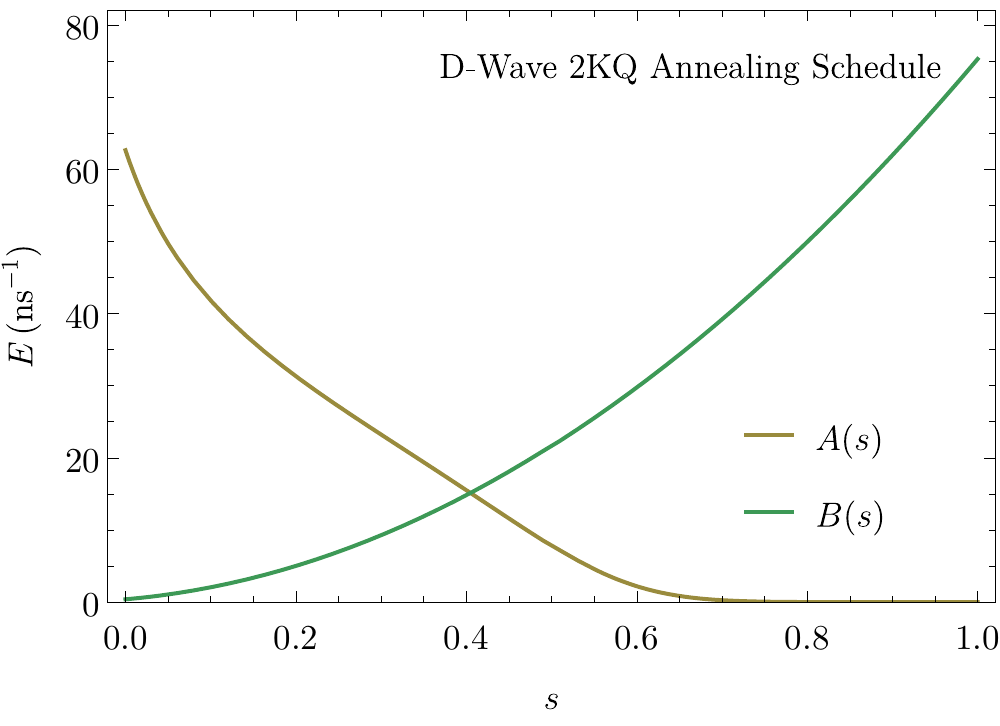}
    \caption{Native annealing schedule of the D-Wave 2000Q low noise processor.}
    \label{fig:dw2kq_schedule}
\end{figure}

We now consider the quantum annealing problem defined by a system
Hamiltonian of the form 
\begin{equation}
    H_{S}(s)=\frac{A(s)}{2}H_{X}+\gamma\frac{B(s)}{2}H_{Z}\,,\label{eq:QA_Hamiltonian}
\end{equation}
where 
\begin{equation}
    H_{X}=-\sum_{i}\sigma_{i}^{x},\qquad H_{Z}=\sum_{i}h_{i}\sigma_{i}^{z}+\sum_{i<j}J_{ij}\sigma_{i}^{z}\sigma_{j}^{z}\ .
    \label{eq:HXHZ}
\end{equation}
and $\gamma\in(0,1]$ is a scaling factor that controls the energy
scale of the computational problem. We refer to $A(s)$ and $B(s)$
as the \emph{physical schedules} of the anneal. They are fixed properties
of the quantum annealing hardware (see Fig.~\ref{fig:dw2kq_schedule}), but $s$ can be made to depend
on $\tau=t/t_{f}$, in the adjustable way that we describe below for the DW2KQ. We call the function $s(\tau):[0,1]\mapsto[0,1]$
the \emph{control schedule}. 

We model the anneal according to the AME~\cite{albashQuantum12}
with the Hamiltonian in Eq.~\eqref{eq:QA_Hamiltonian}.
The AME is derived from the total Hamiltonian $H(t)=H_S(t) + H_B + H_{SB}$, with $H_{SB} =g \sum_i A_i\otimes B_i$ where the $A_i$ ($B_i$) are the dimensionless system (bath) operators, subject to an adiabatic approximation. The resulting Liouvillian
is of the time-dependent Davies form
\begin{subequations}
    \begin{align}
        \mathcal{L}(t) & =-i\left[H_{S}(t)+H_{LS}(t),\bullet\right]+\mathcal{D}(t)\label{eq:AME}\\
        \mathcal{D}(t) & =\sum_{ij\omega}\gamma_{ij}(\omega)\Big(A_{j}(\omega)\bullet A_{i}^{\dagger}(\omega)\nonumber \\
        &\qquad  -\frac{1}{2}\left\{ A_{i}^{\dagger}(\omega)A_{j}(\omega),\bullet\right\} \Big)\ .
    \end{align}
\end{subequations}
Here $H_{LS}(t)$ is the Lamb-shift term, $\gamma_{ij}(\omega)$
is the Fourier transform of the bath correlation function 
\begin{equation}
    G_{ij}(t_{1},t_{2})=\langle B_{i}(t_{1})B_{j}(t_{2})\rangle=G_{ij}(t_{1}-t_{2})\ ,
\end{equation}
$B_i(t) = e^{ iH_B t}B_i e^{- iH_B t}$, $\langle X\rangle = \Tr(\rho_B X)$ where $\rho_B$ is the initial bath state,
and $\omega(t)$ are the Bohr frequencies of $H_{S}(t)$ (to simplify
notation we suppress their explicit time-dependence when convenient), i.e., the differences between the instantaneous eigenenergies of $H_{S}(t)$.
The Lindblad operators $A_{i}(\omega)$ that appear in the Davies
generator (``dissipator'') $\mathcal{D}$ are defined by 
\begin{equation}
    e^{irH_{S}(t)}A_{i}e^{-irH_{S}(t)}=\sum_{\omega}e^{-ir\omega}A_{i}(\omega)\ ,\label{eq:jump}
\end{equation}
i.e.,
\beq
A_{i}(\omega) = \sum_{E_b-E_a=\omega}\Pi_a A_i \Pi_b\ ,
\eeq
where $H_{S} = \sum_a E_a \Pi_a$ and $\Pi_a$ are the eigenprojectors of $H_S$.

The Lamb shift is given by  
\begin{equation}
    H_{LS}=\sum_{ij\omega}S_{ij}(\omega)A_{i}^{\dagger}(\omega)A_{j}(\omega)\ ,\label{eq:Lamb_shift}
\end{equation}
with
\beq
S_{ij}(\omega)=\int_{-\infty}^{+\infty}d\omega'\gamma_{ij}(\omega')\mathcal{P}\left(\frac{1}{\omega-\omega'}\right)\ ,
\eeq
where $\mathcal{P}$ is the Cauchy principal value. It commutes with $H_S$.

This master equation
generates a completely positive map and describes a system with a slowly varying system
Hamiltonian $H_{S}(t)$, weakly interacting with an infinite bath.
Moreover $\mathcal{L}(t)$ has the property that it has the Gibbs
state as a steady state, i.e., $\mathcal{L}(t)\sigma(t)=0$ with $\sigma(t)=e^{-\beta H_{S}(t)}/Z(t)$, $Z(t) = \Tr e^{-\beta H_{S}(t)}$. 

In our simulations we assume longitudinal field coupling, i.e., the system operators in the system-bath interaction Hamiltonian are $A_{i}=\sigma_{i}^{z}$ with $\gamma_{ij}(\omega) = \delta_{ij}\gamma(\omega)$ (independent coupling) 
and the bath has an Ohmic spectral density, so that
\begin{equation}
    \gamma(\omega)=2\pi \eta_0 g^{2}\frac{\omega}{1-e^{-\beta\omega}}\ .
    \label{eq:Ohmic}
\end{equation}

\subsection{A modified BCT}
\label{sec:Theory-B}

Unfortunately, we cannot directly apply Prop.~\ref{prop:BCT} 
to annealers with $A(t_{f})=0$ (such as the DW2KQ) because of the following result (proven in App.~\ref{app:Proof-of-degeneracy}):

\begin{prop}
\label{prop:degeneracy}
Assume that $\mathcal{L}(\tau)$ is the AME Liouvillian [Eq.~\eqref{eq:AME}]
with the Hamiltonian in Eq.~\eqref{eq:QA_Hamiltonian} and 
longitudinal system-bath coupling (i.e., that commutes with $H_Z$).
Then, at the end of the anneal when $A(t_{f})=0$, the zero eigenvalue
of $\mathcal{L}(t_{f})$ is at least $d$-fold degenerate, where $d$ is the system's Hilbert space dimension.
\end{prop}

Since presumably, when $A(t)\neq0$ for $t<t_{f}$ the Lindbladian
$\mathcal{L}(t)$ has a unique steady state $\sigma(t)$ which corresponds
to its zero eigenvalue, Prop.~\ref{prop:degeneracy} means that the Liouvillian
gap closes at $t_{f}$ and we cannot use Prop.~\ref{prop:BCT}.

We now generalize the BC result Prop.~\ref{prop:BCT} to the situation where one enforces the BC conditions at the same point as where the Liouvillian gap closes, e.g., at $t=t_f$.

\begin{prop}
\label{prop:BC_gapless_end} 
Assume that $\left\Vert \mathcal{L}(\tau)\right\Vert _{1,1}$
is summable in $\tau\in[0,1]$, $\mathcal{L}(\tau)$ is differentiable to order $k+2$ in
a neighborhood of $\tau=1$, and generates a trace-preserving and
hermiticity-preserving contractive semigroup, i.e., $\left\Vert e^{r\mathcal{L}(\tau)}\right\Vert _{1,1}\le1$
for all $r\ge0$, $\tau\in\left[0,1\right]$. Suppose $\mathcal{L}(\tau)$
has a unique steady state $\sigma(\tau)$ at each time $\tau\in[0,1]$
except at the point $\tau_{0}=1$ where the Liouvillian gap closes as
$\Delta\simeq v(\tau_0-1)^{\alpha}$ for
a linear schedule (i.e., in the absence of a BC schedule). Let $\rho(\tau)$ be the solution of Eq.~\eqref{eq:MEode}
with initial condition $\rho(0)=\sigma(0)$ and $\sigma(1^{-}):=\lim_{r\to1^{-}}\sigma(r)$.
Assume we enforce boundary cancellation at the end, i.e., $\mathcal{L}(\tau)$
has vanishing derivatives at $\tau=1$ to order $k$: $\left.\partial_{\tau}^{(j)}\mathcal{L}\right|_{\tau=1}=0$
$\forall j=1,2,\ldots,k$ [note that here $\mathcal{L}(s(\tau))$ is written for
simplicity as $\mathcal{L}(\tau)$]. Then boundary cancellation is only mildly effective
in that the adiabatic error satisfies
\bes
\begin{align}
\left\Vert \rho(1)-\sigma(1^{-})\right\Vert _{1} & \leq\frac{C}{t_{f}^{\eta}}\ , \label{eq:adia_error_scaling}\\
\mathrm{with\,\,} & \eta=\frac{k+1}{k\alpha+\alpha+1} <\frac{1}{\alpha}
\label{eq:eta_gapless_end} .
\end{align}
\ees
\end{prop}
The proof is given in App.~\ref{app:Proof-of-Proposition_end}, which also discusses the relaxed smoothness assumptions of Prop.~\ref{prop:BCT} relative to Ref.~\cite{camposvenutiError18}.
The fact that the gap closes for the Lindbladian modeling the D-Wave annealers was already
noted in Ref.~\cite{venutiAdiabaticity16}, and indeed Eq.~\eqref{eq:eta_gapless_end} with $k=0$ was 
derived there.

For completeness we comment also on the case where the Liouvillian gap closes in the middle of the evolution and one enforces boundary cancellation at the end. 
We have performed numerical simulations under these conditions (see App. \ref{app:gap_mid}), all agreeing with a scaling of the adiabatic error as in Eq.~\eqref{eq:adia_error_scaling}, but with an exponent given by $\eta=k+1$, as in Eq.~\eqref{eq:BCT}. Analytical results will be presented elsewhere. 

Summarizing, if the gap closes at the end of the evolution, implementing BC is mildly advantageous in that the best one can do is to change the exponent $\eta$ from $1/(\alpha+1)$ to $1/\alpha$ for an infinitely flat schedule ($k\to\infty$). If, instead, the gap closes in the middle of the evolution, we infer from our numerical results that is has no effect on the scaling of the adiabatic error and implementing BC results in a favorable exponent $\eta=k+1$, just as in the case where the Liouvillian gap does not close at all.

\subsection{Freezing}
\label{sec:freezing}
As the Hamiltonian starts to commute with the system-bath coupling operators near the end of the anneal,
the dynamics of the system will begin to slow down. This gives another significant consideration 
regarding the successful implementation of BC schedules, 
which is that the system state appears to stop evolving at a point $t_0 < t_f$, a phenonemon of quantum annealing called freezing.
To explain freezing, we write the master equation in the instantaneous energy eigenbasis $\{|n\rangle\}$, $n=0,1,\ldots,d-1$
(we omit the time dependence for clarity), i.e., the basis which diagonalizes the Hamiltonian $H_{S}(t)$. Consider the region of the anneal where $A\left(t\right)$ is small but non-zero, and also $B(t) > 0$. It is then natural to assume that the spectrum of
$H_{S}(t)$ is non-degenerate. The density
matrix in the energy basis is $\rho(t)=\sum_{mn}\rho_{mn}|m\rangle\!\langle n|$.
The diagonal elements of $\rho$ in this basis evolve according to
the Pauli master equation~\cite{Pauli-master-equation} (for a modern derivation see, e.g., Ref.~\cite{Lidar:2019aa}). In particular, the ground state probability
$\rho_{00}=\langle0|\rho(t)|0\rangle$ evolves according to 
\bes
\label{eq:freezing}
\begin{align}
\label{eq:freezing-a}
\dot{\rho}_{00} & =\sum_{n}\left(\rho_{nn}W_{0n}-\rho_{00}W_{n0}\right) \\
 & =\sum_{n>0}W_{0n}\left(\rho_{nn}-\rho_{00}e^{-\beta\left(E_{n}-E_{0}\right)}\right)\ ,
 \label{eq:freezing-b}
\end{align}
\ees
where the transition rate matrix $W$ has the following matrix elements:\footnote{The matrix $W$ satisfies detailed balance, i.e., $W_{nm}e^{-\beta E_{m}}=W_{mn}e^{-\beta E_{n}}$,
which follows from an analogous equation for $\gamma(\omega)$, which
in turns follows from the Kubo-Martin-Schwinger (KMS) conditions on the bath correlation function.} 
\begin{equation}
W_{mn}=\sum_{i,j}\gamma_{ij}(E_{n}-E_{m})\langle m|\sigma_{j}^{z}|n\rangle\langle n|\sigma_{i}^{z}|m\rangle.
\label{eq:matrix_M}
\end{equation}

On the basis of Eqs.~\eqref{eq:freezing} and~\eqref{eq:matrix_M},
freezing is seen to be a consequence of the following argument. Since
we are in the adiabatic regime, only the lowest levels are populated,
i.e., $\rho_{nn}\simeq0$ for $n$ greater than some $n_{A}$. Eq.~\eqref{eq:freezing}
is then replaced by 
\bes
\label{eq:freezing-2}
\begin{align}
\label{eq:freezing-2a}
\dot{\rho}_{00}&\simeq\sum_{n=1}^{n_{A}}W_{0n}\left(\rho_{nn}-\rho_{00}e^{-\beta\left(E_{n}-E_{0}\right)}\right) \\
&\qquad -\rho_{00}\sum_{n=n_{A}+1}^{d}W_{0n}e^{-\beta\left(E_{n}-E_{0}\right)}\ ,
\label{eq:freezing-2b}
\end{align}
\ees
where $d$ is the system's Hilbert space dimension. When $A(t)=0$, the Hamiltonian is diagonal in the computational basis. Let the Hamming distance between the ground state
$|0\rangle$ and the excited states $\{|n\rangle\}_{n=1}^{n_{A}}$ when $A(t)=0$
be at least $q$ (note that $q\ge 1$). Using perturbation theory
around $A(t)=0$ it can be shown that 
$\langle0|\sigma_{j}^{z}|n\rangle=O\left(A^{q}\right)$;
see App.~\ref{app:Proof-of-freezing}. Hence, at the time $t_0$ at which  $A$ is sufficiently small (but non-zero)
one has $W_{0n}= O(A^{2q}) \simeq0$ for $n=1,2,\ldots,n_{A}$, and we can neglect the sum in  line~\eqref{eq:freezing-2a}; we determine $t_0$ below.
As for the term in line~\eqref{eq:freezing-2b}, the transition rates between higher excited states
are typically smaller, and moreover, the terms are exponentially suppressed
because of the larger gaps (i.e., for sufficiently small temperatures $E_{n}-E_{0}\gg 1/\beta$, given that $n\ge n_A+1 \ge 2$). As a
consequence, Eq.~\eqref{eq:freezing-2} becomes $\dot{\rho}_{00}\approx0$,
i.e., the ground state population is effectively frozen for $t\ge t_0$. 

The location of the freezing point for the $n$th excited state is determined by the
point where 
the relaxation time for the $n$th excited state, given by $\tau_{\text{rel}}^{(n)} = W_{0n}^{-1}$, becomes longer than the anneal time. As long as just one of these relaxation times, with $n\in\{1,\ldots,n_{A}\}$ is too long, the system will not thermalize. For the system to freeze, i.e., $\dot{\rho}_{00}\approx0$, we need all transitions to cease. Hence, we define the freezing point as the solution $t_0$ of $\min_{n\in\{1,\ldots,n_{A}\}}\tau_{\text{rel}}^{(n)} =t_f$, or:\footnote{A shortcut to deriving Eq.~\eqref{eq:freezing_point} is to interpret $1/\tau_{\text{rel}}^{(n)} = \gamma\left(E_{n}-E_{0}\right)\left|\langle 0|V|n\rangle\right|^{2} = W_{0n}$ as a statement of Fermi's golden rule for the transition rate, with $V = \sum_{j}\sigma_{j}^{z}$ playing the role of the perturbation, and $\gamma$ the density of states.}
\begin{equation}
t_0 := \{\min t\in[0,t_f] \text{ s.t. }\max_{n\in\{1,\ldots,n_{A}\}} W_{0n}(t)=1/t_{f}\}\ .
\label{eq:freezing_point}
\end{equation}
In terms of the control schedule $s(\tau)$, we have $s_0 := s(\tau_0)$, where $\tau_0 = t_0/t_f$. In practice, to determine the freezing point $s_0$ we implement a closely related procedure described in App.~\ref{app:freezing}.

Note that, as a consequence of the perturbative argument, (for sufficiently
small $A$) the rates $W_{0n}$ are decreasing functions of $A$.
So if $A(t)$ is decreasing in $t$, $W_{0n}$ can be considered
zero for $t\ge t^{\ast}$. A similar argument works when substituting $0\leftrightarrow m$,
and one finds that the population in the $m$th excited state, $\rho_{mm}$, is frozen,
albeit with possibly different freezing points than given by Eq.~\eqref{eq:freezing_point}. 

Another consequence of this argument is that the phenomenon of freezing
should be more pronounced (i.e., occur for smaller $t_0$ and resulting in smaller $\dot{\rho}_{00}$ for $t>t_0$) for those problems where the ground state
is separated by a large Hamming distance from the excited states, i.e., a larger tunneling barrier such as the T-gadget compared to the FM-gadget, discussed in Sec.~\ref{sec:8QG} below.
These problems are the ones which are harder to simulate with standard classical
simulated annealing with single spin flip moves~\cite{kirkpatrick_optimization_1983} (cluster flip moves~\cite{PhysRevLett.115.077201} would not necessarily be similarly affected). 

Finally, note that a \emph{simulation} of a master equation to which the considerations above apply, such as the AME (see Sec.~\ref{sec:Theory-AME}), is also expected to exhibit freezing.

\subsection{Boundary cancellation with a ramp at the end}
\label{sec:ramp}

Since the ground state population does not change past the freezing
point, no change in the schedule would be effective if performed
after freezing. In view of these considerations we perform BC before
freezing sets in, which --- for the standard control schedule $s(\tau)=\tau$  (see Fig.~\ref{fig:dw2kq_schedule}) --- happens for $s\approx 0.55$,
depending on the problem. Our strategy will be the following. First,
evolve from $t=0$ to $t_{f}$ with a BC schedule.
To avoid freezing, we ensure that $A(t_{f})\neq0$ at the end of the BC schedule. 
Thus, $t_f$ does \emph{not} correspond to the usual total anneal time for which $A(t_f)=0$. Right after the BC schedule ends, we perform
a linear ramp\footnote{The term ``quench'' is used in the D-Wave documentation instead of ramp~\cite{dwave-manual}.} of duration $t_r = 1\,\mu$s until the schedules reach their final values (in particular $A(t_{f}+t_r)=0$), after which the system is measured in the computational
basis. The state after the entire evolution can be written as 
$\mathcal{E}_{\mathrm{ramp}}\mathcal{E}_{\mathrm{BC}}\rho(0)$,
where $\mathcal{E}_{\mathrm{BC}}$ ($\mathcal{E}_{\mathrm{ramp}}$) is the
evolution through the BC schedule (ramp). However, random local fields
and coupler perturbations [integrated control errors (ICE)] result
in a Hamiltonian that does not behave as intended in the ideal case~\cite{Albash:2019ab},
and these errors have been well documented in the D-Wave processors~\cite{dwave-manual}.
The effect of such random perturbations can be controlled with error
suppression and correction \cite{youngAdiabatic13,pudenzErrorcorrected14,pearsonAnalog19},
which will be employed below for the ferromagnetic chain gadget. Due
to this ICE effect, the measured state is better represented by 
\beq
\rho_{\mathrm{final}}:=\mathsf{E}_{J}\left[\mathcal{E}_{\mathrm{ramp}}\mathcal{E}_{\text{BC}}\rho(0)\right]\ ,
\label{eq:rho_f}
\eeq
where we denoted by $\mathsf{E}_{J}\left[\bullet\right]$ the
average over the noise on the random couplings $J_{ij}$ and fields
$h_{i}$. 

Let us now define $\rho(t_{f}):=\mathcal{E}_{\text{BC}}\rho(0)$, while
$\sigma(t_{f})$ is the Gibbs state at the end of the BC schedule [recall that $\rho(0)= \sigma(0)$],
corresponding to the Hamiltonian in Eq.~\eqref{eq:QA_Hamiltonian}. The
adiabatic theorem in its various forms, including with boundary cancellation,
provides an upper bound on the pre-ramp distance $\|\delta\|_1$, where $\delta:=\rho(t_{f})-\sigma(t_{f})$. Defining 
\beq
\sigma_{\mathrm{final}}:=\mathsf{E}_{J}\left[\mathcal{E}_{\mathrm{ramp}}\sigma(t_{f})\right]\ ,
\label{eq:sigma_f}
\eeq
$\delta$ can be related to $\rho_{\mathrm{final}}$ and $\sigma_{\mathrm{final}}$
via the following bound:
\bes
\label{eq:18}
\begin{align}
\left\Vert \rho_{\mathrm{final}}-\sigma_{\mathrm{final}}\right\Vert _{1} & =\left\Vert \mathsf{E}_{J}[\mathcal{E}_{\mathrm{ramp}}\delta]\right\Vert _{1}\label{eq:bound}\\
 & \le\mathsf{E}_{J}\left[\left\Vert \mathcal{E}_{\mathrm{ramp}}\delta\right\Vert _{1}\right]\label{eq:Jensen}\\
 & \le\mathsf{E}_{J}\left[\left\Vert \delta\right\Vert _{1}\right]\ .
 \label{eq:CPTP}
\end{align}
\ees
Here Eq.~\eqref{eq:Jensen} follows from from Jensen's inequality~\cite{Jensen:1906up} and the fact that every norm is convex (implying $\left\Vert \mathsf{E}_{J}\left[x\right]\right\Vert _{1}\le\mathsf{E}_{J}\left[\left\Vert x\right\Vert _{1}\right]$),
while Eq.~\eqref{eq:CPTP} follows because $\mathcal{E_{\mathrm{ramp}}}$
is a completely positive and trace-preserving (CPTP) map \cite{nielsenQuantum10} (implying $\left\Vert \mathcal{E}_{\mathrm{ramp}}\delta\right\Vert _{1}\le\left\Vert \delta\right\Vert _{1}$).
Note that $\rho_{\mathrm{final}}$ is the empirically measured state,
while $\sigma_{\mathrm{final}}$ differs from the state that we
sought to prepare $\sigma(t_{f})$, by the presence of the extra operations
$\mathsf{E}_{J}\left[\mathcal{E}_{\mathrm{ramp}}\bullet\right]$.

The bound~\eqref{eq:18} implies a similar bound for the ground state probabilities.
Let us define 
\bes
\label{eq:19}
\begin{align}
\label{eq:19a}
P_{\text{GS}} &:=\Tr\left[|\text{GS}\rangle\!\langle \text{GS}|\rho_{\mathrm{final}}\right]\\
\label{eq:19b}
P_{\text{GS}}^{*} &:=\Tr\left[|\text{GS}\rangle\!\langle \text{GS}|\sigma_{\mathrm{final}}\right]\ ,
\end{align}
\ees
where $|\text{GS}\rangle$ is the ground state of the Hamiltonian at the
end of the anneal, i.e., the ground state of $H_{Z}$. 
Since\footnote{Here \unexpanded{$\left\Vert y\right\Vert _{\infty}$} is the operator
norm of $y$, i.e., its maximum singular value, which is $1$ for an orthogonal
projection.} 
\beq
\left|\Tr\left[|\text{GS}\rangle\!\langle \text{GS}|x\right]\right|\le\left\Vert |\text{GS}\rangle\!\langle \text{GS}|\right\Vert _{\infty}\left\Vert x\right\Vert _{1}=\left\Vert x\right\Vert _{1}\ ,\eeq
 it follows that the \emph{adiabatic error} defined as
\begin{equation}
D_{\text{GS}}:=\left|P_{\text{GS}}-P_{\text{GS}}^{*}\right|\ ,
\label{eq:D_GS}
\end{equation}
satisfies
 \beq
D_{\text{GS}} \le \left\Vert \rho_{\mathrm{final}}-\sigma_{\mathrm{final}}\right\Vert _{1}\ .
 \label{eq:16}
 \eeq
The adiabatic error quantifies the difference between the ground state overlaps of the experimentally measured state ($\rho_{\mathrm{final}}$) and the Gibbs state  ($\sigma_{\mathrm{final}}$).

\subsection{An adiabatic error bound that combines everything}
\label{sec:everything}

Combining Eqs.~\eqref{eq:18} and~\eqref{eq:16} with Props.~\ref{prop:BCT} and~\ref{prop:BC_gapless_end} and our numerical evidence,
we obtain 
\begin{align}
\label{eq:21}
D_{\text{GS}} \le \left\Vert \rho_{\mathrm{final}}-\sigma_{\mathrm{final}}\right\Vert _{1} &\le \frac{C}{t_{f}^{\eta}}\ ,
\end{align}
where $C$ is now the noise averaged constant ($\mathsf{E}_{J}\left[\bullet\right]$) and $\eta$ depends
on the physical assumptions. Namely, $\eta=k+1$ if the Liouvillian
either has a non-zero spectral gap throughout the anneal or if BC is enforced after the Liouvillian gap has already closed somewhere along the anneal, while $\eta=(k+1)/(k\alpha+\alpha+1)$
if the Liouvillian gap closes at the same point at which BC is enforced (Prop.~\ref{prop:BC_gapless_end}). 

Of course we may reformulate Eq.~\eqref{eq:21} as a bound on $P_{\text{GS}}$:
\beq
P_{\text{GS}}^* - \frac{C}{t_{f}^{\eta}} \leq P_{\text{GS}} \leq P_{\text{GS}}^* + \frac{C}{t_{f}^{\eta}}\ .
\label{eq:PGS-bound}
\eeq

\subsection{Anomalous heating}
\label{sec:anom-heat}

Our discussion so far has assumed that the effective temperature of the system remains constant as a function of both the anneal parameter $\tau = t/t_f$ and the total anneal time $t_f$. However, there is evidence to suggest that the latter is in fact not the case in the D-Wave devices, i.e., the temperature is $t_f$-dependent. The reason is an unintentional but omnipresent high-energy photon flux that enters the D-Wave chip from higher temperature stages through cryogenic filtering, which accumulates over long anneal times and manifests as an effectively higher on-chip temperature~\cite{anomalous-heating}. This anomalous heating phenomenon will hinder our ability to test the BCP, since it means that in fact $P_{\text{GS}}^*$ is a function of $t_f$, which complicates testing the BCT prediction as summarized by Eq.~\eqref{eq:PGS-bound}. Indeed, for a Gibbs state $\sigma_{\text{final}} = e^{-\beta H_S}/Z$ (with $\beta$, $H_S$, and the partition function $Z$ all evaluated at $t=t_f$), expanding $H_S$ in its eigenbasis as $H_S =\sum_{i=0} E_i \ketb{i}{i}$ we readily find
\beq
P_{\text{GS}}^* = \langle \text{GS}|\sigma_{\mathrm{final}}|\text{GS}\rangle = \frac{1}{1+\sum_{i=1}e^{-\beta(t_f)\Delta_{i}}} \ ,
\eeq
where $\ket{0}$ is the ground state and $\Delta_{i}:= E_i-E_0$.

Assuming for simplicity that $\beta(t_f) = \beta_0 + a/t_f$, i.e., a temperature that depends linearly on $t_f$ with rate $a>0$, and that $1 \ll \beta(t_f)\Delta_1 \ll \beta(t_f)\Delta_i$ for $i\ge 2$, we can write this as 
\beq
P_{\text{GS}}^* \approx 1 - e^{-\beta(0)\Delta_1}e^{-a\Delta_1/t_f} \ ,
\eeq
i.e., the algebraic scaling with $t_f$ of Eq.~\eqref{eq:PGS-bound} becomes obscured by an exponential scaling due to $P_{\text{GS}}^*$.

However, in reality we do not know the functional form of $\beta(t_f)$, and the assumption $1 \ll \beta(t_f)\Delta_1 \ll \beta(t_f)\Delta_i$ may not hold. Therefore, in the analysis of our experimental results in Sec.~\ref{sec:results} below, we instead use an ansatz of the form 
\beq
P_{\text{GS}} = \bar{P}_{\text{GS}}^* + \frac{C'}{t_{f}^{\eta'}}\ ,
\label{eq:PGS-fit}
\eeq
where $\bar{P}_{\text{GS}}^*$ is a free fitting parameter representing an averaged value of the true (unknown) $P_{\text{GS}}^*(t_f)$,  $C'$ becomes another fitting parameter which already accounts for noise averaging as explained in Sec.~\ref{sec:everything}, and 
$\eta'$ plays the role of the effective scaling exponent, i.e., our proxy for $\eta$ in Eq.~\eqref{eq:PGS-bound}.

\section{Methods}
\label{sec:methods}

\subsection{Boundary Cancellation Protocol (BCP)}
\label{sec:bcp}

\begin{figure}[t]
\includegraphics[width=0.95\columnwidth]{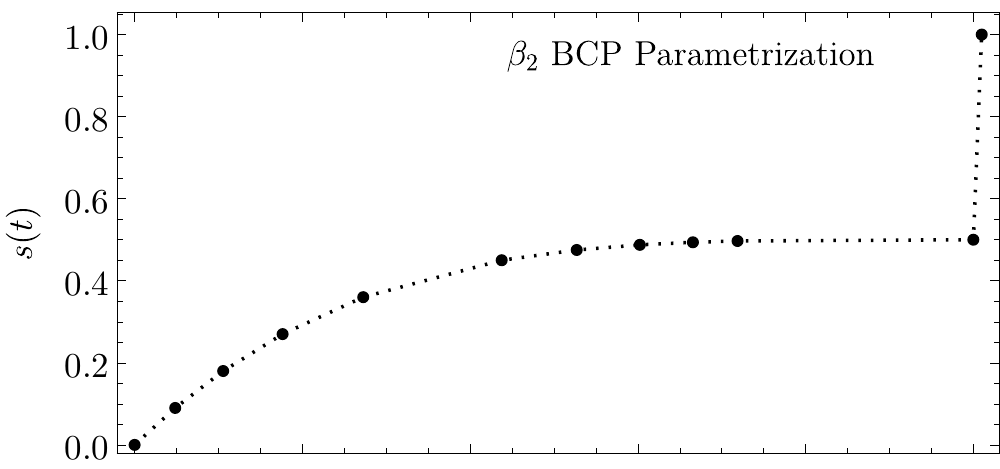}
\includegraphics[width=0.95\columnwidth]{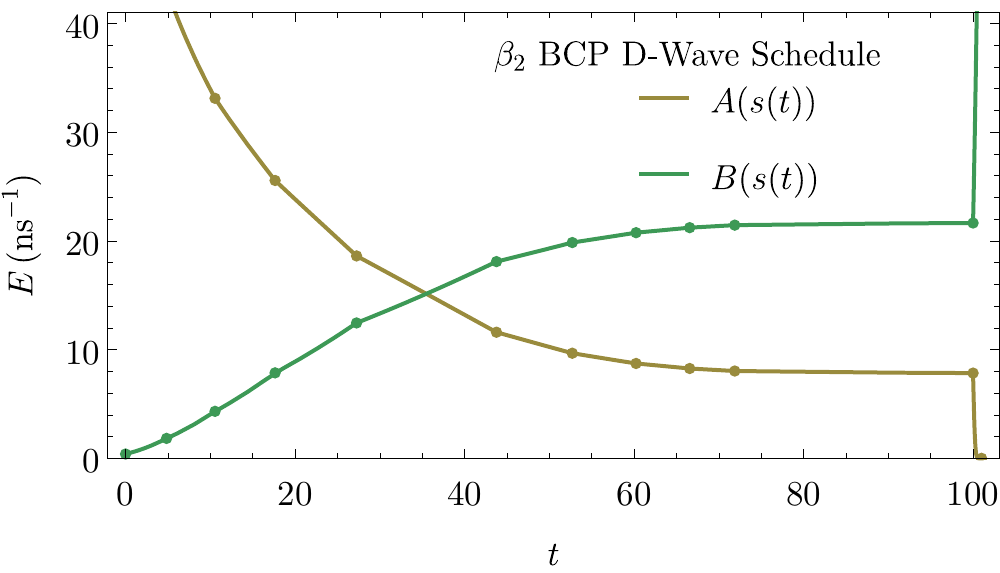} 
\caption{An example of the D-Wave boundary cancellation protocol with the $\beta_{2}$
schedule ramped at $s_{\text{BC}}=0.50$, with the constructed parametric
schedule (top) and the corresponding physical schedules (bottom).
For the schedule shown here we chose an anneal time of $t_{f}=100\,\mu$s. The ramp duration is $t_r=1\mu$s. Contrast with the native schedule of the DW-LN processor shown in Fig.~\ref{fig:dw2kq_schedule}. The most significant impact is on the $B(s)$ schedule, which is not natively flat, in contrast to the $A(s)$ schedule, which natively approaches $0$ in a very flat manner.}
\label{fig:sched}
\end{figure}

\begin{figure}
\centering \subfigure[\ ]{\includegraphics[height=1.84in]{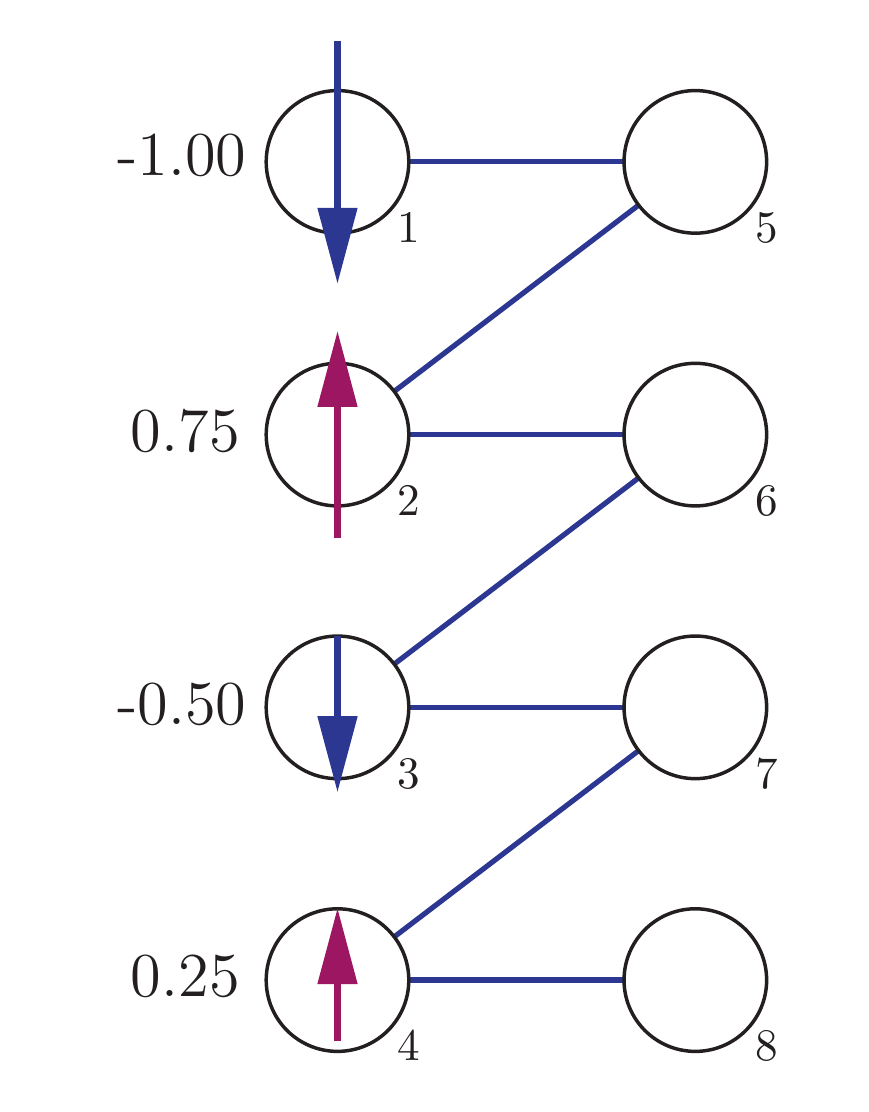}}
\hspace{0.1in} \subfigure[\ ]{\includegraphics[height=1.8in]{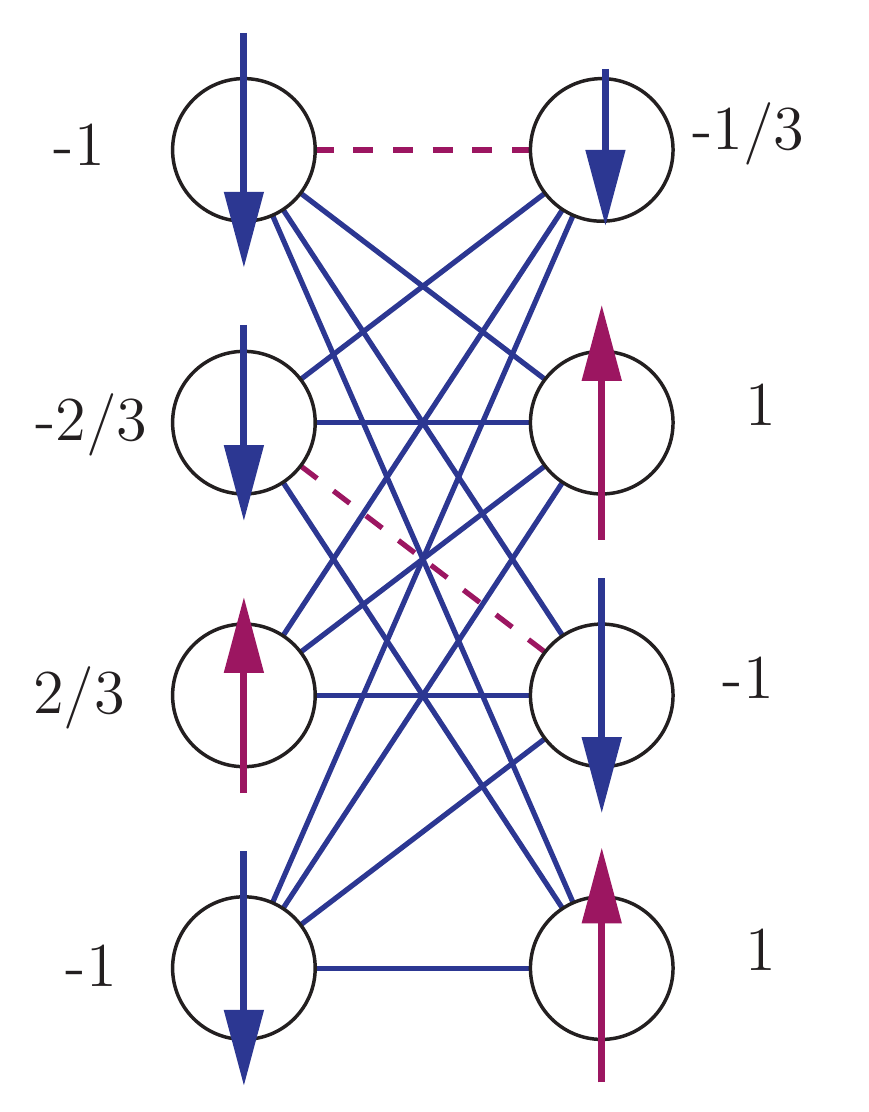}}
\caption{Illustration of the FM-gadget (a) and T-gadget (b) embedded into the
Chimera architecture as represented by an Ising Hamiltonian $H_Z$ [Eq.~\eqref{eq:HXHZ}]. Ferromagnetic
(blue) and anti-ferromagnetic (dashed purple) couplings all have the
same strength of $J_{ij}=-1$ and $J_{ij}=1$, respectively. Local fields $h_i$ are indicated by the
arrows and their value beside them. The ordering of the qubits in
the Chimera architecture is also enumerated in (a).}
\label{fig:fm_chain}
\end{figure}

\begin{figure*}
\centering \subfigure[\ FM-gadget]{\includegraphics[width=0.46\textwidth]{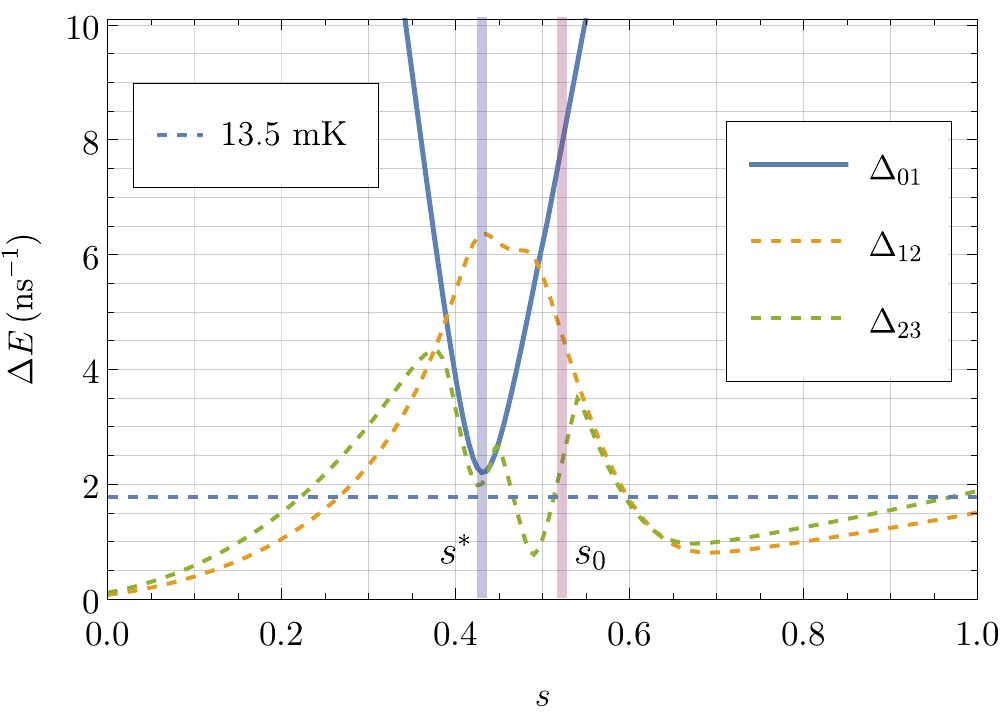}\label{fig:fm_gaps-a}}
\hfill{}\subfigure[\ T-gadget]{\includegraphics[width=0.46\textwidth]{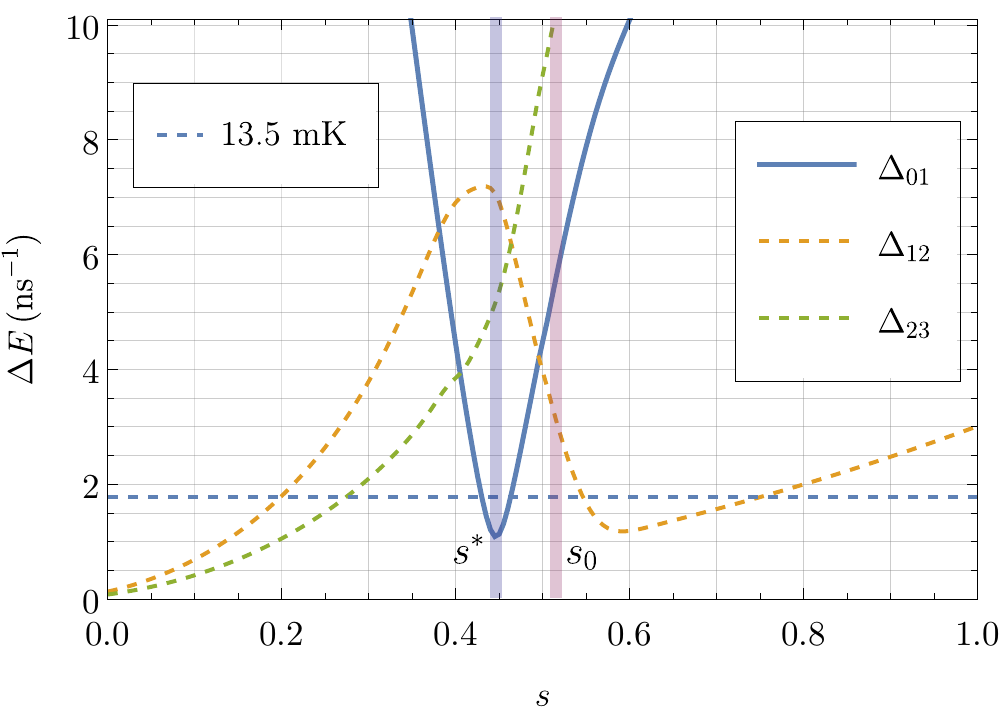}\label{fig:fm_gaps-b}}
\caption{Spectral gaps of (a) the FM-gadget and (b) the T-gadget (with $\gamma=0.5$),
as the schedule is varied according to typical physical schedules
$A(s)$ and $B(s)$ of the D-Wave processor (App.~\ref{app:betasched}),
shown in units of $\mathrm{ns}^{-1}$ ($\hbar=1$). The notation $\Delta_{n,n+1}$
denotes the energy gap between the $n$th and $n+1$th Hamiltonian
eigenstates, with $n=0$ being the ground state. The $13.5\,\mathrm{mK}$
energy scale is also shown, which is the reported dilution fridge temperature of the
DW-LN processor. The minimum gap location is $s^{*}=0.43$ for the
FM-gadget and $s^{*}=0.44$ for the T-gadget. The minimum gap region
is marked by the light blue shading, while the estimated start of
frozen dynamics at $s_0$ is marked by pink shading,
found by solving Eq.~\eqref{eq:freezing_point} (see App.~\ref{app:freezing} for details).
Note that for both gadgets, the first, second, and third excited states
are all initially degenerate in the subspace of a single $\ket{-}$
excitation of the transverse-field ground state. The nonzero excited
state gaps $\Delta_{12}$ and $\Delta_{23}$ of the FM gadget ($\Delta_{12}$ only for the T gadget) at $s=1$ are the result
of symmetry breaking due to a crosstalk term we included in the
Hamiltonian, of strength $\chi = 0.02$ (see App.~\ref{app:xtalk}). The avoided level
crossing of the T-gadget is slightly narrower than that of the FM-gadget.
}
\label{fig:fm_gaps}
\end{figure*}

We performed most of our experiments with the D-Wave 2000Q low noise (DW-LN) processor
accessed through D-Wave Leap. We also performed additional experiments
with the D-Wave 2000Q processor at the NASA Quantum Artificial Intelligence
Laboratory (DW-NA) as well as the D-Wave Advantage (DWA) processor
through D-Wave Leap. 
In the processor specifications, the hardware-determined schedules
$A(s)$ and $B(s)$ are parametrized by the user-defined control schedule
$s(\tau)$. In the standard case the control is linear, i.e., $s(\tau)=\tau$
but in general $s(\tau)$ can be programmed in a piecewise linear
manner as a function of time. The processor permits a maximum of $12$
points to specify the piecewise linear function $s(\tau)$. We take
advantage of this enhanced capacity to approximate a BC schedule.
 The allowed range of programmable anneal times is $t_{f}\in[1,2000]\,\mu$s.

Even though $A(s)$ and $B(s)$ themselves need not satisfy the vanishing
derivative requirement of the BCP, it follows from the chain rule that
$A(s(t))$ and $B(s(t))$ do, as long as the control schedule $s(t)$
satisfies this requirement and $A(s)$ and $B(s)$ are differentiable
to the same order as $s(t)$. To be concrete, we take 
\begin{align}
s(\tau)=\beta_{k}(\tau)\ ,\qquad\tau=t/t_{f}\ ,\label{eq:s=00003Dbeta}
\end{align}
where 
\bes
\begin{align}
\beta_{k}(\tau) &:=\frac{\mathrm{B}_{\tau}(1,k+1)}{\mathrm{B}_{1}(1,k+1)} \\
 &= 1-(1-\tau)^{k+1}  \label{eq:beta_kt}
\end{align}
\ees
is the regularized incomplete beta function of order $k$~\cite{RPL:10,camposvenutiError18},
with
\begin{equation}
\mathrm{B}_{x}(a,b):=\int_{0}^{x}y^{a-1}(1-y)^{b-1}dy
\label{eq:ibf}
\end{equation}
being the incomplete beta function. As is apparent from Eq.~\eqref{eq:beta_kt}, 
the function $\beta_{k}(\tau)$
has exactly $k$ vanishing $\tau$-derivatives at $\tau=1$,
as required by the BCT. We refer to this class of control schedules
simply as \emph{beta schedules}. 

As we discussed above, in contrast to the theoretical setup of Ref.~\cite{camposvenutiError18},
due to freezing there is no practical advantage to flattening the
schedule as $s$ approaches $1$. The effectiveness of the BCP is
most apparent when the flattening of the schedule occurs after the
avoided level crossing of the anneal (at $s=s^*$), but before the dynamics freeze (at $s=s_0$).
Thus, rather than using the standard schedules $\{A(s),B(s)\}$ with
linear $s(\tau)$ that freeze out around $s\simeq0.5$, the beta schedule
needs to be adjusted so that it terminates at an appropriate point $s_{\text{BC}}$ corresponding to $t=t_f$ and satisfying
$0< s^* < s_{\text{BC}}<s_0\le 1$, rather than always at $s=1$, when the open system
dynamics are already frozen out. The anneal is then ramped to the final values $A(t_f+t_r)=0$ and $B(t_f+t_r)$. The precise construction of the
control schedule is given in App.~\ref{app:betasched}. 

Note that the piecewise linear approximation of Eq.~\eqref{eq:s=00003Dbeta}
cannot exactly satisfy the requirement of vanishing derivatives at
the end. However, Ref.~\cite[Prop.~3]{camposvenutiError18}
shows that (at least for the case $k=1$) if one tries to enforce
a vanishing derivative but achieves only approximate vanishing, the
overall adiabatic error is diminished nonetheless. A reasonable expectation
is that the scaling will be milder than the theoretical ideal of $1/t_{f}^{k+1}$ for Liouvillians with a spectral gap (Prop.~\ref{prop:BCT}), i.e., will
be of the form $1/t_{f}^{\eta}$ with $\eta<k+1$ for schedules that
only approximate the desired $\beta_{k}$ schedule. An example of
this schedule construction is shown in Fig.~\ref{fig:sched}.

\subsection{8 Qubit Gadgets}
\label{sec:8QG}

We briefly describe the two gadgets utilized in our experiments, which are both illustrated in Fig.~\ref{fig:fm_chain} along with their couplings and bias values in the Hamiltonian $H_Z$ in Eq.~\eqref{eq:HXHZ}.
The first gadget we use is an 8-qubit ferromagnetic (FM) chain,  depicted in Fig.~\ref{fig:fm_chain}  using its embedding in a single unit cell of the Chimera graph in the D-Wave 2000Q processor, which is a bipartite $K_{4,4}$ graph.
The local fields in this gadget were adjusted to break the most significant degeneracies in the ground subspace and first excited subspace that are present in a simple ferromagnetic chain without any fields. 
This gadget thus enforces a large Hamming distance between its ground
state and any first exited state without introducing additional two-body interactions. The remaining degeneracy in the first excited state is broken by crosstalk interactions in the device, which we model and include in simulations of the Hamiltonian (see App.~\ref{app:xtalk}).
In addition, it is simple to tile and embed the QAC encoding of the FM-gadget (see below) on the Chimera graph.

The second gadget we utilize is the one introduced in Ref.~\cite{AlbashDemonstration18}.
This gadget was thoroughly analyzed there and established to pose
a small-gap tunneling barrier between the ground state and the first
excited state, so we refer to it as the ``T-gadget''. For finer
control of the energy gap, the T-gadget is scaled to $\gamma=0.5$
[recall Eq.~\eqref{eq:QA_Hamiltonian}]. The ground state of
both gadgets is the all-$0$ state, denoted by $|\mathbf{0}\rangle$.
Both gadgets also share the all-$1$
state $|\mathbf{1}\rangle$ as their first excited state.

The spectral gaps of the gadgets are shown in Fig.~\ref{fig:fm_gaps}.
The most crucial distinction between them is that the FM-gadget has
a minimum gap $\Delta_{01}(s^*)$ to the first excited state that is larger than
the dilution fridge operating temperature, while the opposite holds for the T-gadget.
Figure~\ref{fig:fm_gaps} suggests that implementing the BCP with the termination point $s_{\text{BC}}$ somewhere in the range
of $[0.45,0.50]$ will be the most effective for both gadgets, since $s^*\approx 0.43$ and $s_0\approx 0.5$. Before
this range, the BCP spends more time slowing down before the minimum
gap, while after this range it goes through the minimum gap too quickly
and only slows down going into the freezing phase.

\subsection{Encoding with Quantum Annealing Correction}

\label{sec:QAC}

We supplement the BCP with quantum annealing correction (QAC) \cite{pudenzErrorcorrected14}
and compare the scaling improvement versus QAC with a simple linear
anneal. The QAC encoding of the computational Hamiltonian in \eqref{eq:QA_Hamiltonian}
has the form 
\begin{equation}
H_{\text{QAC}}=\overline{H}_{Z}+\lambda H_{P}\label{eq:QAC}
\end{equation}
where $\overline{H}_{Z}$ is an $m$-qubit repetition of the problem
Hamiltonian $H_{Z}$ wherein each $\sigma_{i}^{z}$ and $\sigma_{i}^{z}\sigma_{j}^{z}$
term is replaced by its encoded counterpart $\overline{\sigma_{i}^{z}}=\sum_{\ell=1}^{m}\sigma_{i_{\ell}}^{z}$
and $\overline{\sigma_{i}^{z}\sigma_{j}^{z}}=\sum_{\ell=1}^{m}\sigma_{i_{\ell}}^{z}\sigma_{j_{\ell}}^{z}$,
respectively, and the penalty Hamiltonian imposes an energy cost by
ferromagnetic coupling to an ancilla penalty-qubit, 
\begin{equation}
H_{P}=-\sum_{i}(\sigma_{i_{1}}^{z}+\ldots+\sigma_{i_{m}}^{z})\sigma_{i_{p}}^{z}.
\end{equation}
Here the index $i$ is over the logical qubits, $i_{1}\ldots i_{m}$
are the physical qubits corresponding to the logical qubit $i$, and
$i_{p}$ is the ancilla for qubit $i$. A majority vote over the results
of the measurement on the computational basis of the (odd) $m$ physical
qubits decides the spin value of logical qubit $i$. This strategy
of combining an energy penalty against errors anticommuting with $H_{P}$
(the sum of the stabilizer generators of the bit-flip code)~\cite{jordan2006error,Marvian:2014nr,Bookatz:2014uq,Jiang:2015kx,Marvian-Lidar:16,Marvian:2017aa,Lidar:2019ab},
along with a bit-flip repetition code, has been demonstrated numerous
times to be successful at enhancing the performance of quantum annealing,
despite not being able to protect against pure dephasing errors (which
commute with $H_{P}$)~\cite{pudenzQuantum15,Mishra:2015,MNAL:15,Vinci:2015jt,Matsuura:2016aa,Pearson:2019aa}.
We only consider $m=3$ QAC encoding, which is naturally embedded
onto the Chimera graph of all the DW2KQ processors. As the encoded FM-gadget
can be naturally embedded with no additional minor embedding, we restrict
our attention to it, and do not consider the T-gadget under QAC, which
would require minor embedding. For the same reason we do not consider
the more powerful nested QAC strategy~\cite{vinci2015nested,Vinci:2017ab,Matsuura:2018,Li:2020aa},
which is left for future work. We estimate the optimal penalty strength
$\lambda$ for QAC by finding the value that 
maximizes
the ground state probability of a simple linear anneal. 

\begin{figure*}
\centering \subfigure[\ %
FM-gadget -- $P_{\text{GS}}$]%
{\includegraphics[width=0.87\columnwidth]{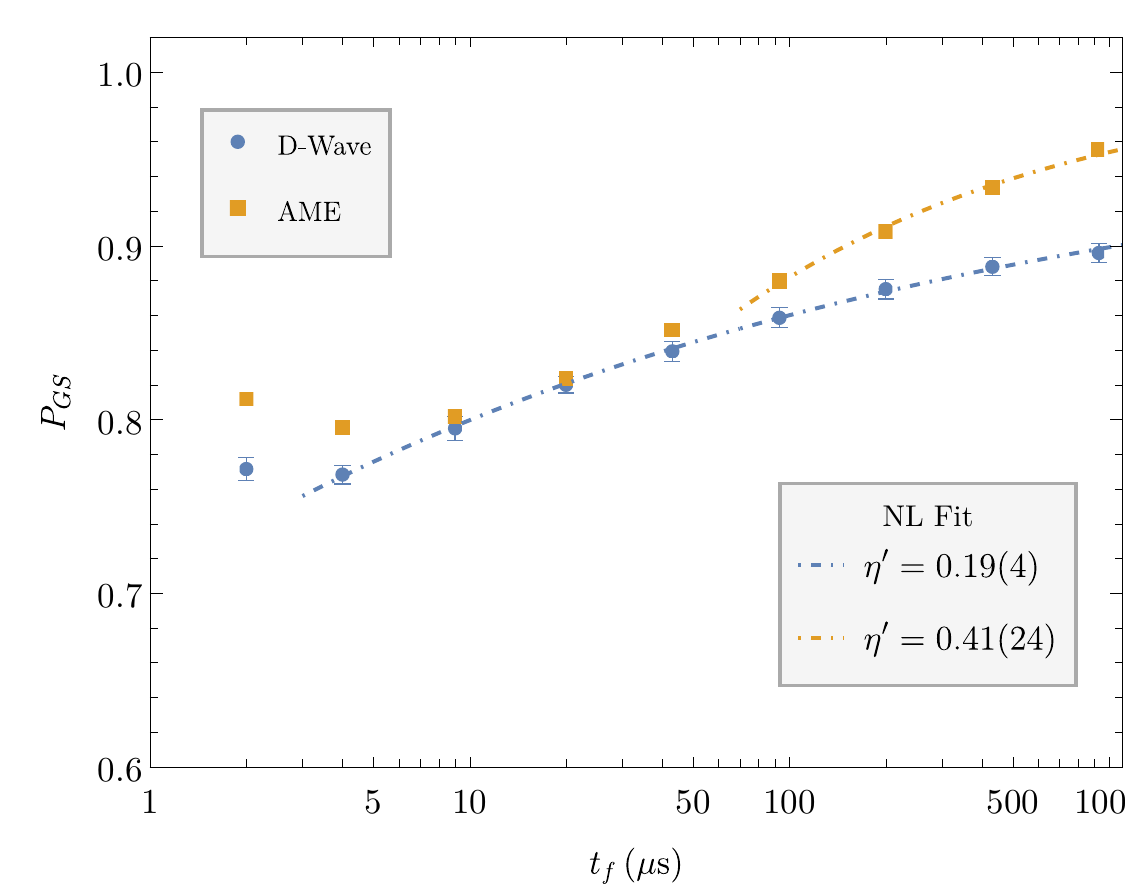}\label{fig:dwsim_lin-a}}
\centering \subfigure[\ %
FM-gadget -- $D_{\text{GS}}$]%
{\includegraphics[width=0.87\columnwidth]{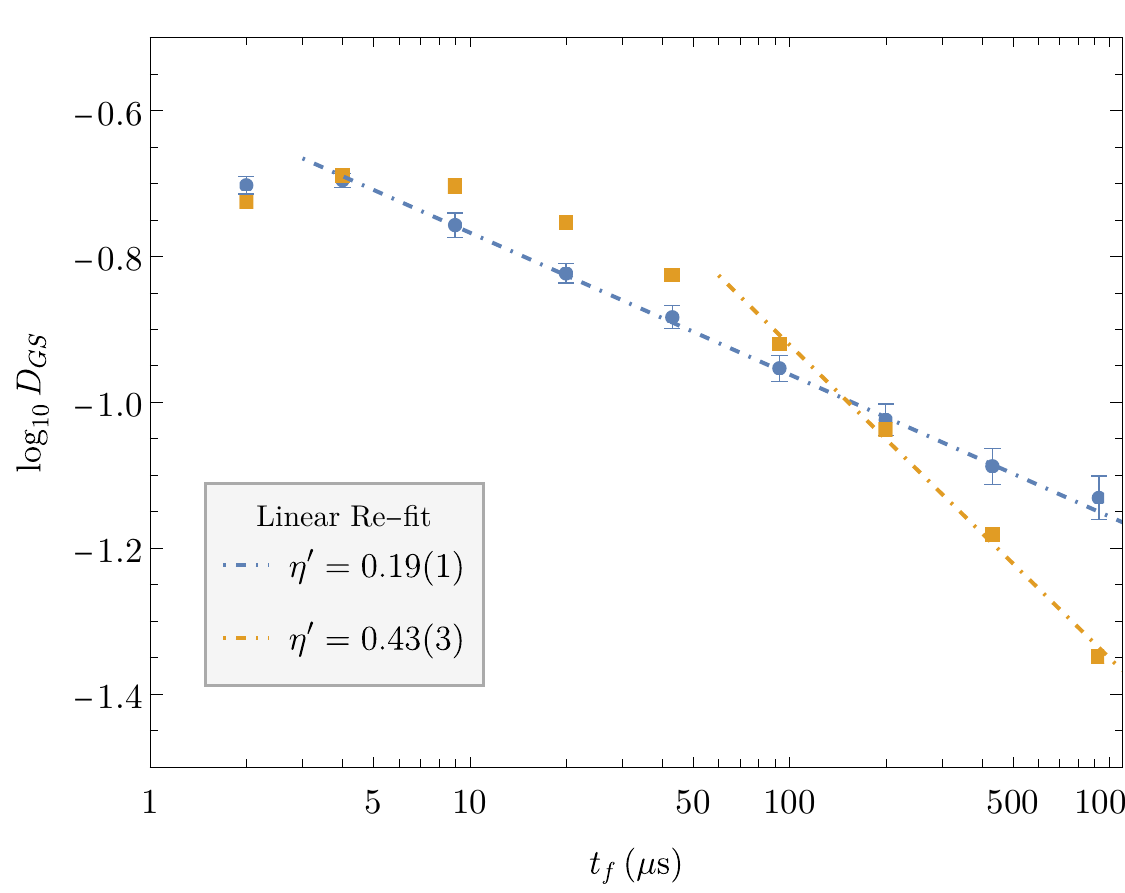}\label{fig:dwsim_lin-b}}
\centering \subfigure[\ %
T-gadget -- $P_{\text{GS}}$]%
{\includegraphics[width=0.87\columnwidth]{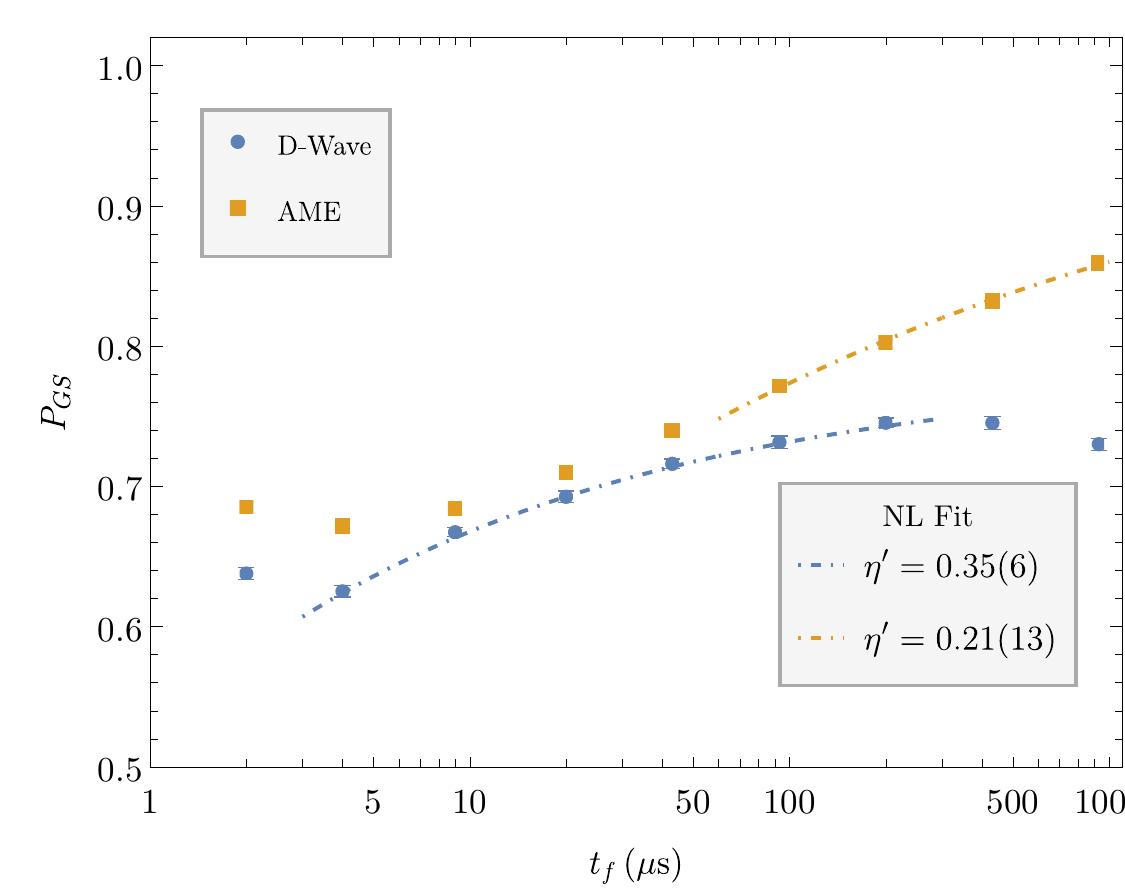}\label{fig:dwsim_lin-c}}
\centering \subfigure[\ %
T-gadget -- $D_{\text{GS}}$]%
{\includegraphics[width=0.87\columnwidth]{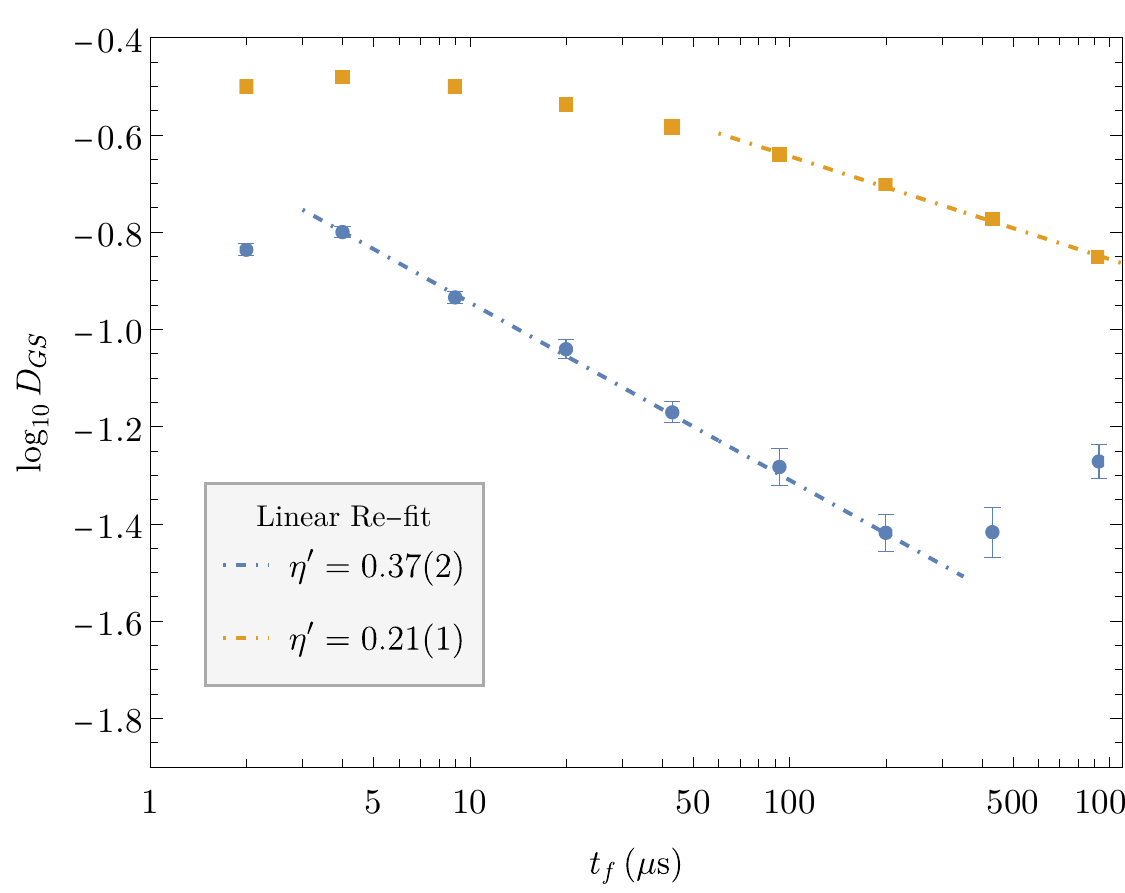}\label{fig:dwsim_lin-d}}
\caption{%
    DW-LN and AME results for $P_{\text{GS}}$ with a simple linear control schedule (without a ramp) for the FM-gadget [(a) and (b)] and the T-gadget [(c) and (d)]. 
    A non-linear fit to \eqref{eq:PGS-fit} is applied on the data and yields estimates for $\eta'$ as described in App.~\ref{app:data}.
    The initial constrained nonlinear fit to  \eqref{eq:PGS-fit} yields values for $\eta'$ and $\bar{P}_{\text{GS}}^*$ that may have high variance.  
    A linear refit on $D_{\text{GS}}$ is done assuming $P_{\text{GS}}^* = \bar{P}_{\text{GS}}^*$ to refine $\eta'$ with a smaller error.
    The values of $\eta'$ and their associated uncertainties we find in this manner are reported in the legends. 
    Error bars on the measured ground state probabilities represent $95\%$ confidence intervals.
    AME simulation parameters for both gadgets: $\eta_0 g^{2}=5.0\times10^{-4}$,
$T=13.5$mK (see Sec.~\ref{sec:Theory-AME}) and crosstalk strength $\chi=0.02$ (see App.~\ref{app:xtalk}). 
}
\label{fig:dwsim_lin}
\end{figure*}

\section{Results}
\label{sec:results}

In this section we report on our results for $P_\text{GS}$, arising from both runs on the DW-LN and numerical simulations. As explained in Sec.~\ref{sec:anom-heat}, we use Eq.~\eqref{eq:PGS-fit} to fit our empirical results. 
Our data collection procedure is described in App.~\ref{app:data}.

\begin{figure*}
\centering \subfigure[\ $s_{\text{BC}}=0.45$]{\includegraphics[width=0.329\textwidth]{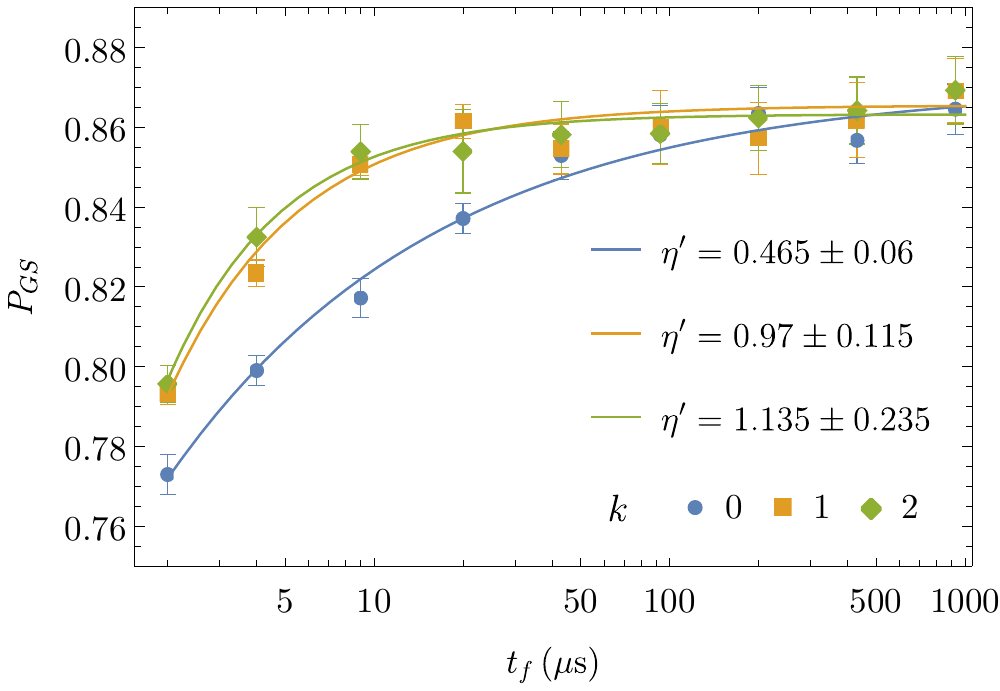}\label{fig:6a}}
\subfigure[\ $s_{\text{BC}}=0.47$]{\includegraphics[width=0.329\textwidth]{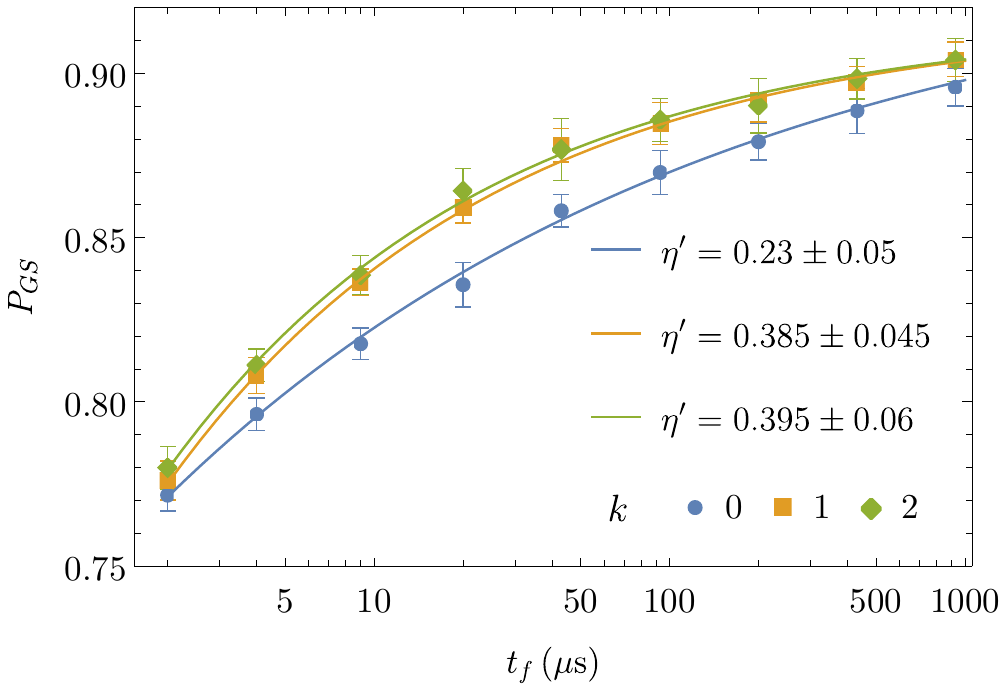}\label{fig:6b}}
\subfigure[\ $s_{\text{BC}}=0.50$]{\includegraphics[width=0.329\textwidth]{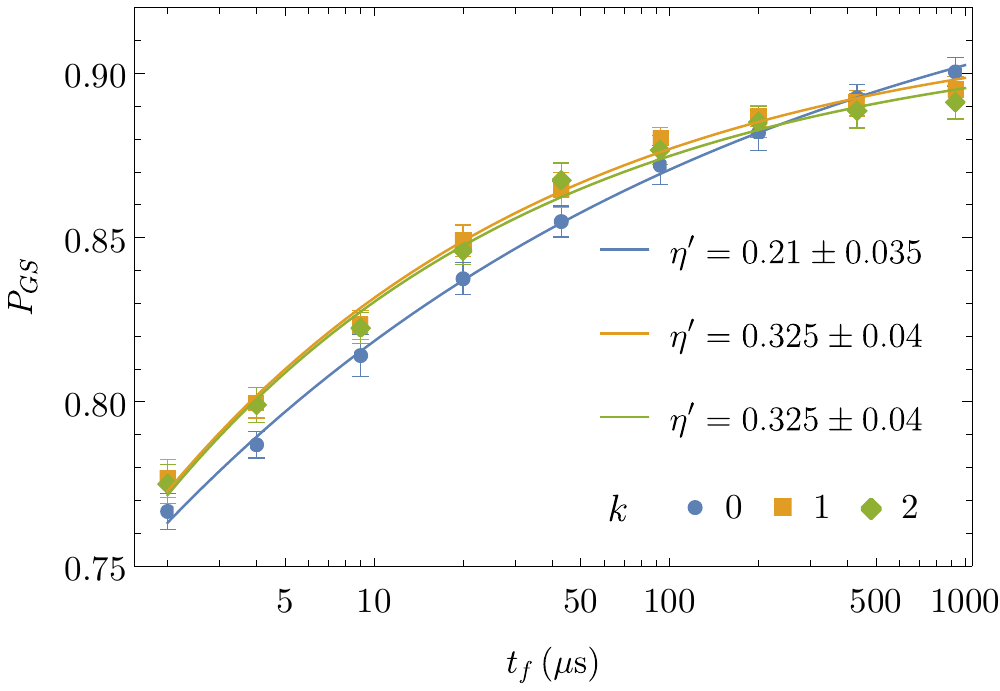}\label{fig:6c}}
\subfigure[\ $s_{\text{BC}}=0.45$]{\includegraphics[width=0.329\textwidth]{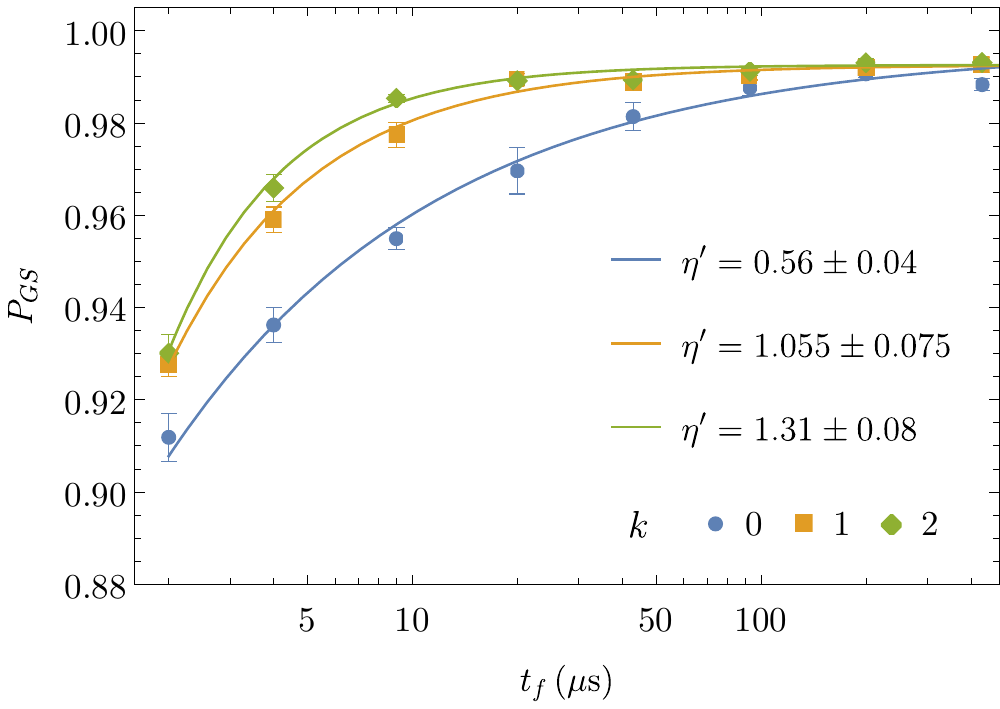}\label{fig:6d}}
\subfigure[\ $s_{\text{BC}}=0.50$]{\includegraphics[width=0.329\textwidth]{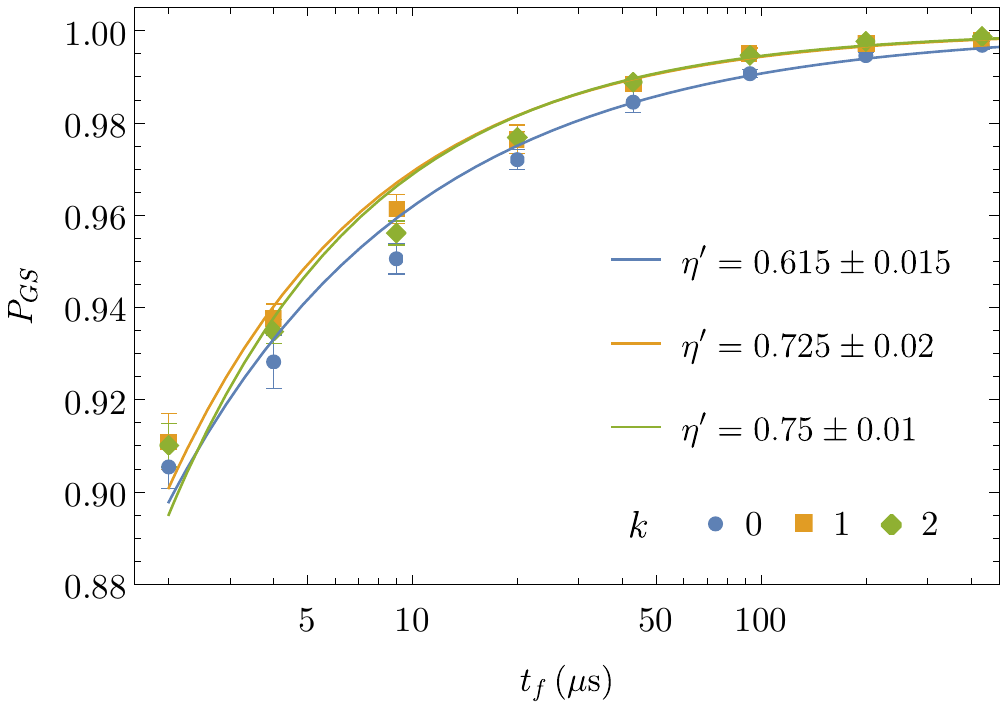}\label{fig:6e}}
\caption{DW-LN adiabatic distance scaling fits for the FM-gadget with BCP at
(a) $s_{\text{BC}}=0.45$, (b) $s_{\text{BC}}=0.47$, and (c) $s_{\text{BC}}=0.50$ past the
minimum gap location of $s^{*}\approx0.43$ but before freeze-out at $s_0\approx0.51$.
(d) and
(e): Adiabatic distance scaling of QAC encoded FM-gadget with BCP
at $s_{\text{BC}}=0.45$ and $s_{\text{BC}}=0.50$ (see Sec.~\ref{sec:QAC-results}).
Symbols represent the experimentally measured data points, solid lines are fits to
$1/t_{f}^{\eta'}$. The $\eta'$ values reported in the legends an well as other parameters are obtained from nonlinear fits to the measured data as explained in App.~\ref{app:data}. Error bars represent $95\%$ confidence intervals.
}
\label{fig:fm_sq_scal}
\end{figure*}

\subsection{Linear control schedule ($k=0$)}
\label{sec:lcs}

We first consider the FM and T-gadgets under the native $A(s)$ and $B(s)$ schedules
with linear control $s(\tau)=\tau$ and no ramp ($s_{\text{BC}}=1$), i.e., a standard quantum annealing protocol. This means that the anneal is stopped deep inside the frozen regime, i.e., freezing is expected to have a noticeable effect. 

In Fig.~\ref{fig:dwsim_lin} we plot, for both the FM-gadget and the T-gadget, the measured ground state probability $P_\text{GS}$
for the DW-LN, along with the same quantity obtained through AME simulations. 
We found $P_{\text{GS}}^*$ for DW-LN by a non-linear fit to the function $C'/t_f^{\eta'} + \bar{P}_\text{GS}^*$, where $C'$ and $\eta'$ are free parameters (left column of Fig.~\ref{fig:dwsim_lin}). In the AME case only the last four data points were used in the fit (since otherwise $t_f$ is too small, as evidenced by the curvature observed for small $t_f$ in the right column), while in the DW-LN case we dropped the first data point for both gadgets for the same reason. In the T-gadget case we also dropped the last two data points, as they are likely corrupted by anomalous heating (see Sec.~\ref{sec:anom-heat}).
    This leads to the values $P_{\text{GS}}^*=0.97$ and $P_{\text{GS}}^*=0.78$ for the FM-gadget [Fig.~\ref{fig:dwsim_lin-a}] and T-gadget [Fig.~\ref{fig:dwsim_lin-c}], respectively.
    For the AME, we took $P_{\text{GS}}^*$ to be $1$, as this is consistent with the BCT for a large gap and the nonlinear fits did not find a smaller value. 
    We then refined the value of $\eta'$ by a linear fit to $\log_{10} (P_{\text{GS}}-P_{\text{GS}}^*)$ as a function of $\log_{10} t_f$, since we expect $P_{\text{GS}}-P_{\text{GS}}^*$ to scale as $C/t_f^\eta$ by Eq.~\eqref{eq:PGS-bound}. Using Eqs.~\eqref{eq:21} and~\eqref{eq:PGS-bound}, we interpret $\log_{10} (P_{\text{GS}}-P_{\text{GS}}^*)$ as a proxy for $\log_{10}D_{\text{GS}}$, plotted in Fig.~\ref{fig:dwsim_lin-b} and Fig.~\ref{fig:dwsim_lin-d} for the FM-gadget and T-gadget respectively. 
    
Several observations are in order from Fig.~\ref{fig:dwsim_lin}. 
\begin{itemize}
\item The $P_{\text{GS}}$ values at the largest $t_f$ are larger for the FM-gadget than for the T-gadget, in both the AME and DW-LN cases. This is consistent with the larger FM-gadget gap (see Fig.~\ref{fig:fm_gaps}) and the larger tunneling barrier of the T-gadget (quantified by the ground state being separated by a larger Hamming distance from the excited states), which causes an earlier onset of freezing and hence ``locks in'' the smaller ground state population of the T-gadget.
\item The FM gadget AME simulation has a larger $\eta'$ ($0.43$) than the T-gadget ($0.21$). This signifies a faster convergence to the adiabatic limit of $P_{\text{GS}}=1$.  This too is consistent with the larger FM-gadget gap and the larger tunneling barrier of the T-gadget. However, the opposite holds for the DW-LN $\eta'$ values ($0.19$ \textit{vs} $0.37$, for the FM and T-gadgets, respectively).
\item The AME data for the T-gadget is in closer agreement with the power-law bound of the adiabatic theorem than the FM-gadget. This is evidenced by the good agreement with the linear fit in Fig.~\ref{fig:dwsim_lin-d}, but the persistence of some curvature at large $t_f$ in Fig.~\ref{fig:dwsim_lin-b}. We may speculate that the FM-gadget will transition to a power-law behavior at even larger $t_f$ than we were able to simulate, though one should keep in mind that in any case, the power-law bound of the adiabatic theorem is not tight.
\item For the DW-LN at large $t_f$, we observe an appreciable reduction in $P_{\text{GS}}$ [Fig.~\ref{fig:dwsim_lin-c}], and a concomitant increase in $D_{\text{GS}}$ [Fig.~\ref{fig:dwsim_lin-d}] in the T-gadget case, and a small positive curvature in the FM-gadget case. This suggests that an anomalous excitation (heating) mechanism is at work in the D-Wave case for $t_f \gtrsim 300\mu$s. The source of this anomaly was discussed in Sec.~\ref{sec:anom-heat}.
The anomalous heating phenomenon hinders our ability to test the BCP, since it means that the effective temperature is itself $t_f$-dependent and increasing. As explained above, this is why we treated $P_{\text{GS}}^*$ as a fitting parameter.
\item The DW-LN FM-gadget scaling [Fig.~\ref{fig:dwsim_lin-b}] is much closer to the expected $D_{\text{GS}} \sim 1/t_f^\eta$ than the T-gadget [Fig.~\ref{fig:dwsim_lin-d}], which is obscured by anomalous heating. It is unclear why this phenomenon should more severely impact the T-gadget. 
\end{itemize}

We report the values of $\eta'$ extracted from our fits in Fig.~\ref{fig:dwsim_lin}.
To understand what value of $\eta'$ to expect, note that according to our model of the D-Wave annealer and in view of Proposition
\ref{prop:degeneracy}, the
Liouvillian gap $\Delta(\tau)$ must close at the end of the anneal. Recall that $\Delta\simeq v(\tau-\tau_{0})^{\alpha}$ with $\tau_0=1$ (or $t=t_f$) being the end of the anneal. 
We do not know the associated value of the exponent $\alpha$; 
however, given that $\mathcal{L}$ is a normal operator, if the Liouvillian depends smoothly on the schedule in a small interval around $t_{f}$, 
a standard result \cite{kato_perturbation_1995} implies that $\alpha$ must
be an integer. On the other hand, $\alpha=1$ is incompatible
with the fact that the lowest eigenvalue of $\mathcal{L}$ is zero,
and that the gap must close smoothly. 
The value observed in the single qubit case is $\alpha=2$~\cite{venutiAdiabaticity16}. According to Prop.~\ref{prop:BC_gapless_end} and more specifically Eq.~\eqref{eq:eta_gapless_end} with $k=0$ (due to the linear schedule), this would correspond to
an exponent $\eta=1/3$, which is also the scaling observed in the
single qubit case~\cite{venutiAdiabaticity16}. 

However, as seen in Fig.~\ref{fig:dwsim_lin-d}, the AME T-gadget case has $\eta' = 0.21(1)$, which -- when equated with $1/(\alpha+1)$ according to Eq.~\eqref{eq:eta_gapless_end} with $k=0$ -- yields $\alpha=4$.\footnote{One might be tempted to associate the flatness of the native $A(s)$ schedule (see Fig.~\ref{fig:dw2kq_schedule}) with the enforcement of a BC-type schedule with a $k\gg 0$ value, which would in turn correspond to $\eta=1/\alpha$ according to Eq.~\eqref{eq:eta_gapless_end}. However, the native $B(s)$ schedule is not flat at all, and it dominates the end of the anneal.} This larger value of $\alpha$ (and corresponding smaller value of $\eta'$) than was found in the single qubit case is not unexpected given the very different spectrum associated with the T-gadget.

As remarked above, in contrast to the T-gadget, the AME value of $\eta'$ for the FM-gadget is at least $0.43$, and is likely to be slightly higher since the largest $t_f$ value for which we were able to run AME simulations appears to be too small for convergence. From Eq.~\eqref{eq:21} we expect $D_{\text{GS}}$ to scale with an exponent $\eta' \ge \eta$, where for $k=0$ we have $1/(\alpha+1) \leq \eta \leq 1/\alpha$ [Eq.~\eqref{eq:eta_gapless_end}]. Thus $\eta'\approx 0.43$ is consistent with these expectations for any integer $\alpha \ge 2$.


The DW-LN scaling is consistent with the expectation $\eta' \ge \eta$ for the T-gadget, for which $\eta'=0.37$. However, anomalous heating makes it difficult to be confident in the extracted $\eta'$ value for this gadget. For the FM-gadget we find a good fit to $D_{\text{GS}} \sim 1/t_f^{\eta'}$ with $\eta'=0.19$, which could be explained by $\alpha=4$, as in the AME T-gadget case. However, the small positive curvature in the DW-LN FM-gadget data at large $t_f$ suggests that the true scaling may be obscured by anomalous heating here as well.

Since anomalous heating plays such a significant role in the T-gadget behavior at large $t_f$, we abandoned it for our experiments involving BCP schedules with $k\ge1$. We focus for the rest of this work on the FM-gadget.


\subsection{BCP schedules with $k\ge1$}

We now turn to actual boundary cancellation schedules. 
In light of our discussion in Sec.~\ref{sec:Theory}, a successful BCP must occur before the closing of the Liouvillian gap and freeze-out, while at the same time the Hamiltonian gap remains large enough to promote relaxation to the ground state.
We thus apply the BCP at a point $s^* < s_{\text{BC}} < s_0$, such that $A(s_{\text{BC}})\neq0$. 
The BC schedule is followed by a ramp of  $1\,\mu\mathrm{s}$ to the
final values of $A$ and $B$. 

Following the same nonlinear fit procedure as in Sec.~\ref{sec:lcs}, we plot the DW-LN $P_{\text{GS}}$ results for the FM-gadget in Fig.~\ref{fig:fm_sq_scal}. More specifically, Fig.~\ref{fig:6a}-\ref{fig:6c} shows the results for $s_{\text{BC}}\in\{0.45,0.47,0.50\}$, while 
Fig.~\ref{fig:6d}-\ref{fig:6e} shows analogous plots after
applying QAC, which we discuss in Sec.~\ref{sec:QAC-results}. 

The first observation that emerges from Fig.~\ref{fig:fm_sq_scal} is that the $\beta_0$ schedule results in a significantly larger (i.e., better) scaling exponent than the $k=0$ schedule of Sec.~\ref{sec:lcs}, for which we found $\eta'=0.19$. The difference between the two is that for the case shown in Fig.~\ref{fig:fm_sq_scal} the schedule includes a ramp at $s=s_{\text{BC}}$. As the ramp point grows from $0.45$ to $0.47$ to $0.50$, the value of $\eta'$ correspondingly declines from $0.47$ to $0.23$ to $0.21$. This confirms that stopping the BC schedule before freeze-out (at $s_0\approx 0.52$) has a significant positive impact on the performance of the BCP.

If the Liouvillian is gapped at the endpoint of the BC schedule, the theoretical expectation for the scaling of the adiabatic error is $\eta=k+1$ for the $\beta_{k}$ schedule (recall Sec.~\ref{sec:everything}). Instead, Figure~\ref{fig:fm_sq_scal} shows that the measured exponents $\eta'$ exhibit a strong dependence on the value of $s_{\text{BC}}$. The most favorable case is $s_{\text{BC}}=0.45$ [Fig.~\ref{fig:6a}], for which the exponents $\eta'$ are $\{0.47\ (k=0),\ 0.97\ (k=1),\ 1.1\ (k=2)\}$, i.e., roughly half that value for $k=0,1$. The $k=2$ exponent is statistically indistinguishable from that for $k=1$. It is likely that this is due to the fact that the relatively small number of schedule interpolation points does not allow us to approximate the ideal $\beta_2$ schedule sufficiently well (see Sec.~\ref{sec:highprecBCP} below).
Other sources of error, such as ICE and anomalous heating, may also play a role.

While this level of agreement with the BCT for the gapped Liouvillian case is far from impressive, the dependence of $\eta'$ on $k$ is less consistent with the BCT prediction
for when the Liouvillian
gap closes at $s_{\text{BC}}$; in this case Prop.~\ref{prop:BC_gapless_end} for $\alpha=1$
yields $\eta=\{0.5,\,0.66,\,0.75\}$ for $k=0,1,2$ with an asymptotic
value of $1$ for a flat schedule ($k\to\infty$), while using $\alpha=2$
yields $\eta=\{0.33,\,0.4,\,0.42\}$ for $k=0,1,2$ and an asymptotic
value of $0.5$. It is thus reasonable to conclude that the predictions of 
Prop.~\ref{prop:BC_gapless_end} are less consistent with the data than those of Prop.~\ref{prop:BCT}, and in this sense we can interpret Fig.~\ref{fig:fm_sq_scal}
as evidence that the Liouvillian describing 
the experiment
does not become gapless at the end of the BC schedule.

The improvement in the scaling of the adiabatic error provided by
the BC schedule is most prominent for $s_{\text{BC}}=0.45$. As we increase
$s_{\text{BC}}$ to $0.47$ and $0.5$ the improvement gradually disappears. We interpret this
as being due to the fact that the freezing point $s_0\approx 0.52$ is being approached, and so any change
made to the schedule can be expected to be effective only on much longer timescales
$t_{f}$. 

We note that in contrast to the scaling behavior, which favors the smallest $s_{\text{BC}}$ value, the asymptotic value of $P_{\text{GS}}$ is highest for the largest $s_{\text{BC}}$ value. This is consistent with the the larger gap at $s_{\text{BC}}=0.50$ [see Fig.~\ref{fig:fm_gaps-a}]. This effect is in fact more significant than the improvement due to the BCP. 

For additional experiments performed using the the NASA Quantum Artificial Intelligence
Laboratory as well as the D-Wave Advantage processor,
see App.~\ref{app:altdw}. These results are in qualitative agreement with those shown in Fig.~\ref{fig:fm_sq_scal}, and do not modify our conclusions in any significant way.

\begin{figure}
\centering 
\includegraphics[width=0.9\columnwidth]{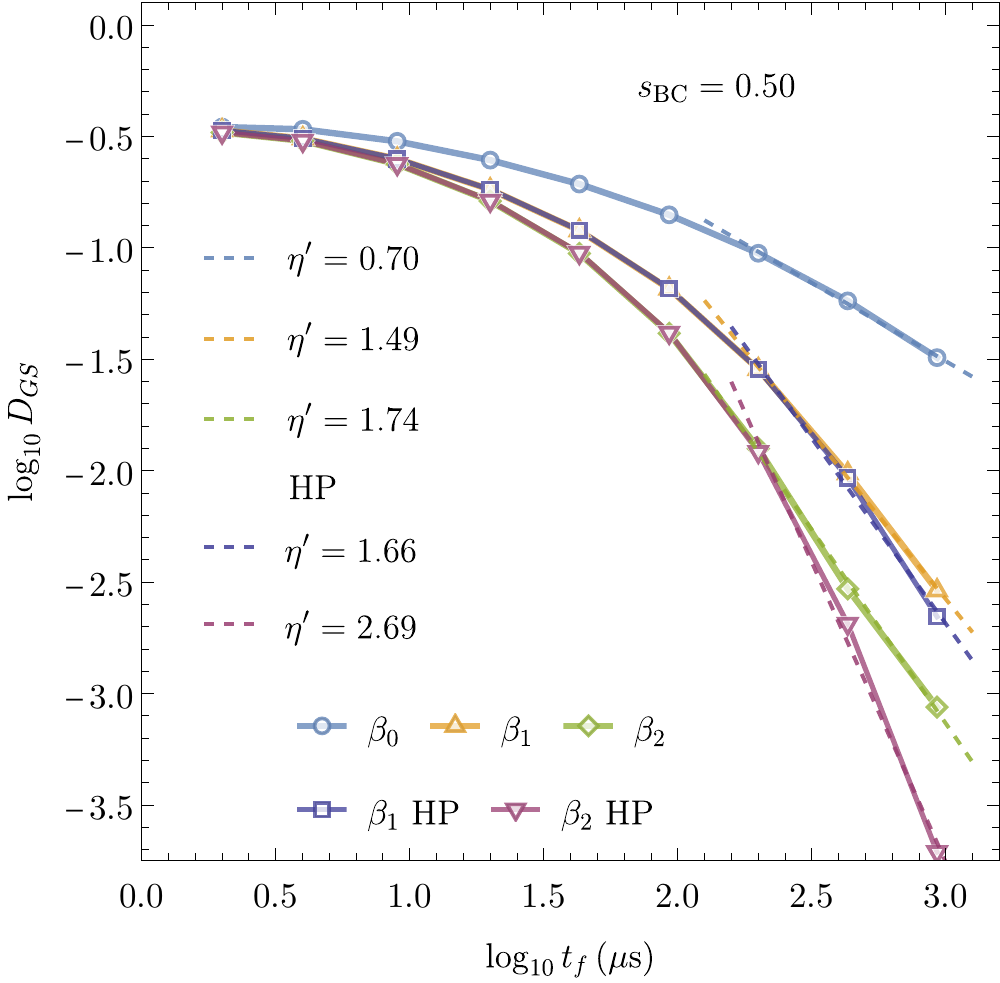}
\caption{Adiabatic error of AME simulations of the D-Wave programmed BCP and
high-precision variants (denoted HP in the legend), for the FM-gadget. Here $s_{\text{BC}}=0.50$, using the schedule of the DW-LN processor
and a crosstalk parameter of $0.02$ (see App.~\ref{app:xtalk}). The adiabatic error is evaluated relative to the Gibbs state ground state probability $\bar{P}_{\text{GS}}^{*}=0.9690$ at $s=0.50$.
}
\label{fig:fm_ame_hp}
\end{figure}

\subsection{High precision BCP}
\label{sec:highprecBCP}

To test the effect of the relatively low precision of the implementation of the BCP schedules (using only $12$ interpolation points), we simulated
the AME for an additional set of higher-precision constructions of the
BC schedule, with $8$ logarithmic partitions up to $s_{\text{BC}}$, rather
than the $4$ allotted for the D-Wave schedule (see the construction
of App.~\ref{app:betasched}). The scaling of the adiabatic error for the FM-gadget is shown in Fig.~\ref{fig:fm_ame_hp}. 
Since the adiabatic error still appears to have some concavity, the scaling exponents should be interpreted as lower bounds for the asymptotic scaling.
We see that the high precision schedule brings the scaling of the adiabatic error significantly closer
to the value predicted by the Prop.~\ref{prop:BCT} version of the BCT within anneal times that are currently accessible to D-Wave devices: we find $\eta'= 1.66,2.69$ for the higher precision schedules, compared to the theoretically expected values of $k+1 = 2,3$, respectively. The lower precision schedules result in $\eta'= 1.49,1.74$, respectively.
While the values of $\eta'$ for the high precision schedules could be refined by simulations at longer anneal times, we do not include them here to avoid floating point error effects. In any case, it would be much more difficult to experimentally verify the adiabatic error to the precision required at longer annealing times.
As such, Fig.~\ref{fig:fm_ame_hp} is likely to be the most practical precision test of boundary cancellation to date on open systems larger than a single qubit.

We also simulated the AME with a precision of $100$ evenly spaced points in the control
schedule, to approximate reaching the limit of the analytically exact beta schedule.
The results are indiscernible from the high-precision piecewise
schedule shown in Fig.~\ref{fig:fm_ame_hp}, indicating that the latter already implements the beta schedule
at the smallest precision that is meaningful within the simulated
annealing times.

We conclude that the relatively poor agreement with the Prop.~\ref{prop:BCT} BCT predictions we observed in Figure~\ref{fig:fm_sq_scal}(a)-(c), is likely to be improved in future higher precision implementations of the beta schedule.

\begin{figure}
\includegraphics[width=0.9\columnwidth]{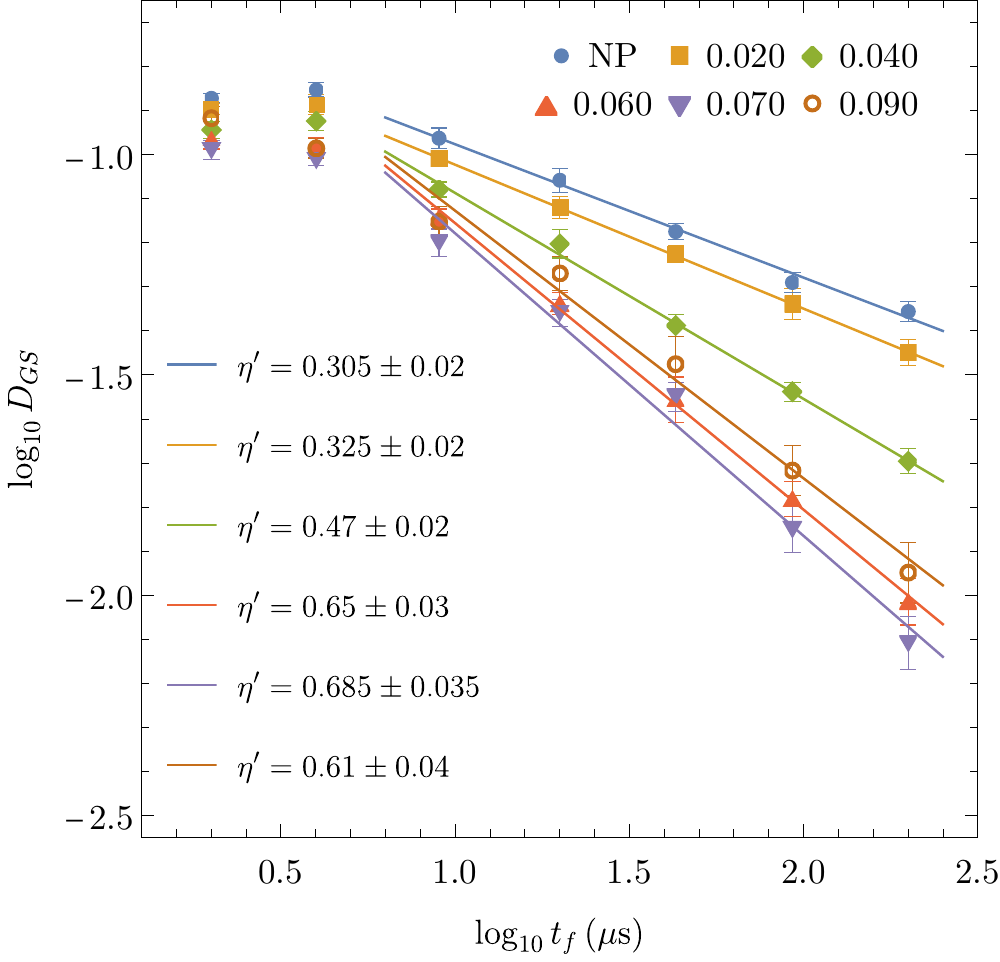}
\caption{Adiabatic error for a linear anneal of the QAC-encoded FM-gadget with
no ramp. Different curves correspond to different values of the penalty
strengths, ranging from no penalty (NP; $\lambda=0$) to $\lambda=0.09$
(see symbol legend). The largest slope $\eta'=0.69\pm0.03$ is obtained
for $\lambda=0.07$.  
We directly assumed a linear model with $\bar{P}_{\text{GS}}^*=1$ for comparability between the different penalties; performing a non-linear fit first does not change the slopes for $\lambda > 0.04$.
}
\label{fig:fm_qac_np}
\end{figure}

\subsection{QAC-Encoded FM Gadget}
\label{sec:QAC-results}

The scaling of the adiabatic error of the QAC-encoded FM-gadget for
various penalty strengths $\lambda$ [recall Eq.~\eqref{eq:QAC}]
is shown in Fig.~\ref{fig:fm_qac_np}, for the case of the standard
linear control schedule with no ramp. We observe that
from the set of values tried, the optimal penalty strength is $\lambda=0.07$.
Using this optimal penalty value, the ground state probabilities of
BCP with QAC are shown in Fig.~\ref{fig:6d}-\ref{fig:6e}. The
QAC-protected $P_{\text{GS}}$ results for both $s_{\text{BC}}$ values shown are significantly higher than their unprotected ($\lambda=0$, NP) counterparts in Fig.~\ref{fig:6a}-\ref{fig:6c}, consistent
with many prior results on the improved performance offered by QAC.

However, we are primarily interested here in QAC's effect on the BCP.
In this regard, at $s_{\text{BC}}=0.45$, there is a notable distinction
between different $k$ values, so that QAC amplifies the effect of the BC schedule. 

At $s_{\text{BC}}=0.50$, QAC-protected BCP still improves over the linear anneal
($k=0$) by $P_{\text{GS}}$ value, but the improvement is less distinct
for different $k$ values. Moreover, $\eta'$ is smaller (at most $0.61$) than that for the optimal-$\lambda$
linear QAC anneal in Fig.~\ref{fig:fm_qac_np} (for which $\eta'=0.69$),
signaling that freezing has already mostly occurred before $s=0.50$,
and that the optimal $\lambda$ value is $s_{\text{BC}}$-dependent.


Our results demonstrate that the combination of BCP
and QAC as error suppression methods is more powerful in terms of
the scaling of the adiabatic distance than either method alone: while
$k=2$ BCP has $\eta'=1.1$ [Fig.~\ref{fig:6a}] and optimal-$\lambda$ linear
($k=0$) QAC has $\eta'=0.69$ (Fig.~\ref{fig:fm_qac_np}), their combination yields $\eta'=1.3$ [Fig.~\ref{fig:6d}] and the largest $P_{GS}$ of any other schedule for $t_f=12\,\mathrm{\mu s}$.

\begin{figure*}
\includegraphics[width=1\textwidth]{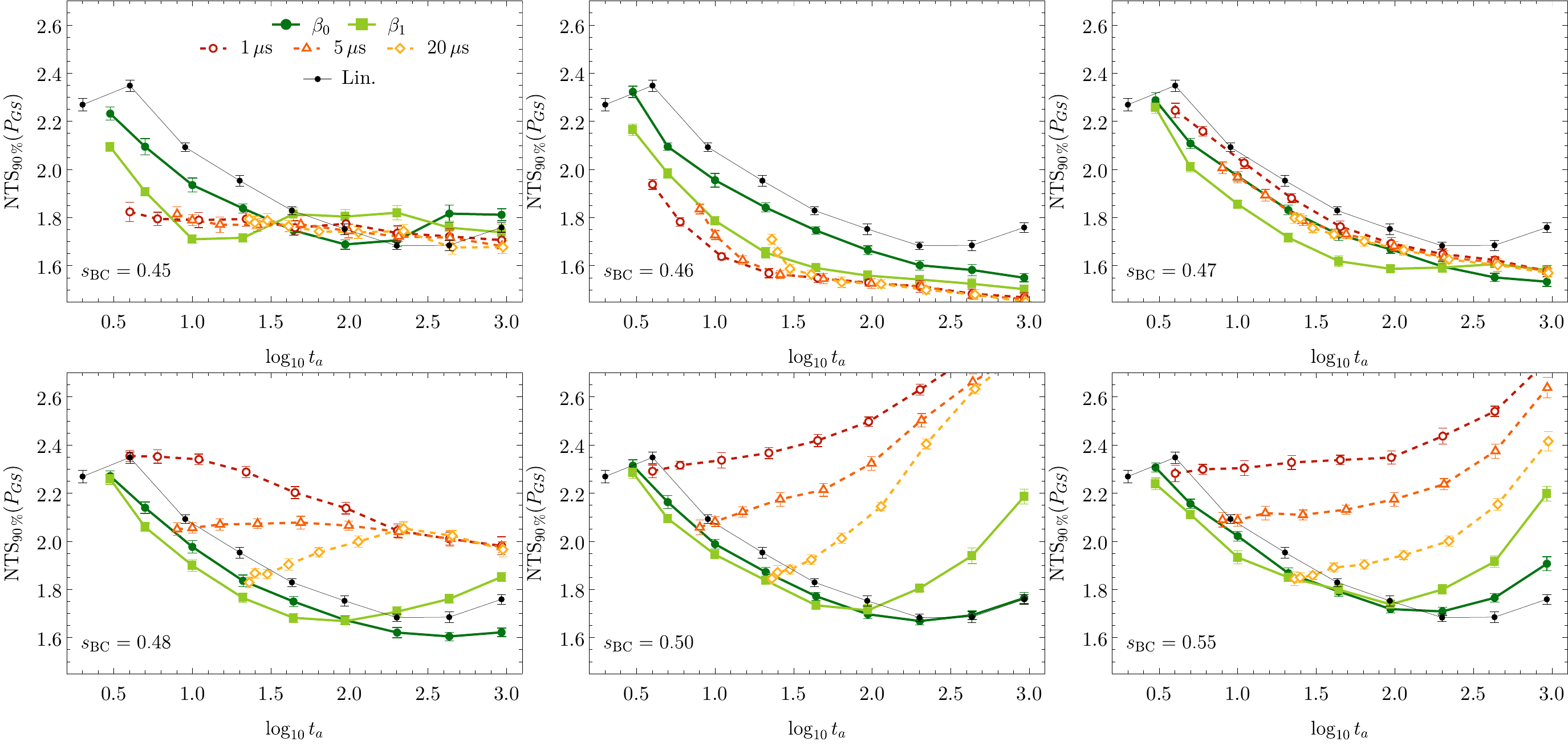} 
\caption{Number of tries to solution for the T-gadget, comparing the pausing
protocols (empty symbols) and $\beta_{0},\beta_{1}$ BCP (all with
a ramp at $s_{\text{BC}}$), along with the linear anneal without a ramp (denoted
``Lin.''; the same data for all panels), for reference. Results shown are for DW-LN, as a function
of the total anneal time of each protocol. Different panels have different
values of $s_{\text{BC}}$, from $0.45$ to $0.55$. The pausing protocol
is shown for different initial anneal times $t_{0}$ of $1$,$5$,
and $20\,\mathrm{\mu s}$ (empty symbols), followed by a pause of
varying length $t_{p}$ and a $1\,\mathrm{\mu s}$ long ramp (so that
the paused portion of the schedule is comparable to BCP). For BCP
(filled symbols), $t_{a}=t_{f}+1\mathrm{\mu s}$. For pausing, $t_{a}=t_{0}+t_{p}+1\mathrm{\mu s}$.
Both schedules are plotted against $t_{a}$ rather than their natural
parameters ($t_{f}$ and $t_{p}$ respectively) for direct comparison. 
\label{fig:t_pau_nts_hi}}
\end{figure*}

\subsection{Comparison of the BCP to the Pause-Ramp protocol}
\label{sec:tries}

A different protocol that attempts to exploit slowing down the anneal is the pausing protocol, which interrupts an ordinary linear anneal by a single pause \cite{marshallPower19,chenWhy20,izquierdo2020ferromagnetically,albash2020comparing,izquierdoAdvantage22}. 
Pausing directly uses thermal relaxation at a single point in the anneal 
to try to increase the ground state probability. As in the BCP, this point should also be after 
an avoided level crossing, but before the open system dynamics freeze~\cite{chenWhy20}. 

We compare the BCP to a variant of the usual pausing schedule~\cite{marshallPower19} referred to here as the pause-ramp (PR) schedule. While the pausing protocol is typically constructed by interrupting a linear anneal with a pause (but no ramp), PR is constructed by making the first linear segment last for $t_0$, pausing for $t_p$, and ramping to the end. As with the BCP, we use a ramp that is $1\,\mathrm{\mu s}$ long in all cases. PR is better suited for comparison with BCP than the usual pausing schedule since they differ only in their equilibrium state preparation and not in their behavior during the frozen phase.

As our metric for comparison we use not the ground state probability but rather the number of tries-to-solution (NTS) at a fixed anneal time (with 90\% confidence):
\begin{equation}
\mathrm{NTS}_{90\%}(P_{\text{GS}}) =
\frac{\log(1-0.90)}{\log(1-P_{\text{GS}} )}\,
\label{eq:NTS}
\end{equation}
This metric is simply another way to study the ground state probability, interpreted as the expected number of tries needed to find the ground state at least once with 90\% confidence, at the given anneal time. We use this rather than the standard time-to-solution (TTS) metric, since the latter requires identifying the optimal anneal time~\cite{speedup,AlbashDemonstration18,Hen:2015rt}. However, neither the FM-gadget nor the T-gadget exhibited an optimal anneal time in our experiments (not shown). 

%

Figure~\ref{fig:t_pau_nts_hi} shows the NTS results for the T-gadget
at various values of $s_{\text{BC}}$. It compares the standard schedule (no BCP or pausing), the BCP protocol at $k=0$ and $k=1$, and the PR schedule with different initial anneal and pause times. Starting at $s_{\text{BC}}=0.45$ (top left)
the gap is small and thermalization is fast. BCP exhibits two distinct
optimal anneal times (minima) for $k=0$ and $k=1$, both of which
result in slightly smaller NTS than pausing. Also, the $k=1$ case
is optimal at smaller $t_{a}$ than $k=0$, which can be interpreted
as as advantage of the higher order protocol. This trend persists
for most of the $s_{\text{BC}}$ values shown. Right at $s_{\text{BC}}=0.46$, pausing
achieves its optimal NTS curve, with a slight advantage over BCP,
and the BCP of each order $k$ is very distinct. However, the advantage
of pausing is only present at $s_{\text{BC}}=0.46$, and disappears already
at $s_{\text{BC}}=0.47$, showing that pausing performance is highly sensitive
to the pause point. Since the BCP slowdown is smooth, the influence
of the optimal thermalization at $s_{\text{BC}}=0.46$ remains present, although
the distinction between BCP orders diminishes as $s_{\text{BC}}$ grows.
Well before $s_{\text{BC}}$ approaches the freeze-out point ($s_0\approx0.51$), 
pausing becomes detrimental for the T-gadget due to excitations (witnessed by the increase in the NTS metric), while BCP mitigates the excitations until the anneal time becomes too long to avoid them.

These results paint a picture of the two protocols as follows. 
The relaxation induced by the BCP depends on
the properties of the Liouvillian over a \emph{neighborhood} $(s_{\text{BC}}-\epsilon,s_{\text{BC}}]$.
If this neighborhood overlaps with the neighborhood of an avoided
level crossing, then the BCP has an opportunity for an advantage,
as the speed at which the crossing is traversed is minimized and detrimental
excitations to the excited state are reduced. In contrast, PR schedules
must try to discontinuously stop the anneal as close as possible to
the crossing, and the time required by a paused schedule could be increased
due to the longer pause time needed to recover the ground state population,
unless the precise optimal pause point is quickly found. 

This suggests an important practical point: optimization
of the BCP can be simpler than that of PR schedules, by the need to optimize just one discrete parameter
$k$ and one continuous parameter $s_{\text{BC}}$. Moreover, $s_{\text{BC}}$ need not be exactly
tuned, since even an approximate value can allow the BCP to
closely approach its optimal performance. On the other hand, pausing requires
$s_{\text{BC}}$ to be optimized very precisely, as well as the anneal and pause
times $t_{0}$ and $t_{p}$.

\section{Discussion and Outlook}
\label{sec:discussion}

We have extended the boundary cancellation theorem for open systems
to the case where the Liouvillian gap vanishes at the end
of the anneal, and derived the asymptotic scaling of the adiabatic
error with the anneal time $t_{f}$. Armed with the corresponding theoretical
expectation of the scaling of the adiabatic error for the gapped and gapless
cases,
we set out
to test the scaling predictions and to improve the success probability of quantum
annealing hardware by implementing boundary cancelling schedules.  The
specific functional form of these schedules induces a smooth slowdown
in accordance with the boundary cancellation theorem. We experimentally tested boundary
cancellation protocols for open systems and 
evaluated their performance and error-suppression characteristics on specifically designed
$8$-qubit gadgets embedded on the DW-LN annealer. 

While a quantitative agreement with the theoretical predictions was not observed, we did demonstrate that as long as the protocol terminates before the onset of freezing, it can increase the ground state population in the examples studied here beyond what is achievable with simple linear anneals, and it does so with shorter anneal times. These results are in qualitative agreement with the theoretical scaling predictions of the BCT.

In conjunction with quantum annealing correction (QAC), the boundary
cancellation protocol is also capable of improved adiabatic error
scaling over what would be achieved with either method alone. While
this does not immediately translate to a ground state solution speedup
within the annealing problems studied here, we have shown that BCP-QAC
is a novel error suppression strategy successfully combining two complimentary
methods: the suppression of environmentally induced logical errors
and the promotion of relaxation via boundary cancellation.

In contrasting the BCP with the pause-ramp protocol, we found that the BCP is significantly less sensitive to the location of the ramp point, and achieves better performance except at the exact ramp point where the pause-ramp protocol is optimal.

With the small system size of $8$ qubits used in this work, it was
possible to collect a large number of annealing samples as well as
validate the protocol behavior against the energy spectrum and open
system simulations. Future work will assess the protocol for larger
system sizes and the impact it has on the scaling of time-to-solution
as a function of problem size. The largest expected improvement in the protocol's performance, based on our simulations, will arise from an increase in the number and resolution of interpolation points along the annealing schedule, which will allow a more faithful experimental implementation of the ideally smooth annealing schedules demanded by the theoretical protocol.

\acknowledgments This research is based upon work (partially) supported
by the Office of the Director of National Intelligence (ODNI), Intelligence
Advanced Research Projects Activity (IARPA) and the Defense Advanced
Research Projects Agency (DARPA), via the U.S. Army Research Office
contract W911NF-17-C-0050. The views and conclusions contained herein
are those of the authors and should not be interpreted as necessarily
representing the official policies or endorsements, either expressed
or implied, of the ODNI, IARPA, DARPA, ARO, or the U.S. Government.
The U.S. Government is authorized to reproduce and distribute reprints
for Governmental purposes notwithstanding any copyright annotation
thereon. The authors acknowledge the Center for Advanced Research
Computing (CARC) at the University of Southern California for providing
computing resources that have contributed to the research results
reported within this publication. URL: \url{https://carc.usc.edu}.

\appendix


\section{Proof of degeneracy when $A(t)=0$}
\label{app:Proof-of-degeneracy}

We assume that $A_{i}=\sigma_{i}^{z}$. When $A(t)=0$,
$H_{S}\propto H_{Z}$ and commutes with $A_{i}$, so that Eq.~\eqref{eq:jump}
yields $A_{i}(\omega)=\sigma_{i}^{z}\delta_{\omega,0}$. Therefore,  the dissipator in Eq.~\eqref{eq:AME} becomes
\begin{align}
\left.\mathcal{D}\right|_{A(t)=0} =\sum_{ij}\gamma_{ij}(0)\Big(\sigma_{j}^{z}\bullet\sigma_{i}^{z}-\frac{1}{2}\left\{ \sigma_{i}^{z}\sigma_{j}^{z},\bullet\right\} \Big).
\end{align}
Now let $|n\rangle$ be an eigenstate of $H_Z$, where $n=1,\dots,d$ and $d$ is the system's Hilbert space dimension. Then $|n\rangle\!\langle n|$
commutes with $\sigma_{i}^{z}$, so that $\mathcal{D}\left(|n\rangle\!\langle n|\right)=0$.
At the same time, $A_{i}(\omega)=\sigma_{i}^{z}\delta_{\omega,0}$
implies that also the Lamb shift is diagonal in the $\sigma^{z}$
basis. Together with the fact that $|n\rangle\!\langle n|$ also commutes
with the system Hamiltonian this implies $\mathcal{L}\left(|n\rangle\!\langle n|\right)=0$,
which means that the kernel of $\mathcal{L}$ is at least $d$-dimensional at $A(t_{f})=0$, i.e., that the zero
eigenvalue of $\mathcal{L}$ is at least $d$-fold degenerate.

\section{Proof of Proposition \ref{prop:BC_gapless_end}}
\label{app:Proof-of-Proposition_end} 

Here we prove Proposition \ref{prop:BC_gapless_end}, which concerns the case where the gap closes at the end of the anneal and we enforce boundary cancellation at this point.

We denote by $\mathcal{L}(s)$ the Liouvillian without the BC schedule (i.e., with a linear schedule around $s=1$) while $\mathcal{L}(s(\tau))$ [or $\mathcal{L}(\tau)$ for simplicity] indicates the Liouvillian with the BC schedule. 
We will use the series developed in \cite{avronAdiabatic12} and
refined in \cite{camposvenutiError18} together with ideas from \cite[App.~J]{venutiAdiabaticity16}. The assumption that 
$\int_0^1 ds \left\Vert \mathcal{L}(s)\right\Vert _{1,1} <\infty$
assures, by Carath\'{e}odory's theorem, that the solution of Eq.~\eqref{eq:MEode} exists and is unique in an extended sense~\cite{hale_ordinary_2009}, while the requirement that
$\mathcal{L}(\tau)$ is $k+2$ differentiable at $k=1$ implies that one can use the series Eq.~\eqref{eq:adia_series1-1} below with $\tau=1$ and $N=k$.


By assumption, we have a boundary cancelling schedule at $s=1$.
Such a schedule is a function $s=s(\tau)$ where $s(1)=1$ and $s^{(j)}(1)=0$
for $j=1,2,\ldots,k$. That is, 
\beq
s(\tau)=1-g(1-\tau)^{k+1}+O(\delta^{k+2})\ , \quad \delta\equiv 1-\tau
\label{eq:ftau}
\eeq
when $\tau\to1^{-}$ for some constant $g$.
Let 
\beq
\mathcal{L}(s)=\sum_{j=1}\lambda_{j}(s)P_{j}(s)+D(s)
\label{eq:Lexpansion}
\eeq
be the instantaneous spectral decomposition of \emph{$\mathcal{L}(s)$}
(without the BC schedule), i.e., $\lambda_{j}(s)$ are eigenvalues, $P_{j}(s)$
are eigenprojectors, and $D(s)$ is a nilpotent term. We denote by $P(s)$ (without a subscript) the projector related to the zero eigenvalue. The assumption
of contractivity implies that $D(s)$ cannot belong to the $\mathcal{L}(s)=0$
subspace. We denote by $\sigma(\tau)$ the unique steady state of $\mathcal{L}(s(\tau))$,
i.e., $\mathcal{L}(s(\tau))\sigma(\tau)=0$, $\Tr(\sigma(\tau))=1$ while $\rho(\tau)$
is the solution of Eq.~\eqref{eq:MEode} with the BC schedule and with the initial
condition $\rho(0)=\sigma(0)$. Note that in the main text $\mathcal{L}(s(\tau))$ is written for simplicity as $\mathcal{L}(\tau)$. 

We use a similar approach (and notation) as in Ref.~\cite{venutiAdiabaticity16}.
Here $\mathcal{E}(\tau):=\mathcal{E}(\tau,0)$ is the evolution superoperator
satisfying $\partial_{\tau}\mathcal{E}(\tau,\tau')=t_{f}\mathcal{L}(s(\tau))\mathcal{E}(\tau,\tau')$
and $\mathcal{E}(\tau_{3},\tau_{2})\mathcal{E}(\tau_{2},\tau_{1})=\mathcal{E}(\tau_{3},\tau_{1})$ together with 
$\mathcal{E}(\tau,\tau) = \1$, i.e.,
\beq
\mathcal{E}(\tau_{2},\tau_{1})=\mathcal{E}^{-1}(\tau_{3},\tau_{2})\mathcal{E}(\tau_{3},\tau_{1}) .
\label{eq:E-inv}
\eeq
Let $V(\tau)$ be the adiabatic intertwiner as defined in \cite{venutiAdiabaticity16} and, following the same notation therein, $W(\tau):=P(\tau)V(\tau)=V(\tau)P(0)$.
We start with Eq.~(D4) of Ref.~\cite{venutiAdiabaticity16} which we
reproduce here for clarity:
\beq
\left[\mathcal{E}^{-1}(\tau)V(\tau)-\1\right]P(0)=\int_{0}^{\tau}\mathcal{E}^{-1}(\tau_{1})\dot{W}(\tau_{1})d\tau_{1},
\label{eq:D4-old}
\eeq
where the dot denotes differentiation with respect to the argument of the corresponding operator. 
Hence
\bes
\begin{align}
&\left[\mathcal{E}^{-1}(1)V(1)-\1\right]P(0) 
 =\left(\int_{0}^{\tau}+\int_{\tau}^{1}\right)\mathcal{E}^{-1}(\tau_{1})\dot{W}(\tau_{1})d\tau_{1}\\
 & \quad=\left(\mathcal{E}^{-1}(\tau)V(\tau)-\1\right)P(0)+ \int_{\tau}^{1}\mathcal{E}^{-1}(\tau_{1})\dot{W}(\tau_{1})d\tau_{1} ,
\end{align}
\ees

Now apply $\mathcal{E}(1) := \mathcal{E}(1,0)$ from the left and use Eq.~\eqref{eq:E-inv} repeatedly, to obtain:
\bes
\begin{align}
&\left(V(1)-\mathcal{E}(1)\right)P(0)  =\left(\mathcal{E}(1,\tau)V(\tau)-\mathcal{E}(1,0)\right)P(0)\notag \\
 & \qquad +\int_{\tau}^{1}\mathcal{E}(1,\tau_{1})\dot{W}(\tau_{1})d\tau_{1}\\
 & \quad =\mathcal{E}(1,\tau)\left(V(\tau)-\mathcal{E}(\tau)\right)P(0) +\int_{\tau}^{1}\mathcal{E}(1,\tau_{1})\dot{W}(\tau_{1})d\tau_{1}.
\end{align}
\ees
Now apply $\sigma(0)$ from the right. Note that $V(\tau)$ is continuous,
being the solution of a differential equation. So $V(1)P(0)\sigma(0)=\lim_{\tau\to1^{-}}\sigma(\tau)=\sigma(1^{-})$
and we obtain finally 
\begin{align}
\sigma(1^{-})-\rho(1) & =\mathcal{E}(1,\tau)\left(\sigma(\tau)-\rho(\tau)\right)\nonumber \\
 &\qquad +\int_{\tau}^{1}\mathcal{E}(1,\tau_{1})\dot{W}(\tau_{1})\sigma(0)d\tau_{1}.\label{eq:starting_point}
\end{align}

Eq.~\eqref{eq:starting_point} is the starting point of our analysis.
Since in Eq.~\eqref{eq:starting_point} $\tau<1$ and the system is
gapped in $\left[0,\tau\right]$, we can use the result of \cite{camposvenutiError18}
to obtain the behavior of $\sigma(\tau)-\rho(\tau)$. Assuming that $\mathcal{L}(\tau)$ is at least $N+2$ times differentiable in a neighborhood of $\tau=1$, one has the following series \cite{camposvenutiError18} with $\epsilon=1/t_{f}$ for $\tau$ in the same neighborhood: 
\bes
\label{eq:recursion}
\begin{align}
\label{eq:adia_series1-1}
\rho(\tau) & =\sigma(\tau)+\sum_{n=1}^{N}\epsilon^{n}b_{n}(\tau)+\epsilon^{N+1}r_{N}(\epsilon,\tau)
\\
b_{1}(\tau) & =S(\tau)\dot{P}(\tau)\sigma(\tau)=S(\tau)\dot{\sigma}(\tau)\\
b_{n+1}(\tau) & =S(\tau)\dot{b}_{n}(\tau),\quad n=1,2,\ldots,N
\label{eq:bn_recursion}
\end{align}
\ees
where the remainder is 
\begin{multline}
r_{N}(\epsilon,\tau)=b_{N+1}(\tau)-\mathcal{E}(\tau)b_{N+1}(0)\\
\quad-\int_{0}^{\tau}\mathcal{E}(\tau,\tau')\dot{b}_{N+1}(\tau')d\tau'
\label{eq:remainder}
\end{multline}
and $S(\tau)$ is the reduced resolvent, defined as: 
\beq
S(\tau):=\lim_{z\to0}\left(\1-P(\tau)\right)\left(\mathcal{L}(\tau)-z\right)^{-1}\left(\1-P(r)\right) .
\eeq

We will keep $N$ free and only fix it at the end when needed. 
The series \eqref{eq:adia_series1-1} cannot be used at  $\tau=1$, because the assumption of a gapped Liouvillian does not hold there. Hence we must keep
$\tau<1$ but investigate the kind of divergences that arise when $\tau\to1^{-}$. 

The other assumption is that, without the BC schedule, the Liouvillian
gap closes at $s=1$ in such a way that $\lambda_{1}(s)\simeq v(1-s)^{\alpha}$
as $s\to1^{-}$. Therefore, using Eqs.~\eqref{eq:ftau} and~\eqref{eq:Lexpansion}, close to $s=1$ the Liouvillian with the BC schedule
behaves as 
\begin{align}
&\mathcal{L}(s(\tau))\simeq \tilde{v}(1-\tau)^{\tilde{\alpha}}P_{1}(s(\tau))\notag \\
&\qquad +\sum_{j>1}\lambda_{j}(s(\tau))P_{j}(s(\tau))+D(s(\tau)),
\end{align}
with $\tilde{v}=vg^{\alpha}$ and $\tilde{\alpha}=\alpha(k+1)$. Accordingly, the
most diverging term of the reduced resolvent $S(\tau)$ behaves as\footnote{Here, and analogously in the following, the symbol '$\sim$' in Eq.~\eqref{eq:S_div}
is intended to mean $\lim_{\tau\to1^{-}}\left\Vert S(\tau)\right\Vert _{1}\delta^{\tilde{\alpha}}=\left\Vert P_{1}(1)\right\Vert _{1}$.
This implies that one can find a positive constant $C$ such that
$\left\Vert S(\tau)\right\Vert _{1}\le C/\delta^{\tilde{\alpha}}$
for sufficiently small $\delta$. This fact will be repeatedly used
in the following. }
\beq
S(\tau) \sim\frac{P_{1}}{\delta^{\tilde{\alpha}}}, \label{eq:S_div}
\eeq
where henceforth we do not always explicitly emphasize the dependence
on $\tau$ when it does not lead to a divergence. 
Now, the derivatives
of the projectors behave as
\begin{align}
\partial_{\tau}P_{j}(s(\tau)) & =\partial_{s}P_{j}(s)\partial_{\tau}s(\tau) \sim   \partial_{s}P_{j} \delta^{k} .
\label{eq:C13}
\end{align}
Hence, we see that when $\tau\to1^{-}$, $b_{1}$ diverges as 
\beq
b_{1}(\tau)\sim\frac{1}{\delta^{\tilde{\alpha}-k}}.
\eeq

When we construct $b_{n}$ according to Eq.~\eqref{eq:bn_recursion},
at each iteration we gain a derivative and a factor of $S$. Thus, defining $\beta_{n}$ via
$b_{n}(\tau)\sim\delta^{-\beta_{n}}$, we obtain $\beta_{n+1}=\beta_{n}+\tilde{\alpha}+1$. This has the solution
\bes
\begin{align}
\beta_{n} &=n\left(\tilde{\alpha}+1\right)-k-1 \\
& =n\left(\alpha k+\alpha+1\right)-k-1.
 \label{eq:betan_new}
\end{align}
\ees
Hence, we can find positive constants $A_{n}$ such that 
\beq
\left\Vert b_{n}(\tau)\right\Vert_1 \le\frac{A_{n}}{\delta^{\beta_{n}}}.
\eeq
Turning to the remainder, we take the trace norm of Eq.~\eqref{eq:remainder} and 
use the triangle inequality:
\begin{align}
\|r_{N}(\epsilon,\tau)\|_1 &\le \|b_{N+1}(\tau)\|_1 + \|\mathcal{E}(\tau)b_{N+1}(0)\|_1\notag \\
&\quad + \int_{0}^{\tau}\|\mathcal{E}(\tau,\tau')\dot{b}_{N+1}(\tau')\|_1 d\tau' .
\label{eq:remainder2}
\end{align}
Next, we use $\left\Vert \mathcal{E}x\right\Vert _{1}\le\left\Vert \mathcal{E}\right\Vert _{1,1}\left\Vert x\right\Vert _{1}$
for a superoperator $\mathcal{E}$ and operator $x$, and the fact that since $\mathcal{E}(\tau_{2},\tau_{1})$ is a completely positive
trace preserving map for $\tau_{2}\ge \tau_{1}$, we have 
\beq
\left\Vert \mathcal{E}(\tau_{2},\tau_{1})\right\Vert _{1,1}=1 \ , \quad \tau_{2}\ge \tau_{1}.
\label{eq:normE}
\eeq
Therefore
\beq
\int_{0}^{\tau}\|\mathcal{E}(\tau,\tau')\dot{b}_{N+1}(\tau')\|_1 d\tau' \leq \int_{0}^{\tau}\|\dot{b}_{N+1}(\tau')\|_1 d\tau'  . \label{eq:integral_divergence}
\eeq
Now using $b_{n}(\tau)\sim\delta^{-\beta_{n}}$ again we obtain $\Vert \dot{b}_{N+1} (\tau')\Vert_1\sim (1-\tau')^{-\beta_{N+1}-1}
$, which integrated from $0$ to $\tau$ implies that  
\beq
\int_{0}^{\tau}\Vert\mathcal{E}(\tau,\tau')\dot{b}_{N+1}(\tau') \Vert_1 d\tau' \le \frac{B}{\delta^{\beta_{N+1}}},
\eeq
for some positive constant $B$. 
All in all, we obtain for the remainder
\beq
\left\Vert r_{N}(\epsilon,\tau)\right\Vert \le\frac{A_{N+1}}{\delta^{\beta_{N+1}}} +\left\Vert b_{N+1}(0)\right\Vert +\frac{B}{\delta^{\beta_{N+1}}} ,
\label{eq:C20} 
\eeq

Here $\left\Vert b_{N+1}(0)\right\Vert $
is just another constant, independent of $\delta$. Combining these results we obtain from Eq.~\eqref{eq:adia_series1-1}, after redefining the constants:
\begin{equation}
\left\Vert \rho(\tau)-\sigma(\tau)\right\Vert \le\sum_{n=1}^{N+1}\epsilon^{n}\frac{A_{n}}{\delta^{\beta_{n}}}+\epsilon^{N+1}A_{N+2}. \label{eq:bound_at_s}
\end{equation}

It remains to estimate the second term on the RHS of \eqref{eq:starting_point}. Using the
mean value theorem, $\dot{W}=\dot{P}W$~\cite[Eq.~(C1)]{venutiAdiabaticity16}, and $W(\tau)\sigma(0) = \sigma(\tau)$, we have:
\begin{align}
\int_{\tau}^{1}\mathcal{E}(1,\tau_{1})\dot{W}(\tau_{1})\sigma(0)d\tau_{1} & =\delta\mathcal{E}(1,\tilde{\tau})\dot{P}(\tilde{\tau})\sigma(\tilde{\tau}) ,
\label{eq:integral}
\end{align}
where $\tilde{\tau}\in\left[\tau,1\right]$, and we used $\delta = 1-\tau$. Hence, when taking the norm of Eq.~\eqref{eq:integral}, using Eqs.~\eqref{eq:C13} and~\eqref{eq:normE}, we obtain, for some positive constant $C$:

\begin{equation}
\left\Vert \int_{\tau}^{1}\mathcal{E}(1,\tau_{1})\dot{W}(\tau_{1})\sigma(0)d\tau_{1} \right\Vert \le C \delta^{k+1}. \label{eq:integral_bound}
 \end{equation}
 
Using Eqs.~\eqref{eq:bound_at_s} and~\eqref{eq:integral_bound}, we finally obtain after taking the norm of Eq.~\eqref{eq:starting_point}:
\begin{align}
\left\Vert \rho(1)-\sigma(1^{-})\right\Vert  & \le\sum_{n=1}^{N+1}\epsilon^{n}\frac{A_{n}}{\delta^{\beta_{n}}}+\epsilon^{N+1}A_{N+2}+C\delta^{k+1}\label{eq:bound_final}\\
 & =:f(\delta,\epsilon).\nonumber 
\end{align}
In this derivation $\tau=1-\delta$ is sufficiently close to one but otherwise free. Thus, we can now minimize $f$
as a function of $\delta$. Differentiating $f$ we obtain:
\bes
\begin{align}
\partial_{\delta}f & =-\sum_{n=1}^{N+1}\epsilon^{n}\frac{\beta_{n}A_{n}}{\delta^{\beta_{n}+1}}+(k+1)C\delta^{k}=0\\
\hookrightarrow\delta^{-k}\partial_{\delta}f & =-\sum_{n=1}^{N+1}\epsilon^{n}\frac{\beta_{n}A_{n}}{\delta^{n\left(\alpha k+\alpha+1\right)}}+(k+1)C = 0.
\label{eq:minimization}
\end{align}
\ees
We now make the ansatz $\delta=\nu\epsilon^{\gamma}$. We wish to
find a $\gamma$ such that the solution $\nu$ of Eq.~\eqref{eq:minimization} does not depend on
$\epsilon$. This is satisfied if all the summands in Eq.~\eqref{eq:minimization}
do not scale with $\epsilon$, i.e., are constant, which holds for $n-\gamma n\left(\alpha k+\alpha+1\right)=0$. This means that $\gamma=1/\left(\alpha k+\alpha+1\right)$. The value of
$\nu$ can be found by solving
\begin{equation}
\sum_{n=1}^{N+1}\frac{\beta_{n}A_{n}}{\nu^{n\left(\alpha k+\alpha+1\right)}}= (k+1) C . \label{eq:nu_equation}
\end{equation}
Picking $N=1$ in Eq.~\eqref{eq:nu_equation}  one can show that the resulting quadratic equation in $x:=\nu^{-(\alpha k +\alpha +1)}$ always has a positive solution, which means that Eq.~\eqref{eq:nu_equation} always has a real solution $\nu$.

\begin{figure}
\includegraphics[width=0.9\columnwidth]{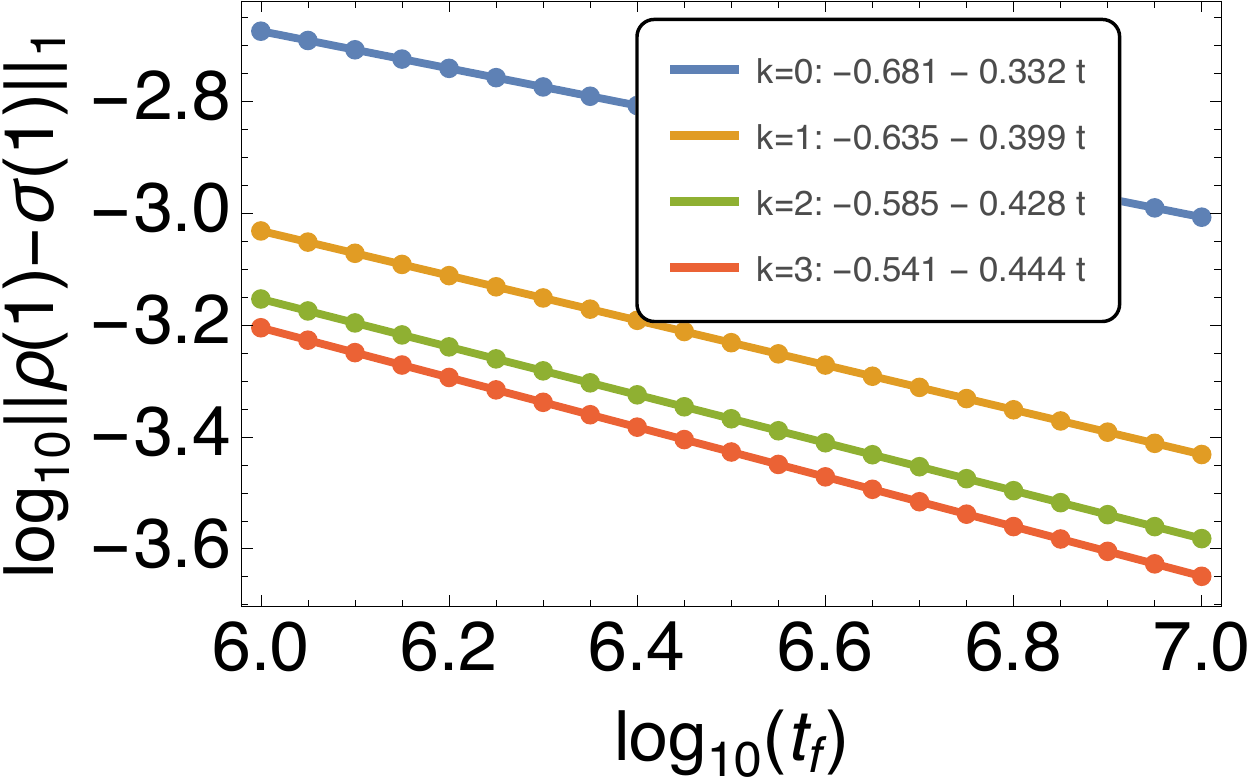}
\caption{Test of the bound given in Proposition~\ref{prop:BC_gapless_end}, for
the case of a single qubit subject to evolution under the AME. The Hamiltonian is $H(\tau)=\omega_{x}(1-\beta_k(\tau))\sigma^{x}+\omega_{z}\beta_k(\tau)\sigma^{z}$
with BC schedule $\beta_k(\tau)=\mathrm{B}_{\tau}(1,k+1)/\mathrm{B}_{1}(1,k+1)$ [$\mathrm{B}$ is the incomplete beta function, Eq.~\eqref{eq:ibf}]
which has $k$ vanishing derivatives at $\tau=1$ and is linear at
$\tau=0$. The system-bath coupling is via $\sigma^{z}$, which assures
that the Liouvillian is gapless at $\tau=1$ with a gap-closing exponent
$\alpha=2$ \cite{venutiAdiabaticity16}. The bath spectral density
is Ohmic [Eq.~\eqref{eq:Ohmic}]; the 
parameters are $g=1,\,\beta=1,$ $\omega_{x}=\omega_{z}=-1/2$ in arbitrary
units. Dots are results of solving for the evolution while the continuous lines are
linear fits. The theoretical expectation given by Eq.~\eqref{eq:eta_BCT_gapless_end}
is $\eta=\{0.333,\,0.4,\,0.428,\,0.444\}$ for $k=\{0,1,2,3\}$, respectively, and is an excellent agreement with the the last column of the legend (obtained via a linear fit).
\label{fig:check_prop_5}}
\end{figure}

Substituting  $\delta=\nu\epsilon^{\gamma}$ with $\gamma=1/\left(\alpha k+\alpha+1\right)$ into Eq.~\eqref{eq:bound_final},
we see that all terms scale as $\epsilon^{\eta}$ with the exponent
\begin{equation}
\eta=\frac{k+1}{\alpha k+\alpha+1}.
\label{eq:eta_BCT_gapless_end}
\end{equation}
 In other words:
\begin{equation}
\left\Vert \rho(1)-\sigma(1^{-})\right\Vert \le C'\epsilon^{\eta}, \label{eq:bound_BCT_gapless_end}
\end{equation}
 as reported in the main text. See Figure \ref{fig:check_prop_5}
for numerical simulations confirming Eqs.~\eqref{eq:eta_BCT_gapless_end} and~\eqref{eq:bound_BCT_gapless_end}.

\section{Gap closing in the middle: Numerical check}
\label{app:gap_mid}

\begin{figure}
\subfigure[]{\includegraphics[width=0.9\columnwidth]{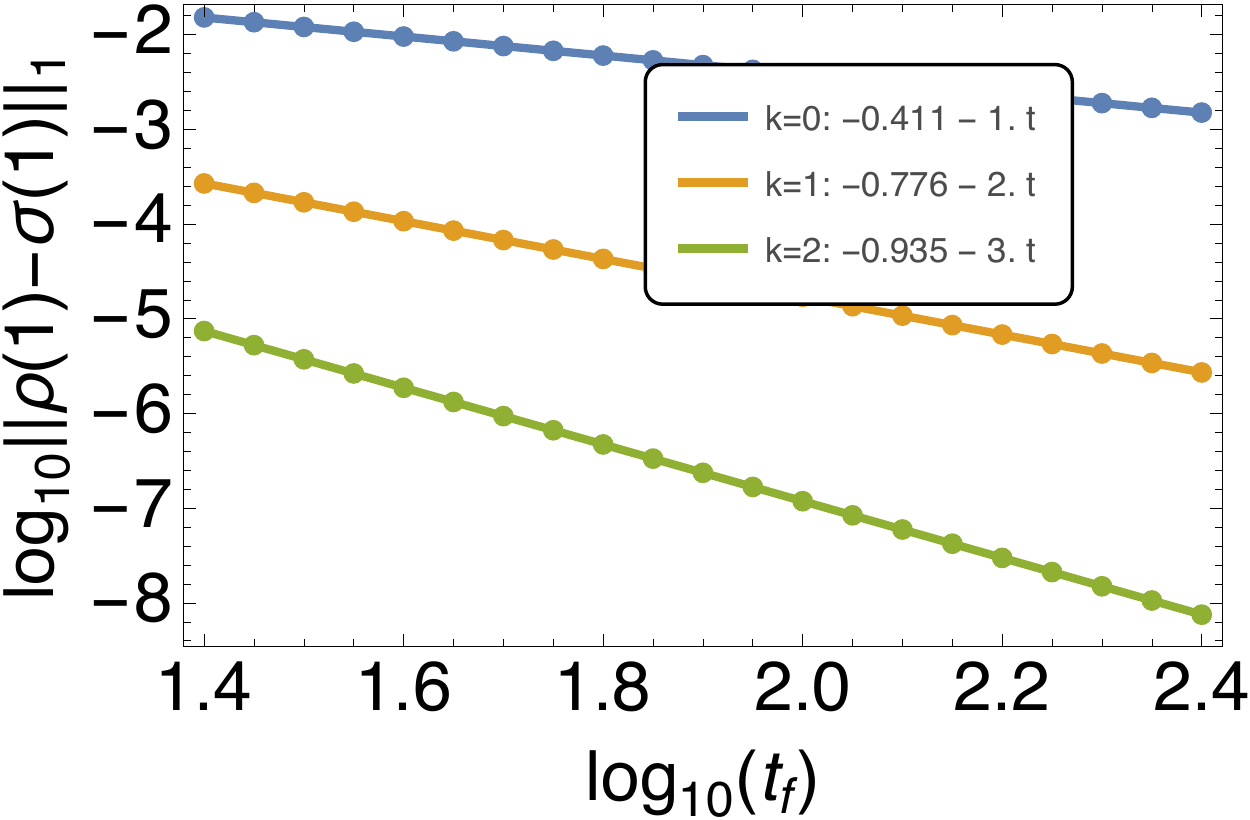}\label{fig:check_prop_4-a}}
\subfigure[]{\includegraphics[width=0.9\columnwidth]{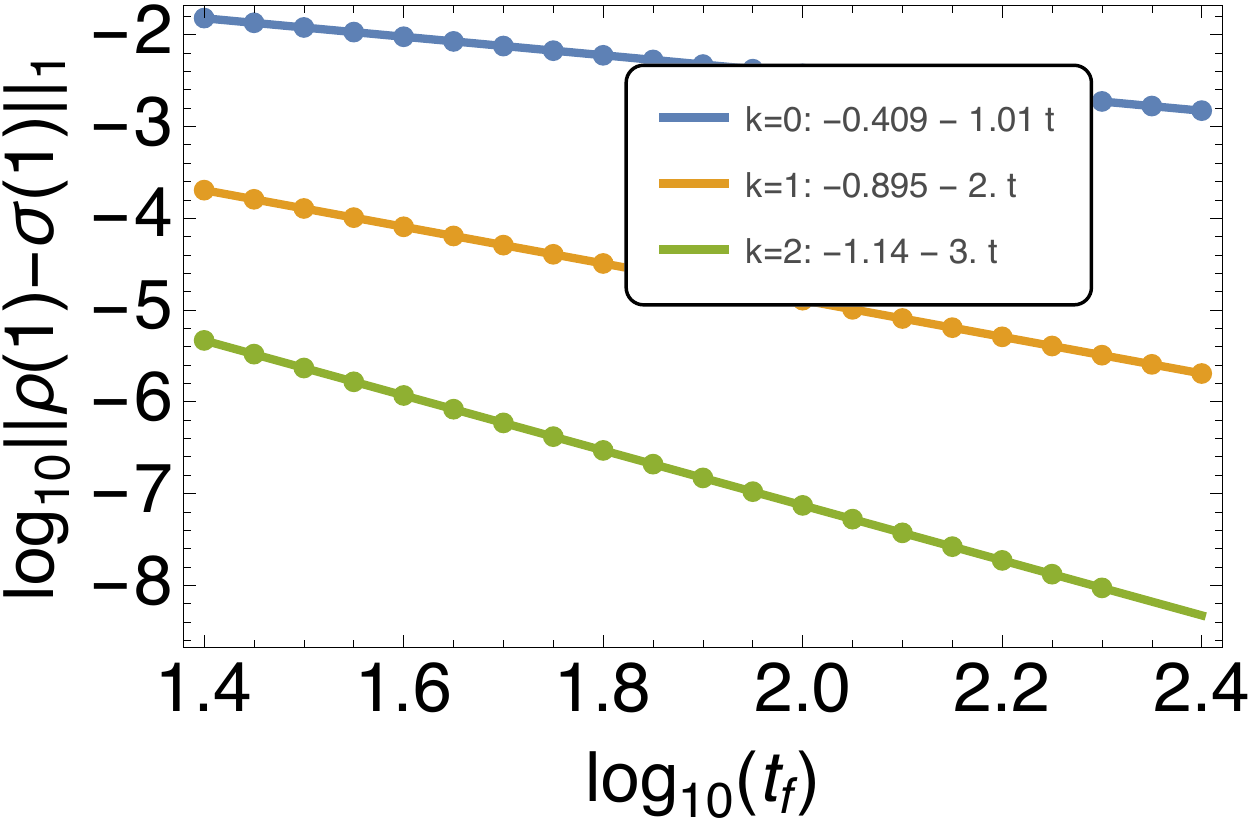}\label{fig:check_prop_4-b}}
\caption{Numerical simulations for the case where the Liouvillian gap closes in the middle of the evolution and one performs BC at the end. 
Results are for  the single qubit AME with the Hamiltonians given in (a) Eq.~\eqref{eq:ham_check1} and (b) Eq.~\eqref{eq:ham_check2}. 
The bath spectral density is Ohmic [Eq.~\eqref{eq:Ohmic}] and the 
parameters are $g=1,\,\beta=1,$  in arbitrary units. Dots are results of solving for the evolution while the continuous lines are
linear fits. The fits are in an excellent agreement with $\eta=k+1$, as shown by the the last column of the two legends.
}
\label{fig:check_prop_4}
\end{figure}

Here we present numerical simulations for the case where the Liouvillian gap closes in the middle of the evolution and one performs boundary cancellation at the end, as discussed in Sec.~\ref{sec:Theory-B} of the main text. 

We first consider the following Hamiltonian:
\bes
\begin{align}
H(\tau) & =\sigma^{y}+(\tau-1)^{k+1}X+\left(\tau-1\right)^{k+2}Y \label{eq:ham_check1}\\
X:= & \left(-1\right)^{k}\left[\sigma^{x}-2^{k+2}\sigma^{z}+\left(2^{k+2}-1\right)\sigma^{y}\right]\\
Y:= & 2\left(-1\right)^{k}\left[\sigma^{x}-2^{k+1}\sigma^{z}+\left(2^{k+1}-1\right)\sigma^{y}\right].
\end{align}
\ees
One can check that $H\left(0\right)=\sigma^{x}$, $H\left(1/2\right)=\sigma^{z}$,
$H(1)=\sigma^{y}$ and that the first $k$ derivatives of $H(\tau)$
at $\tau=1$ are zero. Hence the Liouvillian gap closes with exponent $\alpha=2$
at $\tau=1/2$ and there is a BC schedule at the end. 
Numerical results confirming the expected scaling $\eta = k+1$ are presented in Fig.~\ref{fig:check_prop_4-a}. 

Another possibility is to use the following Hamiltonian, which performs the traditional interpolation from a transverse to a longitudinal field:
\bes
\begin{align}
H\left(\tau\right) & =s\left(\tau\right)\sigma^{x}+\left[1-s\left(\tau\right)\right]\sigma^{z} \label{eq:ham_check2}\\
s(\tau) & :=\left(2\left(2\frac{\mathrm{B}_{(1+\tau)/2}(1,k+1)}{\mathrm{B}_{1}(1,k+1)}-1\right)-1\right)^{2}. \label{eq:sched}
\end{align}
\ees

The schedule in Eq.~\eqref{eq:sched}  starts with $\sigma^{x}$ at $s(0)=1$ and ends with $\sigma^{z}$ at $s(1)=1$, where it has $k$ vanishing derivatives. The Liouvillian gap closes with $\alpha=4$ at some point $s^*_k$. Numerical results again confirm the expected scaling $\eta = k+1$, as shown in Fig.~\ref{fig:check_prop_4-b}.

All our numerical results (we have also performed numerics for up to 4 qubit Hamiltonians, not shown here) predict the following scaling for the adiabatic error in this setting:
\bes
\begin{align}
\left\Vert \rho(1)-\sigma(1)\right\Vert _{1}&\leq \frac{C}{t_{f}^{\eta}}\ , \\
\mathrm{with\,\,} & \eta = k+1\  . 
\end{align}
\ees
In other words, the adiabatic error is insensitive to the closing of the gap in the middle of the evolution. 
Note that this result is different from the analogous one in the closed system setting where the adiabatic error scales with an exponent $\eta = 1/(\alpha +1)$, where $\alpha$ in this case is the exponent that determines the closing of the Hamiltonian (not Liouvillian) gap. 


\section{Proof that $\langle0|\sigma_{j}^{z}|n\rangle=O\left(A^{q}\right)$}\label{app:Proof-of-freezing}

In the main text we stated that $\langle0|\sigma_{j}^{z}|n\rangle=O\left(A^{q}\right)$, and hence that the transition matrix elements $W_{0n}=O\left(A^{2q}\right)$ are suppressed for $n\ge 1$.
Generalizing beyond the scenario in the main text, we now allow for degeneracy in the eigenstates of $H_{Z}$. 

We assume the transverse field is not zero and investigate
the behavior of the rates when $A\to0$. For this value of $A(t)\neq0$,
let $H_{S}=\sum_{n}E_{n}\tilde{P}_{n}$ be the spectral decomposition
of the system Hamiltonian, where $\tilde{P}_{n}$ are the (possibly degenerate)
spectral projectors. The Bohr frequencies have the form $\omega=E_{b}-E_{a}$
and we can label the jump operators $A_{j}(\omega)$ as 
\begin{equation}
A_{ab,j}(t)=\tilde{P}_{a}(t)\sigma_{j}^{z}\tilde{P}_{b}(t).
\end{equation}
We omit the time dependence from here on. 

Let us consider the operators $A_{ab,j}$ perturbatively around $A(t)=0$.
We denote by $P_{a}$ the (unperturbed) spectral projectors
at $A(t)=0$, which are diagonal in the computational basis. By assumption, the degeneracy of $\tilde{P}_{a}$
does not change along the anneal, so that the $\tilde{P}_{a}$
are smoothly connected with the $P_{a}$. We use the Kato perturbation
theory formalism~\cite{kato_perturbation_1995}, and denote the perturbation by
$V:=H_{X}$. Moreover, for level $a$, we define
the reduced resolvent as 
\begin{equation}
R_{a}=\sum_{n\neq a}\frac{P_{n}}{E_{n}-E_{a}}.
\end{equation}
For fixed $a\neq b$ we expand the jump operators in powers of the
transverse field:
\begin{equation}
A_{ab,j}=\tilde{P}_{a}\sigma_{j}^{z}\tilde{P}_{b}:=\sum_{n=0}A^{n}C_{n}.
\end{equation}
At zeroth order we have (since $a\neq b$)
\begin{equation}
C_{0}=P_{a}\sigma_{j}^{z}P_{b} = P_{a}P_{b}\sigma_{j}^{z}=0.
\end{equation}
At first order, perturbation theory gives 
\bes
\begin{align}
C_{1} & =-P_{a}\sigma_{j}^{z}\left(P_{b}VR_{b}+R_{b}VP_{b}\right)\notag \\
&\qquad -\left(P_{a}VR_{a}+R_{a}VP_{a}\right)\sigma_{j}^{z}P_{b}\\
 & =-P_{a}\sigma_{j}^{z}R_{b}VP_{b}-P_{a}VR_{a}\sigma_{j}^{z}P_{b}\\
 & =-\sigma_{j}^{z}R_{b}P_{a}VP_{b}-P_{a}VP_{b}R_{a}\sigma_{j}^{z}\,.
\end{align}
\ees
Now recall that $V=H_{X}=2S_{x}=(S_{+}+S_{-})$ where $S_{x}$ ($S_{\pm}$)
is the total spin-$x$ operator (ladder operator) and hence $V$ changes
the Hamming distance by one. As a consequence, $P_{a}VP_{b}=0$ if
$\text{HD}(a,b)\neq1$ where we denoted by $\text{HD}(a,b)$ the smallest Hamming
distance among the representative states of the projectors $P_{a}$
and $P_{b}$. I.e.,~if $P_{a}=\sum_{\mu=1}^{d_{a}}|\mu a\rangle\langle\mu a|$,
$P_{b}=\sum_{\nu=1}^{d_{b}}|\nu b\rangle\langle\nu b|$ for some Ising
states $|\mu a\rangle,\,|\nu b\rangle$, 
\begin{equation}
\text{HD}(a,b):=\min_{\mu,\nu}\text{HD}\left(|\mu a\rangle,|\nu b\rangle\right).
\end{equation}

At second order we obtain
\begin{align}
C_{2} & =\left(P_{a}VR_{a}+R_{a}VP_{a}\right)\sigma_{j}^{z}\left(P_{b}VR_{b}+R_{b}VP_{b}\right)+\nonumber \\
 & P_{a}\sigma_{j}^{z}\left(R_{b}VP_{b}VR_{b}+R_{b}VR_{b}VP_{b}-R_{b}^{2}VP_{b}VP_{b}\right)+\nonumber \\
 & \left(P_{a}VR_{a}VR_{a}+R_{a}VP_{a}VR_{a}-P_{a}VP_{a}VR_{a}^{2}\right)\sigma_{j}^{z}P_{b}\,.
\end{align}
The formulas become increasingly complicated, but we see that at order
$k$, $C_{k}$ is sum of terms of the form $O_{z}^{(1)}VO_{z}^{(2)}\cdots VO_{z}^{(k+1)}$
where there are $k$ $V$'s and we denote by $O_{z}^{(j)}$ any
operator diagonal in the same basis as $\sigma^{z}$. Two of the $O_{z}^{(i)}$ ($i=1,2,\ldots,k+1$)
comprise $P_{a}$ and $P_{b}$. Let $d_{s}$ be the number of $V$'s sandwiched
between $P_{a}$ and $P_{b}$. Then this term is zero if $\text{HD}(a,b)>d_{s}$.
Since at order $k$, $d_{s}$ is necessarily $\le k$,
we see that $C_{k}$ is zero if $\text{HD}(a,b)>k$. Alternatively, note that
the effect of each $O_{z}^{(j)}$ on computational basis states is simply
that of creating a multiplicative constant but otherwise it leaves
computational basis states unchanged. So, apart from a multiplicative constant
each term in $C_{k}$ has the form $|\mu a\rangle\langle\mu a|V^{d_{s}}|\nu b\rangle\langle\nu b|$.
Summarizing, if $\text{HD}(a,b)=q$, $A_{ab,j}=O\left(A^{q}\right).$ Since
in the Pauli matrix $W_{ab}$  there appear two Lindblad operators per Bohr frequency
($a,b$ pair) we have just shown that  $W_{ab}=O\left(A^{2q}\right)$. In particular, this is true if the level $a=0$
and $b=n$ are non-degenerate as it is the case in the main text. 

\begin{figure}
\includegraphics[width=0.9\columnwidth]{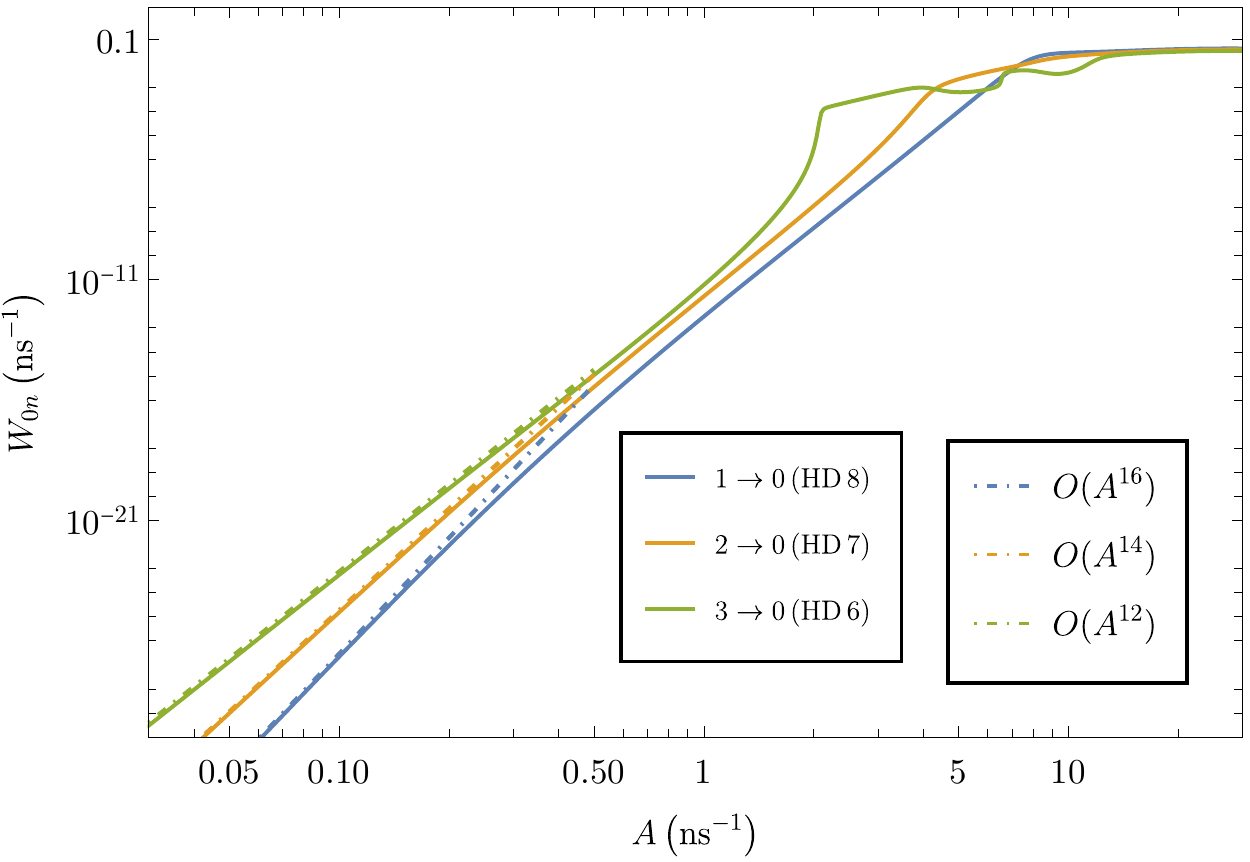}
\caption{The transition matrix elements $W_{0n}$ of the FM-gadget for $n=1,2,3$ (solid), for which $q=8,7,6$, respectively, as a function of $A$, along with $A^{2q}$ shifted to most closely align with the $W_{0n}(A)$ curves (dashed). The agreement is clearly visible for sufficiently small $A$. 
The transverse field schedule used is $A(s)={30\,\mathrm{ns}^{-1}}(1-s)$ while $B(s) = {10\,\mathrm{ns}^{-1}}$ is held constant. This allows the scaling to be seen clearly for small $A$ without changing the overall energy scale of the Hamiltonian.
}
\label{fig:A2q}
\end{figure}

Figure~\ref{fig:A2q} shows that this prediction is borne out in numerical simulations. 

\begin{figure*}
\subfigure[\ ]{\includegraphics[width=0.48\textwidth]{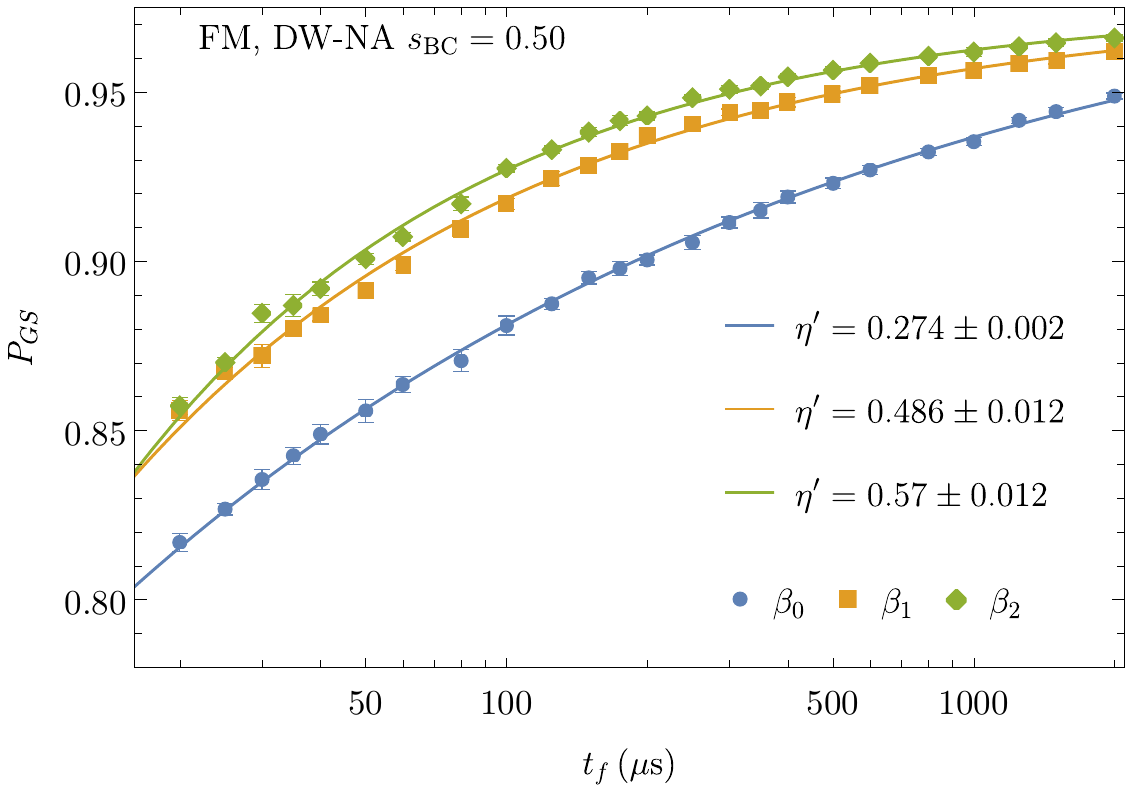}}
\subfigure[\ ]{\includegraphics[width=0.48\textwidth]{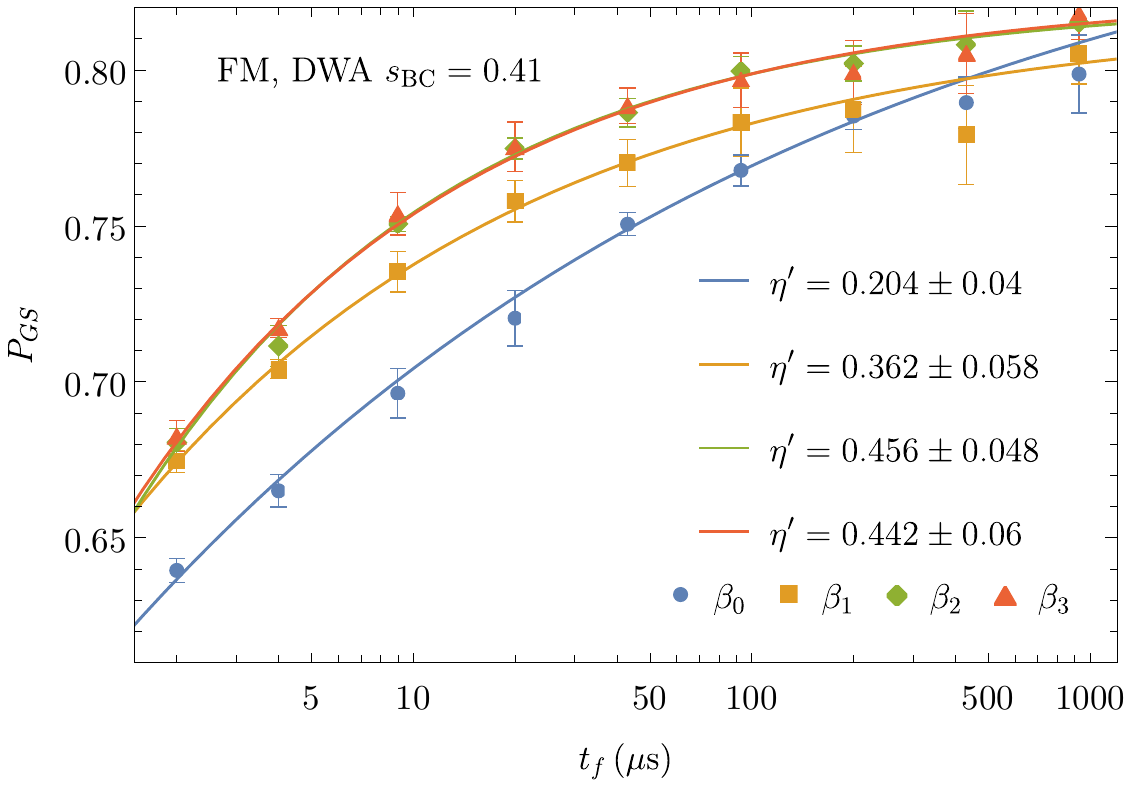}}
\caption{Ground state probabilities and fits for the BCP applied to the FM gadget
using (a) DW2KQ at NASA QuAIL and (b) D-Wave Advantage at the appropriate
values of $s_{\text{BC}}$ where the fits are most distinguishable. 
}
\label{fig:altdw} 
\end{figure*}

\section{Determination of the Freezing Point}
\label{app:freezing}

\begin{figure*}
    \subfigure{\includegraphics[width=0.48\textwidth]{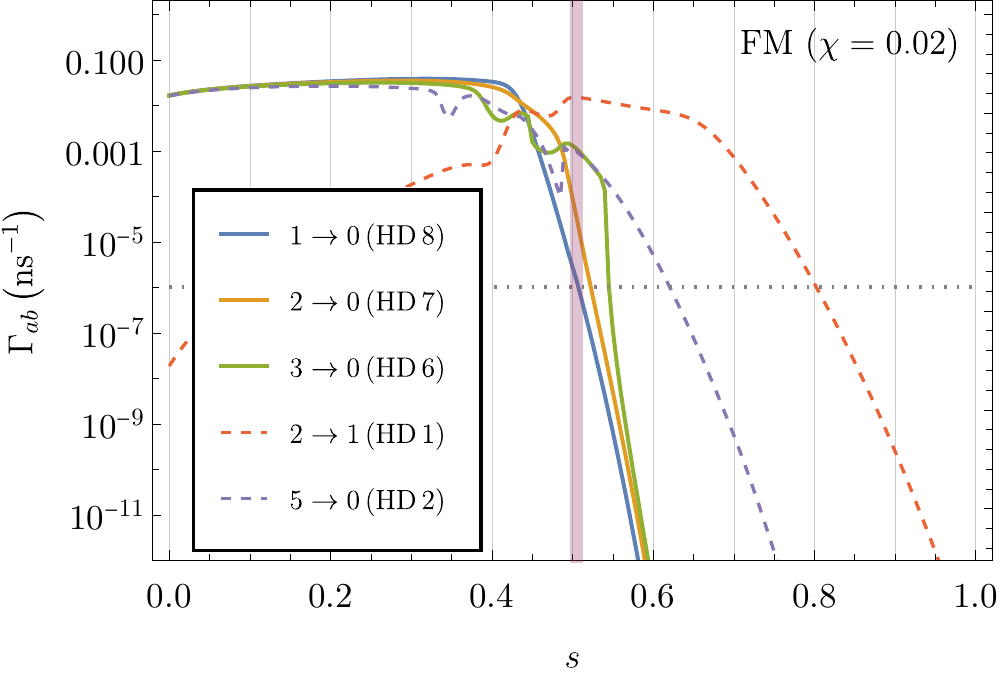}}
    \subfigure{\includegraphics[width=0.48\textwidth]{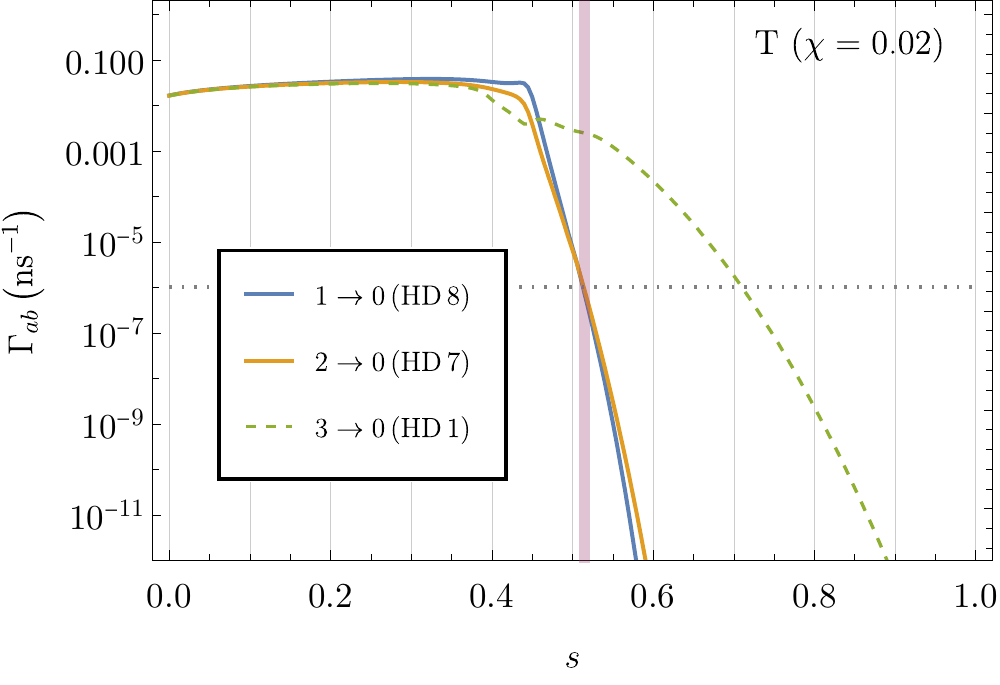}}
    \caption{Thermal transition rates of the adiabatic master
        equation (with $\eta_0 g^2=5.0\times10^{-4}$ )
        for some pairs of energy eigenstates
        $a\to b$ as the DW-LN schedule progresses linearly for the FM-gadget
        (top) and the T-gadget (bottom). The horizontal dotted line denotes
        the rate $1/t_{f}$, which is on the order of magnitude of the smallest
        rate that can be reliably observed within an anneal time of $t_{f}=10^3\,\mathrm{\mu s}$.
        Below this line, thermal relaxation events rapidly become extremely
        rare, hence the onset of frozen dynamics during the anneal. The Hamming
        distances between the final computational states corresponding to
        the eigenstate transitions are also listed. When an excited state
        is only a small Hamming distance away from the ground state, the transition
        freezes later in the anneal since only a much smaller number of qubits
        need to tunnel through. This is the case for the $2\to1$ and $3\to1$
        transitions in the FM-gadget, and the $3\to0$ transition in the T-gadget.}
    \label{fig:adb_therm} 
\end{figure*}

The resulting transition rates
between energy eigenstates in the AME are shown in Fig.~\ref{fig:adb_therm}
for both the FM and the T-gadgets. Since the magnitude of available
annealing times are limited to $10^{3}\,\mathrm{\mu s}$, we expect
that a transition will become too rare to observe once it drops below
$1/t_{f}\approx10^{-6}$\;GHz. With the majority of the population
in the ground state and first excited state, we therefore expect freezing
to occur when $\Gamma_{10}$ goes below this value. This intersection
determines the freeze-out points highlighted in Fig.~\ref{fig:fm_gaps}.

\section{Schedule Construction}
\label{app:betasched}

The annealing schedule of the DW-LN used in our main experiments
is shown in Fig.~\ref{fig:dw2kq_schedule}. All simulations are based
on this schedule and its operating temperature at the time of sample
collection (13.5 $\mathrm{mK}$).

First, suppose $s^{*}(\tau)=\beta_{k}(\tau)$ is a beta-schedule terminating
at $s=1$ and let $\tau^{*}(s)$ denote its inverse function. The
construction of a piecewise-linear approximation to $s^{*}(\tau)$
is as follows: 
\begin{enumerate}
\item The initial point $(\tau_{0},s_{0})=(0,0)$ and the final point $(\tau_{10},s_{10})=(1,1)$. 
\item A separation point $(\tau_{5},s_{5})=(\tau^{*}(s_{c}),s_{c})$ for
some choice of $s_{c}$. 
\item Linearly partition $s_{1},\ldots s_{4}$, i.e., assign $s_{j}=js_{c}/5$
and set $\tau_{j}=\tau^{*}(s_{j})$ for $j=1\ldots4$. 
\item Logarithmically partition $s_{6},\ldots,s_{9}$ between $s_{c}$ and
$1$, i.e., assign $s_{j+1}=1-(1-s_{j})/2$) and set $\tau_{j}=\tau^{*}(s_{j})$
for $j=6\ldots9$. 
\end{enumerate}
Thus, for a schedule that terminates at the ramp point $s_{\text{BC}}$, we
have the following $12$-point scheme: 
\begin{enumerate}
\item Construct the first $11$ points according to the above procedure,
and then rescale each $s_{j}\to s_{\text{BC}}s_{j}$. 
\item The final $12$th point is $(\tau_{11},s_{11})=(\tau_{10}+\delta,1)$,
where $\delta$ is constrained by the maximum ramp slope of the hardware,
i.e., $\delta=1\mu\textrm{s}/t_{f}$. 
\end{enumerate}
Note that when the ramp is included $s=1$ corresponds to $\tau=1+\delta$
and $s_{\text{BC}}<1$. Aside from the order $k$, there are the three annealing
parameters $s_{c},s_{\text{BC}},$ and $t_{f}$ that must be specified for
the protocol. We took $s_{c}=0.9$ for the construction of the beta
schedules.

Adjusting the $s_{c}$ parameter to $0.95$ to improve the precision
of the analytic behavior of the beta schedule did not have a substantial
effect on the ground state probabilities of the empirical D-Wave results.
It is likely that such a precision exceeds what can be reliably and
repeatedly implemented on the hardware.

\section{Crosstalk}
\label{app:xtalk}

It is known that, due to the difficulty of hardware fabrication of
the D-Wave processors, programming the couplings and field to have
certain values induces non-zero fields and couplings also on other
links of the graph, despite designs meant to minimize such crosstalk~\cite{Harris:2009uw}. We use the ferromagnetic
crosstalk model defined by the following transformations~\cite{albashConsistency15}:
\bes
\label{eq:xtalk}
\begin{align}
h_{i} & \mapsto h_{i}-\chi\sum_{k\neq i}J_{ik}h_{k}\\
J_{ij} & \mapsto J_{ij}+\chi\sum_{k\neq i,j}J_{ik}J_{jk}
\end{align}
\ees
 where $\chi$ is the crosstalk strength. We assume a strength
of $\chi=0.02$ throughout. We have checked that optimizing the value of $\chi$
further does not substantially affect our numerical results. 
This crosstalk has the effect of breaking the degeneracy of the first and second excited states shown in Fig.~\ref{fig:fm_gaps}.

We note
that the transformation~\eqref{eq:xtalk} commutes with gauge (or spin-reversal) transformations
of the Hamiltonian~\cite{q-sig,Job:2017aa}, so population shifts
due to crosstalk do not depend on the gauge.

\section{Data Collection and Fitting}
\label{app:data}

The probabilities at a given anneal time are obtained from a sample
of 20 submissions (10 for QAC), where each submission programs 5 random
global gauges of the gadget repeated over 31 unit cells, and samples
each gauge 200 times (see Ref.~\cite{Job:2017aa} for a review of
the gauging procedure and other best practices). To mitigate time-fluctuating
systematic biases, submissions for the same anneal time are not consecutively
submitted (anneal times are cycled for every schedule). 
Estimates for the ground state probability are taken from the average of the
ground state probabilities of the 20 submissions. Error bars for each measured ground state probabilities are derived from twice the standard deviation over all submissions.
Either by non-linear fits on the transformed data, or by linear fits on the log-log transformed data assuming a fixed $\bar{P}_{\text{GS}}^*$, 
we find the
constant $C$ and exponent $\eta$ such that the $\chi^{2}$ with
$D_{\mathrm{\text{GS}}}$ is minimized. 
Nonlinear fits are performed as a function of $1/t_{f}$. 
The estimates of $\eta'$ for the D-Wave data are obtained by  performing fits on 100 bootstrap samples of the collected ground state probabilities.
We use the median values of the parameter distributions as the estimators for the parameters as they are robust to the outliers in bootstrapping that occur. The error estimate is derived from half of the range of the 90\% confidence interval of the parameter distribution. 
The non-linear fits are evaluated using the Mathematica function \texttt{NonlinearModelFit} with the method \texttt{NMinimize}(\texttt{NelderMead}). The non-linear $\chi^2$ objective was weighed by the variances of the ground state probabilities.

\section{Results for alternative D-Wave QPUs}
\label{app:altdw}

We performed similar BCP quantum annealing experiments with the DW2KQ
at NASA QuAIL (DW-NA) and with the D-Wave Advantage (DWA) annealers,
while the main results used the low-noise DW2KQ processor through
D-Wave Leap (DW-LN) The scaling behavior of the ground state probabilities
at appropriate values of $s_{\text{BC}}$ is shown in Fig.~\ref{fig:altdw}.
These results should be contrasted with Fig.~\ref{fig:fm_sq_scal}(a),
for the DW-LN. All three QPUs have different schedules and thermal
properties that affect the qualitative behavior of the BCP, so it
is difficult to make a direct comparison between the devices. However,
$\eta$ is generally larger in DW-LN compared to either DW-NA or
DWA when $s_{\text{BC}}$ is in the appropriate region for each annealer.
Furthermore, DWA appears to exhibit more significant time-dependent
deviations, as can be noted in the $\beta_{2}$ schedule at $t_{f}=431$
in Fig.~\ref{fig:altdw}.


\begin{thebibliography}{69}%
    \makeatletter
    \providecommand \@ifxundefined [1]{%
        \@ifx{#1\undefined}
    }%
    \providecommand \@ifnum [1]{%
        \ifnum #1\expandafter \@firstoftwo
        \else \expandafter \@secondoftwo
        \fi
    }%
    \providecommand \@ifx [1]{%
        \ifx #1\expandafter \@firstoftwo
        \else \expandafter \@secondoftwo
        \fi
    }%
    \providecommand \natexlab [1]{#1}%
    \providecommand \enquote  [1]{``#1''}%
    \providecommand \bibnamefont  [1]{#1}%
    \providecommand \bibfnamefont [1]{#1}%
    \providecommand \citenamefont [1]{#1}%
    \providecommand \href@noop [0]{\@secondoftwo}%
    \providecommand \href [0]{\begingroup \@sanitize@url \@href}%
    \providecommand \@href[1]{\@@startlink{#1}\@@href}%
    \providecommand \@@href[1]{\endgroup#1\@@endlink}%
    \providecommand \@sanitize@url [0]{\catcode `\\12\catcode `\$12\catcode
        `\&12\catcode `\#12\catcode `\^12\catcode `\_12\catcode `\%12\relax}%
    \providecommand \@@startlink[1]{}%
    \providecommand \@@endlink[0]{}%
    \providecommand \url  [0]{\begingroup\@sanitize@url \@url }%
    \providecommand \@url [1]{\endgroup\@href {#1}{\urlprefix }}%
    \providecommand \urlprefix  [0]{URL }%
    \providecommand \Eprint [0]{\href }%
    \providecommand \doibase [0]{http://dx.doi.org/}%
    \providecommand \selectlanguage [0]{\@gobble}%
    \providecommand \bibinfo  [0]{\@secondoftwo}%
    \providecommand \bibfield  [0]{\@secondoftwo}%
    \providecommand \translation [1]{[#1]}%
    \providecommand \BibitemOpen [0]{}%
    \providecommand \bibitemStop [0]{}%
    \providecommand \bibitemNoStop [0]{.\EOS\space}%
    \providecommand \EOS [0]{\spacefactor3000\relax}%
    \providecommand \BibitemShut  [1]{\csname bibitem#1\endcsname}%
    \let\auto@bib@innerbib\@empty
    \bibitem [{\citenamefont {Farhi}\ \emph {et~al.}(2000)\citenamefont {Farhi},
        \citenamefont {Goldstone}, \citenamefont {Gutmann},\ and\ \citenamefont
        {Sipser}}]{farhiQuantum00}%
    \BibitemOpen
    \bibfield  {author} {\bibinfo {author} {\bibfnamefont {E.}~\bibnamefont
            {Farhi}}, \bibinfo {author} {\bibfnamefont {J.}~\bibnamefont {Goldstone}},
        \bibinfo {author} {\bibfnamefont {S.}~\bibnamefont {Gutmann}}, \ and\
        \bibinfo {author} {\bibfnamefont {M.}~\bibnamefont {Sipser}},\ }\href@noop {}
    {\bibfield  {journal} {\bibinfo  {journal} {arXiv:quant-ph/0001106}\ }
        (\bibinfo {year} {2000})},\ \Eprint {http://arxiv.org/abs/quant-ph/0001106}
    {arXiv:quant-ph/0001106} \BibitemShut {NoStop}%
    \bibitem [{\citenamefont {Kadowaki}\ and\ \citenamefont
        {Nishimori}(1998)}]{kadowakiQuantum98}%
    \BibitemOpen
    \bibfield  {author} {\bibinfo {author} {\bibfnamefont {T.}~\bibnamefont
            {Kadowaki}}\ and\ \bibinfo {author} {\bibfnamefont {H.}~\bibnamefont
            {Nishimori}},\ }\href {\doibase 10.1103/PhysRevE.58.5355} {\bibfield
        {journal} {\bibinfo  {journal} {Physical Review E}\ }\textbf {\bibinfo
            {volume} {58}},\ \bibinfo {pages} {5355} (\bibinfo {year}
        {1998})}\BibitemShut {NoStop}%
    \bibitem [{\citenamefont {Amin}\ \emph {et~al.}(2018)\citenamefont {Amin},
        \citenamefont {Andriyash}, \citenamefont {Rolfe}, \citenamefont
        {Kulchytskyy},\ and\ \citenamefont {Melko}}]{Amin:2016}%
    \BibitemOpen
    \bibfield  {author} {\bibinfo {author} {\bibfnamefont {M.~H.}\ \bibnamefont
            {Amin}}, \bibinfo {author} {\bibfnamefont {E.}~\bibnamefont {Andriyash}},
        \bibinfo {author} {\bibfnamefont {J.}~\bibnamefont {Rolfe}}, \bibinfo
        {author} {\bibfnamefont {B.}~\bibnamefont {Kulchytskyy}}, \ and\ \bibinfo
        {author} {\bibfnamefont {R.}~\bibnamefont {Melko}},\ }\href {\doibase
        10.1103/PhysRevX.8.021050} {\bibfield  {journal} {\bibinfo  {journal}
            {Physical Review X}\ }\textbf {\bibinfo {volume} {8}},\ \bibinfo {pages}
        {021050} (\bibinfo {year} {2018})}\BibitemShut {NoStop}%
    \bibitem [{\citenamefont {Kato}(1950)}]{katoAdiabatic50}%
    \BibitemOpen
    \bibfield  {author} {\bibinfo {author} {\bibfnamefont {T.}~\bibnamefont
            {Kato}},\ }\href {\doibase 10.1143/JPSJ.5.435} {\bibfield  {journal}
        {\bibinfo  {journal} {Journal of the Physical Society of Japan}\ }\textbf
        {\bibinfo {volume} {5}},\ \bibinfo {pages} {435} (\bibinfo {year}
        {1950})}\BibitemShut {NoStop}%
    \bibitem [{\citenamefont {Jansen}\ \emph {et~al.}(2007)\citenamefont {Jansen},
        \citenamefont {Ruskai},\ and\ \citenamefont {Seiler}}]{jansenBounds07}%
    \BibitemOpen
    \bibfield  {author} {\bibinfo {author} {\bibfnamefont {S.}~\bibnamefont
            {Jansen}}, \bibinfo {author} {\bibfnamefont {M.-B.}\ \bibnamefont {Ruskai}},
        \ and\ \bibinfo {author} {\bibfnamefont {R.}~\bibnamefont {Seiler}},\ }\href
    {\doibase 10.1063/1.2798382} {\bibfield  {journal} {\bibinfo  {journal}
            {Journal of Mathematical Physics}\ }\textbf {\bibinfo {volume} {48}},\
        \bibinfo {pages} {102111} (\bibinfo {year} {2007})}\BibitemShut {NoStop}%
    \bibitem [{\citenamefont {Joye}(2007)}]{joye_general_2007}%
    \BibitemOpen
    \bibfield  {author} {\bibinfo {author} {\bibfnamefont {A.}~\bibnamefont
            {Joye}},\ }\href {http://link.springer.com/article/10.1007/s00220-007-0299-y}
    {\bibfield  {journal} {\bibinfo  {journal} {Commun. Math. Phys.}\ }\textbf
        {\bibinfo {volume} {275}},\ \bibinfo {pages} {139} (\bibinfo {year}
        {2007})}\BibitemShut {NoStop}%
    \bibitem [{\citenamefont {Garrido}\ and\ \citenamefont
        {Sancho}(1962)}]{garridoDegree62}%
    \BibitemOpen
    \bibfield  {author} {\bibinfo {author} {\bibfnamefont {L.~M.}\ \bibnamefont
            {Garrido}}\ and\ \bibinfo {author} {\bibfnamefont {F.~J.}\ \bibnamefont
            {Sancho}},\ }\href {\doibase 10.1016/0031-8914(62)90109-X} {\bibfield
        {journal} {\bibinfo  {journal} {Physica}\ }\textbf {\bibinfo {volume} {28}},\
        \bibinfo {pages} {553} (\bibinfo {year} {1962})}\BibitemShut {NoStop}%
    \bibitem [{\citenamefont {Nenciu}(1993)}]{nenciu_linear_1993}%
    \BibitemOpen
    \bibfield  {author} {\bibinfo {author} {\bibfnamefont {G.}~\bibnamefont
            {Nenciu}},\ }\href {http://link.springer.com/article/10.1007/BF02096616}
    {\bibfield  {journal} {\bibinfo  {journal} {Commun. Math. Phys.}\ }\textbf
        {\bibinfo {volume} {152}},\ \bibinfo {pages} {479} (\bibinfo {year}
        {1993})}\BibitemShut {NoStop}%
    \bibitem [{\citenamefont {Hagedorn}\ and\ \citenamefont
        {Joye}(2002)}]{Hagedorn:2002kx}%
    \BibitemOpen
    \bibfield  {author} {\bibinfo {author} {\bibfnamefont {G.~A.}\ \bibnamefont
            {Hagedorn}}\ and\ \bibinfo {author} {\bibfnamefont {A.}~\bibnamefont
            {Joye}},\ }\href
    {http://www.sciencedirect.com/science/article/pii/S0022247X01977650}
    {\bibfield  {journal} {\bibinfo  {journal} {Journal of Mathematical Analysis
                and Applications}\ }\textbf {\bibinfo {volume} {267}},\ \bibinfo {pages}
        {235} (\bibinfo {year} {2002})}\BibitemShut {NoStop}%
    \bibitem [{\citenamefont {Lidar}\ \emph {et~al.}(2009)\citenamefont {Lidar},
        \citenamefont {Rezakhani},\ and\ \citenamefont {Hamma}}]{lidarAdiabatic09}%
    \BibitemOpen
    \bibfield  {author} {\bibinfo {author} {\bibfnamefont {D.~A.}\ \bibnamefont
            {Lidar}}, \bibinfo {author} {\bibfnamefont {A.~T.}\ \bibnamefont
            {Rezakhani}}, \ and\ \bibinfo {author} {\bibfnamefont {A.}~\bibnamefont
            {Hamma}},\ }\href {\doibase 10.1063/1.3236685} {\bibfield  {journal}
        {\bibinfo  {journal} {Journal of Mathematical Physics}\ }\textbf {\bibinfo
            {volume} {50}},\ \bibinfo {pages} {102106} (\bibinfo {year}
        {2009})}\BibitemShut {NoStop}%
    \bibitem [{\citenamefont {Wiebe}\ and\ \citenamefont
        {Babcock}(2012)}]{wiebeImproved12}%
    \BibitemOpen
    \bibfield  {author} {\bibinfo {author} {\bibfnamefont {N.}~\bibnamefont
            {Wiebe}}\ and\ \bibinfo {author} {\bibfnamefont {N.~S.}\ \bibnamefont
            {Babcock}},\ }\href {\doibase 10.1088/1367-2630/14/1/013024} {\bibfield
        {journal} {\bibinfo  {journal} {New Journal of Physics}\ }\textbf {\bibinfo
            {volume} {14}},\ \bibinfo {pages} {013024} (\bibinfo {year}
        {2012})}\BibitemShut {NoStop}%
    \bibitem [{\citenamefont {Ge}\ \emph {et~al.}(2016)\citenamefont {Ge},
        \citenamefont {Moln{\'a}r},\ and\ \citenamefont {Cirac}}]{Ge:2015wo}%
    \BibitemOpen
    \bibfield  {author} {\bibinfo {author} {\bibfnamefont {Y.}~\bibnamefont
            {Ge}}, \bibinfo {author} {\bibfnamefont {A.}~\bibnamefont {Moln{\'a}r}}, \
        and\ \bibinfo {author} {\bibfnamefont {J.~I.}\ \bibnamefont {Cirac}},\ }\href
    {http://link.aps.org/doi/10.1103/PhysRevLett.116.080503} {\bibfield
        {journal} {\bibinfo  {journal} {Physical Review Letters}\ }\textbf {\bibinfo
            {volume} {116}},\ \bibinfo {pages} {080503} (\bibinfo {year}
        {2016})}\BibitemShut {NoStop}%
    \bibitem [{\citenamefont {Campos~Venuti}\ and\ \citenamefont
        {Lidar}(2018)}]{camposvenutiError18}%
    \BibitemOpen
    \bibfield  {author} {\bibinfo {author} {\bibfnamefont {L.}~\bibnamefont
            {Campos~Venuti}}\ and\ \bibinfo {author} {\bibfnamefont {D.~A.}\ \bibnamefont
            {Lidar}},\ }\href {\doibase 10.1103/PhysRevA.98.022315} {\bibfield  {journal}
        {\bibinfo  {journal} {Physical Review A}\ }\textbf {\bibinfo {volume} {98}},\
        \bibinfo {pages} {022315} (\bibinfo {year} {2018})}\BibitemShut {NoStop}%
    \bibitem [{\citenamefont {Albash}\ and\ \citenamefont
        {Lidar}(2015)}]{albashDecoherence15}%
    \BibitemOpen
    \bibfield  {author} {\bibinfo {author} {\bibfnamefont {T.}~\bibnamefont
            {Albash}}\ and\ \bibinfo {author} {\bibfnamefont {D.~A.}\ \bibnamefont
            {Lidar}},\ }\href {\doibase 10.1103/PhysRevA.91.062320} {\bibfield  {journal}
        {\bibinfo  {journal} {Physical Review A}\ }\textbf {\bibinfo {volume} {91}},\
        \bibinfo {pages} {062320} (\bibinfo {year} {2015})}\BibitemShut {NoStop}%
    \bibitem [{\citenamefont {Lidar}(2019{\natexlab{a}})}]{Lidar:2019ab}%
    \BibitemOpen
    \bibfield  {author} {\bibinfo {author} {\bibfnamefont {D.~A.}\ \bibnamefont
            {Lidar}},\ }\href {\doibase 10.1103/PhysRevA.100.022326} {\bibfield
        {journal} {\bibinfo  {journal} {Physical Review A}\ }\textbf {\bibinfo
            {volume} {100}},\ \bibinfo {pages} {022326} (\bibinfo {year}
        {2019}{\natexlab{a}})}\BibitemShut {NoStop}%
    \bibitem [{\citenamefont {Johnson}\ \emph {et~al.}(2011)\citenamefont
        {Johnson}, \citenamefont {Amin}, \citenamefont {Gildert}, \citenamefont
        {Lanting}, \citenamefont {Hamze}, \citenamefont {Dickson}, \citenamefont
        {Harris}, \citenamefont {Berkley}, \citenamefont {Johansson}, \citenamefont
        {Bunyk}, \citenamefont {Chapple}, \citenamefont {Enderud}, \citenamefont
        {Hilton}, \citenamefont {Karimi}, \citenamefont {Ladizinsky}, \citenamefont
        {Ladizinsky}, \citenamefont {Oh}, \citenamefont {Perminov}, \citenamefont
        {Rich}, \citenamefont {Thom}, \citenamefont {Tolkacheva}, \citenamefont
        {Truncik}, \citenamefont {Uchaikin}, \citenamefont {Wang}, \citenamefont
        {Wilson},\ and\ \citenamefont {Rose}}]{johnsonQuantum11}%
    \BibitemOpen
    \bibfield  {author} {\bibinfo {author} {\bibfnamefont {M.~W.}\ \bibnamefont
            {Johnson}}, \bibinfo {author} {\bibfnamefont {M.~H.~S.}\ \bibnamefont
            {Amin}}, \bibinfo {author} {\bibfnamefont {S.}~\bibnamefont {Gildert}},
        \bibinfo {author} {\bibfnamefont {T.}~\bibnamefont {Lanting}}, \bibinfo
        {author} {\bibfnamefont {F.}~\bibnamefont {Hamze}}, \bibinfo {author}
        {\bibfnamefont {N.}~\bibnamefont {Dickson}}, \bibinfo {author} {\bibfnamefont
            {R.}~\bibnamefont {Harris}}, \bibinfo {author} {\bibfnamefont {A.~J.}\
            \bibnamefont {Berkley}}, \bibinfo {author} {\bibfnamefont {J.}~\bibnamefont
            {Johansson}}, \bibinfo {author} {\bibfnamefont {P.}~\bibnamefont {Bunyk}},
        \bibinfo {author} {\bibfnamefont {E.~M.}\ \bibnamefont {Chapple}}, \bibinfo
        {author} {\bibfnamefont {C.}~\bibnamefont {Enderud}}, \bibinfo {author}
        {\bibfnamefont {J.~P.}\ \bibnamefont {Hilton}}, \bibinfo {author}
        {\bibfnamefont {K.}~\bibnamefont {Karimi}}, \bibinfo {author} {\bibfnamefont
            {E.}~\bibnamefont {Ladizinsky}}, \bibinfo {author} {\bibfnamefont
            {N.}~\bibnamefont {Ladizinsky}}, \bibinfo {author} {\bibfnamefont
            {T.}~\bibnamefont {Oh}}, \bibinfo {author} {\bibfnamefont {I.}~\bibnamefont
            {Perminov}}, \bibinfo {author} {\bibfnamefont {C.}~\bibnamefont {Rich}},
        \bibinfo {author} {\bibfnamefont {M.~C.}\ \bibnamefont {Thom}}, \bibinfo
        {author} {\bibfnamefont {E.}~\bibnamefont {Tolkacheva}}, \bibinfo {author}
        {\bibfnamefont {C.~J.~S.}\ \bibnamefont {Truncik}}, \bibinfo {author}
        {\bibfnamefont {S.}~\bibnamefont {Uchaikin}}, \bibinfo {author}
        {\bibfnamefont {J.}~\bibnamefont {Wang}}, \bibinfo {author} {\bibfnamefont
            {B.}~\bibnamefont {Wilson}}, \ and\ \bibinfo {author} {\bibfnamefont
            {G.}~\bibnamefont {Rose}},\ }\href {\doibase 10.1038/nature10012} {\bibfield
        {journal} {\bibinfo  {journal} {Nature}\ }\textbf {\bibinfo {volume} {473}},\
        \bibinfo {pages} {194} (\bibinfo {year} {2011})}\BibitemShut {NoStop}%
    \bibitem [{\citenamefont {{D-Wave Systems Inc.}}(2018)}]{DW2KQ}%
    \BibitemOpen
    \bibfield  {author} {\bibinfo {author} {\bibnamefont {{D-Wave Systems
                    Inc.}}},\ }\href
    {https://www.dwavesys.com/sites/default/files/D-Wave%202000Q%20Tech%20Collateral_1029F.pdf}
    {\enquote {\bibinfo {title} {{The D-Wave 2000Q Quantum Computer Technology
                    Overview}},}\ } (\bibinfo {year} {2018})\BibitemShut {NoStop}%
    \bibitem [{\citenamefont {Albash}\ \emph {et~al.}(2012)\citenamefont {Albash},
        \citenamefont {Boixo}, \citenamefont {Lidar},\ and\ \citenamefont
        {Zanardi}}]{albashQuantum12}%
    \BibitemOpen
    \bibfield  {author} {\bibinfo {author} {\bibfnamefont {T.}~\bibnamefont
            {Albash}}, \bibinfo {author} {\bibfnamefont {S.}~\bibnamefont {Boixo}},
        \bibinfo {author} {\bibfnamefont {D.~A.}\ \bibnamefont {Lidar}}, \ and\
        \bibinfo {author} {\bibfnamefont {P.}~\bibnamefont {Zanardi}},\ }\href
    {\doibase 10.1088/1367-2630/14/12/123016} {\bibfield  {journal} {\bibinfo
            {journal} {New Journal of Physics}\ }\textbf {\bibinfo {volume} {14}},\
        \bibinfo {pages} {123016} (\bibinfo {year} {2012})}\BibitemShut {NoStop}%
    \bibitem [{\citenamefont {Davies}(1974)}]{davies_markovian_1974}%
    \BibitemOpen
    \bibfield  {author} {\bibinfo {author} {\bibfnamefont {E.~B.}\ \bibnamefont
            {Davies}},\ }\href {http://projecteuclid.org/euclid.cmp/1103860160}
    {\bibfield  {journal} {\bibinfo  {journal} {Comm. Math. Phys.}\ }\textbf
        {\bibinfo {volume} {39}},\ \bibinfo {pages} {91} (\bibinfo {year}
        {1974})}\BibitemShut {NoStop}%
    \bibitem [{\citenamefont {Albash}\ \emph
        {et~al.}(2015{\natexlab{a}})\citenamefont {Albash}, \citenamefont {Vinci},
        \citenamefont {Mishra}, \citenamefont {Warburton},\ and\ \citenamefont
        {Lidar}}]{albashConsistency15}%
    \BibitemOpen
    \bibfield  {author} {\bibinfo {author} {\bibfnamefont {T.}~\bibnamefont
            {Albash}}, \bibinfo {author} {\bibfnamefont {W.}~\bibnamefont {Vinci}},
        \bibinfo {author} {\bibfnamefont {A.}~\bibnamefont {Mishra}}, \bibinfo
        {author} {\bibfnamefont {P.~A.}\ \bibnamefont {Warburton}}, \ and\ \bibinfo
        {author} {\bibfnamefont {D.~A.}\ \bibnamefont {Lidar}},\ }\href {\doibase
        10.1103/PhysRevA.91.042314} {\bibfield  {journal} {\bibinfo  {journal}
            {Physical Review A}\ }\textbf {\bibinfo {volume} {91}},\ \bibinfo {pages}
        {042314} (\bibinfo {year} {2015}{\natexlab{a}})}\BibitemShut {NoStop}%
    \bibitem [{\citenamefont {Albash}\ \emph
        {et~al.}(2015{\natexlab{b}})\citenamefont {Albash}, \citenamefont {Hen},
        \citenamefont {Spedalieri},\ and\ \citenamefont
        {Lidar}}]{albashReexamination15}%
    \BibitemOpen
    \bibfield  {author} {\bibinfo {author} {\bibfnamefont {T.}~\bibnamefont
            {Albash}}, \bibinfo {author} {\bibfnamefont {I.}~\bibnamefont {Hen}},
        \bibinfo {author} {\bibfnamefont {F.~M.}\ \bibnamefont {Spedalieri}}, \ and\
        \bibinfo {author} {\bibfnamefont {D.~A.}\ \bibnamefont {Lidar}},\ }\href
    {\doibase 10.1103/PhysRevA.92.062328} {\bibfield  {journal} {\bibinfo
            {journal} {Physical Review A}\ }\textbf {\bibinfo {volume} {92}},\ \bibinfo
        {pages} {062328} (\bibinfo {year} {2015}{\natexlab{b}})}\BibitemShut
    {NoStop}%
    \bibitem [{\citenamefont {Bando}\ \emph {et~al.}(2022)\citenamefont {Bando},
        \citenamefont {Yip}, \citenamefont {Chen}, \citenamefont {Lidar},\ and\
        \citenamefont {Nishimori}}]{bando2021breakdown}%
    \BibitemOpen
    \bibfield  {author} {\bibinfo {author} {\bibfnamefont {Y.}~\bibnamefont
            {Bando}}, \bibinfo {author} {\bibfnamefont {K.-W.}\ \bibnamefont {Yip}},
        \bibinfo {author} {\bibfnamefont {H.}~\bibnamefont {Chen}}, \bibinfo {author}
        {\bibfnamefont {D.~A.}\ \bibnamefont {Lidar}}, \ and\ \bibinfo {author}
        {\bibfnamefont {H.}~\bibnamefont {Nishimori}},\ }\href {\doibase
        10.1103/PhysRevApplied.17.054033} {\bibfield  {journal} {\bibinfo  {journal}
            {Physical Review Applied}\ }\textbf {\bibinfo {volume} {17}},\ \bibinfo
        {pages} {054033} (\bibinfo {year} {2022})}\BibitemShut {NoStop}%
    \bibitem [{\citenamefont {Amin}(2015)}]{Amin:2015qf}%
    \BibitemOpen
    \bibfield  {author} {\bibinfo {author} {\bibfnamefont {M.~H.}\ \bibnamefont
            {Amin}},\ }\href {https://link.aps.org/doi/10.1103/PhysRevA.92.052323}
    {\bibfield  {journal} {\bibinfo  {journal} {Physical Review A}\ }\textbf
        {\bibinfo {volume} {92}},\ \bibinfo {pages} {052323} (\bibinfo {year}
        {2015})}\BibitemShut {NoStop}%
    \bibitem [{\citenamefont {Boothby}\ \emph {et~al.}(2020)\citenamefont
        {Boothby}, \citenamefont {Bunyk}, \citenamefont {Raymond},\ and\
        \citenamefont {Roy}}]{boothby2020nextgeneration}%
    \BibitemOpen
    \bibfield  {author} {\bibinfo {author} {\bibfnamefont {K.}~\bibnamefont
            {Boothby}}, \bibinfo {author} {\bibfnamefont {P.}~\bibnamefont {Bunyk}},
        \bibinfo {author} {\bibfnamefont {J.}~\bibnamefont {Raymond}}, \ and\
        \bibinfo {author} {\bibfnamefont {A.}~\bibnamefont {Roy}},\ }\href@noop {}
    {\enquote {\bibinfo {title} {Next-generation topology of d-wave quantum
                processors},}\ } (\bibinfo {year} {2020}),\ \Eprint
    {http://arxiv.org/abs/2003.00133} {arXiv:2003.00133 [quant-ph]} \BibitemShut
    {NoStop}%
    \bibitem [{\citenamefont {Albash}\ and\ \citenamefont
        {Lidar}(2018)}]{AlbashDemonstration18}%
    \BibitemOpen
    \bibfield  {author} {\bibinfo {author} {\bibfnamefont {T.}~\bibnamefont
            {Albash}}\ and\ \bibinfo {author} {\bibfnamefont {D.~A.}\ \bibnamefont
            {Lidar}},\ }\href {\doibase 10.1103/PhysRevX.8.031016} {\bibfield  {journal}
        {\bibinfo  {journal} {Physical Review X}\ }\textbf {\bibinfo {volume} {8}},\
        \bibinfo {pages} {031016} (\bibinfo {year} {2018})}\BibitemShut {NoStop}%
    \bibitem [{\citenamefont {Pudenz}\ \emph {et~al.}(2014)\citenamefont {Pudenz},
        \citenamefont {Albash},\ and\ \citenamefont
        {Lidar}}]{pudenzErrorcorrected14}%
    \BibitemOpen
    \bibfield  {author} {\bibinfo {author} {\bibfnamefont {K.~L.}\ \bibnamefont
            {Pudenz}}, \bibinfo {author} {\bibfnamefont {T.}~\bibnamefont {Albash}}, \
        and\ \bibinfo {author} {\bibfnamefont {D.~A.}\ \bibnamefont {Lidar}},\ }\href
    {\doibase 10.1038/ncomms4243} {\bibfield  {journal} {\bibinfo  {journal}
            {Nature Communications}\ }\textbf {\bibinfo {volume} {5}},\ \bibinfo {pages}
        {3243} (\bibinfo {year} {2014})}\BibitemShut {NoStop}%
    \bibitem [{\citenamefont {Young}\ \emph {et~al.}(2013)\citenamefont {Young},
        \citenamefont {{Blume-Kohout}},\ and\ \citenamefont
        {Lidar}}]{youngAdiabatic13}%
    \BibitemOpen
    \bibfield  {author} {\bibinfo {author} {\bibfnamefont {K.~C.}\ \bibnamefont
            {Young}}, \bibinfo {author} {\bibfnamefont {R.}~\bibnamefont
            {{Blume-Kohout}}}, \ and\ \bibinfo {author} {\bibfnamefont {D.~A.}\
            \bibnamefont {Lidar}},\ }\href {\doibase 10.1103/PhysRevA.88.062314}
    {\bibfield  {journal} {\bibinfo  {journal} {Physical Review A}\ }\textbf
        {\bibinfo {volume} {88}},\ \bibinfo {pages} {062314} (\bibinfo {year}
        {2013})}\BibitemShut {NoStop}%
    \bibitem [{\citenamefont {Pearson}\ \emph
        {et~al.}(2019{\natexlab{a}})\citenamefont {Pearson}, \citenamefont {Mishra},
        \citenamefont {Hen},\ and\ \citenamefont {Lidar}}]{pearsonAnalog19}%
    \BibitemOpen
    \bibfield  {author} {\bibinfo {author} {\bibfnamefont {A.}~\bibnamefont
            {Pearson}}, \bibinfo {author} {\bibfnamefont {A.}~\bibnamefont {Mishra}},
        \bibinfo {author} {\bibfnamefont {I.}~\bibnamefont {Hen}}, \ and\ \bibinfo
        {author} {\bibfnamefont {D.~A.}\ \bibnamefont {Lidar}},\ }\href {\doibase
        10.1038/s41534-019-0210-7} {\bibfield  {journal} {\bibinfo  {journal} {npj
                Quantum Information}\ }\textbf {\bibinfo {volume} {5}},\ \bibinfo {pages} {1}
        (\bibinfo {year} {2019}{\natexlab{a}})}\BibitemShut {NoStop}%
    \bibitem [{\citenamefont {Matsuura}\ \emph {et~al.}(2017)\citenamefont
        {Matsuura}, \citenamefont {Nishimori}, \citenamefont {Vinci}, \citenamefont
        {Albash},\ and\ \citenamefont {Lidar}}]{Matsuura:2016aa}%
    \BibitemOpen
    \bibfield  {author} {\bibinfo {author} {\bibfnamefont {S.}~\bibnamefont
            {Matsuura}}, \bibinfo {author} {\bibfnamefont {H.}~\bibnamefont {Nishimori}},
        \bibinfo {author} {\bibfnamefont {W.}~\bibnamefont {Vinci}}, \bibinfo
        {author} {\bibfnamefont {T.}~\bibnamefont {Albash}}, \ and\ \bibinfo {author}
        {\bibfnamefont {D.~A.}\ \bibnamefont {Lidar}},\ }\href
    {https://link.aps.org/doi/10.1103/PhysRevA.95.022308} {\bibfield  {journal}
        {\bibinfo  {journal} {Physical Review A}\ }\textbf {\bibinfo {volume} {95}},\
        \bibinfo {pages} {022308} (\bibinfo {year} {2017})}\BibitemShut {NoStop}%
    \bibitem [{\citenamefont {Vinci}\ and\ \citenamefont
        {Lidar}(2018)}]{Vinci:2017ab}%
    \BibitemOpen
    \bibfield  {author} {\bibinfo {author} {\bibfnamefont {W.}~\bibnamefont
            {Vinci}}\ and\ \bibinfo {author} {\bibfnamefont {D.~A.}\ \bibnamefont
            {Lidar}},\ }\href {https://link.aps.org/doi/10.1103/PhysRevA.97.022308}
    {\bibfield  {journal} {\bibinfo  {journal} {Physical Review A}\ }\textbf
        {\bibinfo {volume} {97}},\ \bibinfo {pages} {022308} (\bibinfo {year}
        {2018})}\BibitemShut {NoStop}%
    \bibitem [{dwa()}]{dwave-manual}%
    \BibitemOpen
    \href {https://docs.dwavesys.com/docs/latest/doc_qpu.html} {\emph {\bibinfo
            {title} {{D-Wave: Technical Description of the QPU}}}}\BibitemShut {NoStop}%
    \bibitem [{\citenamefont {{D-Wave Systems Inc.}}(2021)}]{DW-manual}%
    \BibitemOpen
    \bibfield  {author} {\bibinfo {author} {\bibnamefont {{D-Wave Systems
                    Inc.}}},\ }\href
    {https://docs.ocean.dwavesys.com/_/downloads/system/en/stable/pdf/} {\enquote
        {\bibinfo {title} {{dwave-system Documentation Release 1.6.0}},}\ } (\bibinfo
    {year} {2021})\BibitemShut {NoStop}%
    \bibitem [{\citenamefont {{Venuti}}\ \emph {et~al.}(2016)\citenamefont
        {{Venuti}}, \citenamefont {Albash}, \citenamefont {Lidar},\ and\
        \citenamefont {Zanardi}}]{venutiAdiabaticity16}%
    \BibitemOpen
    \bibfield  {author} {\bibinfo {author} {\bibfnamefont {L.~C.}~\bibnamefont
            {{Venuti}}}, \bibinfo {author} {\bibfnamefont {T.}~\bibnamefont
            {Albash}}, \bibinfo {author} {\bibfnamefont {D.~A.}\ \bibnamefont {Lidar}}, \
        and\ \bibinfo {author} {\bibfnamefont {P.}~\bibnamefont {Zanardi}},\ }\href
    {\doibase 10.1103/PhysRevA.93.032118} {\bibfield  {journal} {\bibinfo
            {journal} {Physical Review A}\ }\textbf {\bibinfo {volume} {93}},\ \bibinfo
        {pages} {032118} (\bibinfo {year} {2016})}\BibitemShut {NoStop}%
    \bibitem [{\citenamefont {Pauli}(1928)}]{Pauli-master-equation}%
    \BibitemOpen
    \bibfield  {author} {\bibinfo {author} {\bibfnamefont {W.}~\bibnamefont
            {Pauli}},\ }\href@noop {} {\enquote {\bibinfo {title} {{\"Uber das H-Theorem
                    vom Anwachsen der Entropie vom Standpunkt der neuen Quantenmechanik, in
                    Festschrift zum 60. Geburtstage A. Sommerfeld, p.30}},}\ } (\bibinfo {year}
    {Hirzel, Leipzig, 1928})\BibitemShut {NoStop}%
    \bibitem [{\citenamefont {Lidar}(2019{\natexlab{b}})}]{Lidar:2019aa}%
    \BibitemOpen
    \bibfield  {author} {\bibinfo {author} {\bibfnamefont {D.~A.}\ \bibnamefont
            {Lidar}},\ }\href {https://arxiv.org/abs/1902.00967} {\bibfield  {journal}
        {\bibinfo  {journal} {arXiv preprint arXiv:1902.00967}\ } (\bibinfo {year}
        {2019}{\natexlab{b}})}\BibitemShut {NoStop}%
    \bibitem [{\citenamefont {Kirkpatrick}\ \emph {et~al.}(1983)\citenamefont
        {Kirkpatrick}, \citenamefont {Gelatt},\ and\ \citenamefont
        {Vecchi}}]{kirkpatrick_optimization_1983}%
    \BibitemOpen
    \bibfield  {author} {\bibinfo {author} {\bibfnamefont {S.}~\bibnamefont
            {Kirkpatrick}}, \bibinfo {author} {\bibfnamefont {C.~D.}\ \bibnamefont
            {Gelatt}}, \ and\ \bibinfo {author} {\bibfnamefont {M.~P.}\ \bibnamefont
            {Vecchi}},\ }\href {http://science.sciencemag.org/content/220/4598/671}
    {\bibfield  {journal} {\bibinfo  {journal} {Science}\ }\textbf {\bibinfo
            {volume} {220}},\ \bibinfo {pages} {671} (\bibinfo {year}
        {1983})}\BibitemShut {NoStop}%
    \bibitem [{\citenamefont {Zhu}\ \emph {et~al.}(2015)\citenamefont {Zhu},
        \citenamefont {Ochoa},\ and\ \citenamefont
        {Katzgraber}}]{PhysRevLett.115.077201}%
    \BibitemOpen
    \bibfield  {author} {\bibinfo {author} {\bibfnamefont {Z.}~\bibnamefont
            {Zhu}}, \bibinfo {author} {\bibfnamefont {A.~J.}\ \bibnamefont {Ochoa}}, \
        and\ \bibinfo {author} {\bibfnamefont {H.~G.}\ \bibnamefont {Katzgraber}},\
    }\href {\doibase 10.1103/PhysRevLett.115.077201} {\bibfield  {journal}
        {\bibinfo  {journal} {Phys. Rev. Lett.}\ }\textbf {\bibinfo {volume} {115}},\
        \bibinfo {pages} {077201} (\bibinfo {year} {2015})}\BibitemShut {NoStop}%
    \bibitem [{\citenamefont {Albash}\ \emph {et~al.}(2019)\citenamefont {Albash},
        \citenamefont {Martin-Mayor},\ and\ \citenamefont {Hen}}]{Albash:2019ab}%
    \BibitemOpen
    \bibfield  {author} {\bibinfo {author} {\bibfnamefont {T.}~\bibnamefont
            {Albash}}, \bibinfo {author} {\bibfnamefont {V.}~\bibnamefont
            {Martin-Mayor}}, \ and\ \bibinfo {author} {\bibfnamefont {I.}~\bibnamefont
            {Hen}},\ }\href {https://doi.org/10.1088/2058-9565/ab13ea} {\bibfield
        {journal} {\bibinfo  {journal} {Quantum Sci. Technol.}\ }\textbf {\bibinfo
            {volume} {4}},\ \bibinfo {pages} {02LT03} (\bibinfo {year}
        {2019})}\BibitemShut {NoStop}%
    \bibitem [{\citenamefont {Jensen}(1906)}]{Jensen:1906up}%
    \BibitemOpen
    \bibfield  {author} {\bibinfo {author} {\bibfnamefont {J.~L. W.~V.}\
            \bibnamefont {Jensen}},\ }\href {https://doi.org/10.1007/BF02418571}
    {\bibfield  {journal} {\bibinfo  {journal} {Acta Mathematica}\ }\textbf
        {\bibinfo {volume} {30}},\ \bibinfo {pages} {175} (\bibinfo {year}
        {1906})}\BibitemShut {NoStop}%
    \bibitem [{\citenamefont {Nielsen}\ and\ \citenamefont
        {Chuang}(2010)}]{nielsenQuantum10}%
    \BibitemOpen
    \bibfield  {author} {\bibinfo {author} {\bibfnamefont {M.~A.}\ \bibnamefont
            {Nielsen}}\ and\ \bibinfo {author} {\bibfnamefont {I.~L.}\ \bibnamefont
            {Chuang}},\ }\href@noop {} {\emph {\bibinfo {title} {Quantum Computation and
                Quantum Information}}},\ \bibinfo {edition} {10th}\ ed.\ (\bibinfo
    {publisher} {{Cambridge University Press}},\ \bibinfo {address} {{Cambridge ;
            New York}},\ \bibinfo {year} {2010})\BibitemShut {NoStop}%
    \bibitem [{ano()}]{anomalous-heating}%
    \BibitemOpen
    \href
    {https://docs.dwavesys.com/docs/latest/c_qpu_errors.html#high-energy-photon-flux}
    {\enquote {\bibinfo {title} {High-energy photon flux},}\ }\bibinfo
    {howpublished} {D-Wave System Documentation, Other Error Sources}\BibitemShut
    {NoStop}%
    \bibitem [{\citenamefont {Rezakhani}\ \emph {et~al.}(2010)\citenamefont
        {Rezakhani}, \citenamefont {Pimachev},\ and\ \citenamefont {Lidar}}]{RPL:10}%
    \BibitemOpen
    \bibfield  {author} {\bibinfo {author} {\bibfnamefont {A.~T.}\ \bibnamefont
            {Rezakhani}}, \bibinfo {author} {\bibfnamefont {A.~K.}\ \bibnamefont
            {Pimachev}}, \ and\ \bibinfo {author} {\bibfnamefont {D.~A.}\ \bibnamefont
            {Lidar}},\ }\href {http://link.aps.org/doi/10.1103/PhysRevA.82.052305}
    {\bibfield  {journal} {\bibinfo  {journal} {Phys. Rev. A}\ }\textbf {\bibinfo
            {volume} {82}},\ \bibinfo {pages} {052305} (\bibinfo {year}
        {2010})}\BibitemShut {NoStop}%
    \bibitem [{\citenamefont {Jordan}\ \emph {et~al.}(2006)\citenamefont {Jordan},
        \citenamefont {Farhi},\ and\ \citenamefont {Shor}}]{jordan2006error}%
    \BibitemOpen
    \bibfield  {author} {\bibinfo {author} {\bibfnamefont {S.~P.}\ \bibnamefont
            {Jordan}}, \bibinfo {author} {\bibfnamefont {E.}~\bibnamefont {Farhi}}, \
        and\ \bibinfo {author} {\bibfnamefont {P.~W.}\ \bibnamefont {Shor}},\ }\href
    {http://link.aps.org/doi/10.1103/PhysRevA.74.052322} {\bibfield  {journal}
        {\bibinfo  {journal} {{Phys. Rev. A}}\ }\textbf {\bibinfo {volume} {74}},\
        \bibinfo {pages} {052322} (\bibinfo {year} {2006})}\BibitemShut {NoStop}%
    \bibitem [{\citenamefont {Marvian}\ and\ \citenamefont
        {Lidar}(2014)}]{Marvian:2014nr}%
    \BibitemOpen
    \bibfield  {author} {\bibinfo {author} {\bibfnamefont {I.}~\bibnamefont
            {Marvian}}\ and\ \bibinfo {author} {\bibfnamefont {D.~A.}\ \bibnamefont
            {Lidar}},\ }\href {http://link.aps.org/doi/10.1103/PhysRevLett.113.260504}
    {\bibfield  {journal} {\bibinfo  {journal} {Phys. Rev. Lett.}\ }\textbf
        {\bibinfo {volume} {113}},\ \bibinfo {pages} {260504} (\bibinfo {year}
        {2014})}\BibitemShut {NoStop}%
    \bibitem [{\citenamefont {Bookatz}\ \emph {et~al.}(2015)\citenamefont
        {Bookatz}, \citenamefont {Farhi},\ and\ \citenamefont
        {Zhou}}]{Bookatz:2014uq}%
    \BibitemOpen
    \bibfield  {author} {\bibinfo {author} {\bibfnamefont {A.~D.}\ \bibnamefont
            {Bookatz}}, \bibinfo {author} {\bibfnamefont {E.}~\bibnamefont {Farhi}}, \
        and\ \bibinfo {author} {\bibfnamefont {L.}~\bibnamefont {Zhou}},\ }\href
    {http://link.aps.org/doi/10.1103/PhysRevA.92.022317} {\bibfield  {journal}
        {\bibinfo  {journal} {{Phys. Rev. A}}\ }\textbf {\bibinfo {volume} {92}},\
        \bibinfo {pages} {022317} (\bibinfo {year} {2015})}\BibitemShut {NoStop}%
    \bibitem [{\citenamefont {Jiang}\ and\ \citenamefont
        {Rieffel}(2017)}]{Jiang:2015kx}%
    \BibitemOpen
    \bibfield  {author} {\bibinfo {author} {\bibfnamefont {Z.}~\bibnamefont
            {Jiang}}\ and\ \bibinfo {author} {\bibfnamefont {E.~G.}\ \bibnamefont
            {Rieffel}},\ }\href {\doibase 10.1007/s11128-017-1527-9} {\bibfield
        {journal} {\bibinfo  {journal} {{Quant. Inf. Proc.}}\ }\textbf {\bibinfo
            {volume} {16}},\ \bibinfo {pages} {89} (\bibinfo {year} {2017})}\BibitemShut
    {NoStop}%
    \bibitem [{\citenamefont {Marvian}\ and\ \citenamefont
        {Lidar}(2017{\natexlab{a}})}]{Marvian-Lidar:16}%
    \BibitemOpen
    \bibfield  {author} {\bibinfo {author} {\bibfnamefont {M.}~\bibnamefont
            {Marvian}}\ and\ \bibinfo {author} {\bibfnamefont {D.~A.}\ \bibnamefont
            {Lidar}},\ }\href {https://link.aps.org/doi/10.1103/PhysRevLett.118.030504}
    {\bibfield  {journal} {\bibinfo  {journal} {Phys. Rev. Lett.}\ }\textbf
        {\bibinfo {volume} {118}},\ \bibinfo {pages} {030504} (\bibinfo {year}
        {2017}{\natexlab{a}})}\BibitemShut {NoStop}%
    \bibitem [{\citenamefont {Marvian}\ and\ \citenamefont
        {Lidar}(2017{\natexlab{b}})}]{Marvian:2017aa}%
    \BibitemOpen
    \bibfield  {author} {\bibinfo {author} {\bibfnamefont {M.}~\bibnamefont
            {Marvian}}\ and\ \bibinfo {author} {\bibfnamefont {D.~A.}\ \bibnamefont
            {Lidar}},\ }\href {http://link.aps.org/doi/10.1103/PhysRevA.95.032302}
    {\bibfield  {journal} {\bibinfo  {journal} {Physical Review A}\ }\textbf
        {\bibinfo {volume} {95}},\ \bibinfo {pages} {032302} (\bibinfo {year}
        {2017}{\natexlab{b}})}\BibitemShut {NoStop}%
    \bibitem [{\citenamefont {Pudenz}\ \emph {et~al.}(2015)\citenamefont {Pudenz},
        \citenamefont {Albash},\ and\ \citenamefont {Lidar}}]{pudenzQuantum15}%
    \BibitemOpen
    \bibfield  {author} {\bibinfo {author} {\bibfnamefont {K.~L.}\ \bibnamefont
            {Pudenz}}, \bibinfo {author} {\bibfnamefont {T.}~\bibnamefont {Albash}}, \
        and\ \bibinfo {author} {\bibfnamefont {D.~A.}\ \bibnamefont {Lidar}},\ }\href
    {\doibase 10.1103/PhysRevA.91.042302} {\bibfield  {journal} {\bibinfo
            {journal} {Physical Review A}\ }\textbf {\bibinfo {volume} {91}},\ \bibinfo
        {pages} {042302} (\bibinfo {year} {2015})}\BibitemShut {NoStop}%
    \bibitem [{\citenamefont {Mishra}\ \emph {et~al.}(2015)\citenamefont {Mishra},
        \citenamefont {Albash},\ and\ \citenamefont {Lidar}}]{Mishra:2015}%
    \BibitemOpen
    \bibfield  {author} {\bibinfo {author} {\bibfnamefont {A.}~\bibnamefont
            {Mishra}}, \bibinfo {author} {\bibfnamefont {T.}~\bibnamefont {Albash}}, \
        and\ \bibinfo {author} {\bibfnamefont {D.~A.}\ \bibnamefont {Lidar}},\ }\href
    {\doibase 10.1007/s11128-015-1201-z} {\bibfield  {journal} {\bibinfo
            {journal} {Quant. Inf. Proc.}\ }\textbf {\bibinfo {volume} {15}},\ \bibinfo
        {pages} {609} (\bibinfo {year} {2015})}\BibitemShut {NoStop}%
    \bibitem [{\citenamefont {Matsuura}\ \emph {et~al.}(2016)\citenamefont
        {Matsuura}, \citenamefont {Nishimori}, \citenamefont {Albash},\ and\
        \citenamefont {Lidar}}]{MNAL:15}%
    \BibitemOpen
    \bibfield  {author} {\bibinfo {author} {\bibfnamefont {S.}~\bibnamefont
            {Matsuura}}, \bibinfo {author} {\bibfnamefont {H.}~\bibnamefont {Nishimori}},
        \bibinfo {author} {\bibfnamefont {T.}~\bibnamefont {Albash}}, \ and\ \bibinfo
        {author} {\bibfnamefont {D.~A.}\ \bibnamefont {Lidar}},\ }\href
    {http://link.aps.org/doi/10.1103/PhysRevLett.116.220501} {\bibfield
        {journal} {\bibinfo  {journal} {Physical Review Letters}\ }\textbf {\bibinfo
            {volume} {116}},\ \bibinfo {pages} {220501} (\bibinfo {year}
        {2016})}\BibitemShut {NoStop}%
    \bibitem [{\citenamefont {Vinci}\ \emph {et~al.}(2015)\citenamefont {Vinci},
        \citenamefont {Albash}, \citenamefont {Paz-Silva}, \citenamefont {Hen},\ and\
        \citenamefont {Lidar}}]{Vinci:2015jt}%
    \BibitemOpen
    \bibfield  {author} {\bibinfo {author} {\bibfnamefont {W.}~\bibnamefont
            {Vinci}}, \bibinfo {author} {\bibfnamefont {T.}~\bibnamefont {Albash}},
        \bibinfo {author} {\bibfnamefont {G.}~\bibnamefont {Paz-Silva}}, \bibinfo
        {author} {\bibfnamefont {I.}~\bibnamefont {Hen}}, \ and\ \bibinfo {author}
        {\bibfnamefont {D.~A.}\ \bibnamefont {Lidar}},\ }\href
    {http://link.aps.org/doi/10.1103/PhysRevA.92.042310} {\bibfield  {journal}
        {\bibinfo  {journal} {{Phys. Rev. A}}\ }\textbf {\bibinfo {volume} {92}},\
        \bibinfo {pages} {042310} (\bibinfo {year} {2015})}\BibitemShut {NoStop}%
    \bibitem [{\citenamefont {Pearson}\ \emph
        {et~al.}(2019{\natexlab{b}})\citenamefont {Pearson}, \citenamefont {Mishra},
        \citenamefont {Hen},\ and\ \citenamefont {Lidar}}]{Pearson:2019aa}%
    \BibitemOpen
    \bibfield  {author} {\bibinfo {author} {\bibfnamefont {A.}~\bibnamefont
            {Pearson}}, \bibinfo {author} {\bibfnamefont {A.}~\bibnamefont {Mishra}},
        \bibinfo {author} {\bibfnamefont {I.}~\bibnamefont {Hen}}, \ and\ \bibinfo
        {author} {\bibfnamefont {D.~A.}\ \bibnamefont {Lidar}},\ }\href {\doibase
        10.1038/s41534-019-0210-7} {\bibfield  {journal} {\bibinfo  {journal} {npj
                Quantum Information}\ }\textbf {\bibinfo {volume} {5}},\ \bibinfo {pages}
        {107} (\bibinfo {year} {2019}{\natexlab{b}})}\BibitemShut {NoStop}%
    \bibitem [{\citenamefont {Vinci}\ \emph {et~al.}(2016)\citenamefont {Vinci},
        \citenamefont {Albash},\ and\ \citenamefont {Lidar}}]{vinci2015nested}%
    \BibitemOpen
    \bibfield  {author} {\bibinfo {author} {\bibfnamefont {W.}~\bibnamefont
            {Vinci}}, \bibinfo {author} {\bibfnamefont {T.}~\bibnamefont {Albash}}, \
        and\ \bibinfo {author} {\bibfnamefont {D.~A.}\ \bibnamefont {Lidar}},\ }\href
    {http://dx.doi.org/10.1038/npjqi.2016.17} {\bibfield  {journal} {\bibinfo
            {journal} {npj Quant. Inf.}\ }\textbf {\bibinfo {volume} {2}},\ \bibinfo
        {pages} {16017} (\bibinfo {year} {2016})}\BibitemShut {NoStop}%
    \bibitem [{\citenamefont {Matsuura}\ \emph {et~al.}(2019)\citenamefont
        {Matsuura}, \citenamefont {Nishimori}, \citenamefont {Vinci},\ and\
        \citenamefont {Lidar}}]{Matsuura:2018}%
    \BibitemOpen
    \bibfield  {author} {\bibinfo {author} {\bibfnamefont {S.}~\bibnamefont
            {Matsuura}}, \bibinfo {author} {\bibfnamefont {H.}~\bibnamefont {Nishimori}},
        \bibinfo {author} {\bibfnamefont {W.}~\bibnamefont {Vinci}}, \ and\ \bibinfo
        {author} {\bibfnamefont {D.~A.}\ \bibnamefont {Lidar}},\ }\href {\doibase
        10.1103/PhysRevA.99.062307} {\bibfield  {journal} {\bibinfo  {journal}
            {Physical Review A}\ }\textbf {\bibinfo {volume} {99}},\ \bibinfo {pages}
        {062307} (\bibinfo {year} {2019})}\BibitemShut {NoStop}%
    \bibitem [{\citenamefont {Li}\ \emph {et~al.}(2020)\citenamefont {Li},
        \citenamefont {Albash},\ and\ \citenamefont {Lidar}}]{Li:2020aa}%
    \BibitemOpen
    \bibfield  {author} {\bibinfo {author} {\bibfnamefont {R.~Y.}\ \bibnamefont
            {Li}}, \bibinfo {author} {\bibfnamefont {T.}~\bibnamefont {Albash}}, \ and\
        \bibinfo {author} {\bibfnamefont {D.~A.}\ \bibnamefont {Lidar}},\ }\href
    {\doibase 10.1088/2058-9565/ab9aab} {\bibfield  {journal} {\bibinfo
            {journal} {Quantum Science and Technology}\ }\textbf {\bibinfo {volume}
            {5}},\ \bibinfo {pages} {045010} (\bibinfo {year} {2020})}\BibitemShut
    {NoStop}%
    \bibitem [{\citenamefont {Kato}(1995)}]{kato_perturbation_1995}%
    \BibitemOpen
    \bibfield  {author} {\bibinfo {author} {\bibfnamefont {T.}~\bibnamefont
            {Kato}},\ }\href@noop {} {\emph {\bibinfo {title} {Perturbation {Theory} for
                {Linear} {Operators}}}}\ (\bibinfo  {publisher} {Springer},\ \bibinfo {year}
    {1995})\BibitemShut {NoStop}%
    \bibitem [{\citenamefont {Marshall}\ \emph {et~al.}(2019)\citenamefont
        {Marshall}, \citenamefont {Venturelli}, \citenamefont {Hen},\ and\
        \citenamefont {Rieffel}}]{marshallPower19}%
    \BibitemOpen
    \bibfield  {author} {\bibinfo {author} {\bibfnamefont {J.}~\bibnamefont
            {Marshall}}, \bibinfo {author} {\bibfnamefont {D.}~\bibnamefont
            {Venturelli}}, \bibinfo {author} {\bibfnamefont {I.}~\bibnamefont {Hen}}, \
        and\ \bibinfo {author} {\bibfnamefont {E.~G.}\ \bibnamefont {Rieffel}},\
    }\href {\doibase 10.1103/PhysRevApplied.11.044083} {\bibfield  {journal}
        {\bibinfo  {journal} {Physical Review Applied}\ }\textbf {\bibinfo {volume}
            {11}},\ \bibinfo {pages} {044083} (\bibinfo {year} {2019})}\BibitemShut
    {NoStop}%
    \bibitem [{\citenamefont {Chen}\ and\ \citenamefont {Lidar}(2020)}]{chenWhy20}%
    \BibitemOpen
    \bibfield  {author} {\bibinfo {author} {\bibfnamefont {H.}~\bibnamefont
            {Chen}}\ and\ \bibinfo {author} {\bibfnamefont {D.~A.}\ \bibnamefont
            {Lidar}},\ }\href {\doibase 10.1103/PhysRevApplied.14.014100} {\bibfield
        {journal} {\bibinfo  {journal} {Physical Review Applied}\ }\textbf {\bibinfo
            {volume} {14}},\ \bibinfo {pages} {014100} (\bibinfo {year}
        {2020})}\BibitemShut {NoStop}%
    \bibitem [{\citenamefont {Gonzalez~Izquierdo}\ \emph
        {et~al.}(2021)\citenamefont {Gonzalez~Izquierdo}, \citenamefont {Grabbe},
        \citenamefont {Hadfield}, \citenamefont {Marshall}, \citenamefont {Wang},\
        and\ \citenamefont {Rieffel}}]{izquierdo2020ferromagnetically}%
    \BibitemOpen
    \bibfield  {author} {\bibinfo {author} {\bibfnamefont {Z.}~\bibnamefont
            {Gonzalez~Izquierdo}}, \bibinfo {author} {\bibfnamefont {S.}~\bibnamefont
            {Grabbe}}, \bibinfo {author} {\bibfnamefont {S.}~\bibnamefont {Hadfield}},
        \bibinfo {author} {\bibfnamefont {J.}~\bibnamefont {Marshall}}, \bibinfo
        {author} {\bibfnamefont {Z.}~\bibnamefont {Wang}}, \ and\ \bibinfo {author}
        {\bibfnamefont {E.}~\bibnamefont {Rieffel}},\ }\href {\doibase
        10.1103/PhysRevApplied.15.044013} {\bibfield  {journal} {\bibinfo  {journal}
            {Physical Review Applied}\ }\textbf {\bibinfo {volume} {15}},\ \bibinfo
        {pages} {044013} (\bibinfo {year} {2021})}\BibitemShut {NoStop}%
    \bibitem [{\citenamefont {Albash}\ and\ \citenamefont
        {Marshall}(2021)}]{albash2020comparing}%
    \BibitemOpen
    \bibfield  {author} {\bibinfo {author} {\bibfnamefont {T.}~\bibnamefont
            {Albash}}\ and\ \bibinfo {author} {\bibfnamefont {J.}~\bibnamefont
            {Marshall}},\ }\href
    {https://link.aps.org/doi/10.1103/PhysRevApplied.15.014029} {\bibfield
        {journal} {\bibinfo  {journal} {Physical Review Applied}\ }\textbf {\bibinfo
            {volume} {15}},\ \bibinfo {pages} {014029} (\bibinfo {year}
        {2021})}\BibitemShut {NoStop}%
    \bibitem [{\citenamefont {Izquierdo}\ \emph {et~al.}(2022)\citenamefont
        {Izquierdo}, \citenamefont {Grabbe}, \citenamefont {Idris}, \citenamefont
        {Wang}, \citenamefont {Marshall},\ and\ \citenamefont
        {Rieffel}}]{izquierdoAdvantage22}%
    \BibitemOpen
    \bibfield  {author} {\bibinfo {author} {\bibfnamefont {Z.~G.}\ \bibnamefont
            {Izquierdo}}, \bibinfo {author} {\bibfnamefont {S.}~\bibnamefont {Grabbe}},
        \bibinfo {author} {\bibfnamefont {H.}~\bibnamefont {Idris}}, \bibinfo
        {author} {\bibfnamefont {Z.}~\bibnamefont {Wang}}, \bibinfo {author}
        {\bibfnamefont {J.}~\bibnamefont {Marshall}}, \ and\ \bibinfo {author}
        {\bibfnamefont {E.}~\bibnamefont {Rieffel}},\ }\href {\doibase
        10.48550/arXiv.2205.12936} {\  (\bibinfo {year} {2022}),\
        10.48550/arXiv.2205.12936},\ \Eprint {http://arxiv.org/abs/2205.12936}
    {arXiv:2205.12936 [quant-ph]} \BibitemShut {NoStop}%
    \bibitem [{\citenamefont {Ronnow}\ \emph {et~al.}(2014)\citenamefont {Ronnow},
        \citenamefont {Wang}, \citenamefont {Job}, \citenamefont {Boixo},
        \citenamefont {Isakov}, \citenamefont {Wecker}, \citenamefont {Martinis},
        \citenamefont {Lidar},\ and\ \citenamefont {Troyer}}]{speedup}%
    \BibitemOpen
    \bibfield  {author} {\bibinfo {author} {\bibfnamefont {T.~F.}\ \bibnamefont
            {Ronnow}}, \bibinfo {author} {\bibfnamefont {Z.}~\bibnamefont {Wang}},
        \bibinfo {author} {\bibfnamefont {J.}~\bibnamefont {Job}}, \bibinfo {author}
        {\bibfnamefont {S.}~\bibnamefont {Boixo}}, \bibinfo {author} {\bibfnamefont
            {S.~V.}\ \bibnamefont {Isakov}}, \bibinfo {author} {\bibfnamefont
            {D.}~\bibnamefont {Wecker}}, \bibinfo {author} {\bibfnamefont {J.~M.}\
            \bibnamefont {Martinis}}, \bibinfo {author} {\bibfnamefont {D.~A.}\
            \bibnamefont {Lidar}}, \ and\ \bibinfo {author} {\bibfnamefont
            {M.}~\bibnamefont {Troyer}},\ }\href
    {http://science.sciencemag.org/content/345/6195/420} {\bibfield  {journal}
        {\bibinfo  {journal} {Science}\ }\textbf {\bibinfo {volume} {345}},\ \bibinfo
        {pages} {420} (\bibinfo {year} {2014})}\BibitemShut {NoStop}%
    \bibitem [{\citenamefont {Hen}\ \emph {et~al.}(2015)\citenamefont {Hen},
        \citenamefont {Job}, \citenamefont {Albash}, \citenamefont {Ronnow},
        \citenamefont {Troyer},\ and\ \citenamefont {Lidar}}]{Hen:2015rt}%
    \BibitemOpen
    \bibfield  {author} {\bibinfo {author} {\bibfnamefont {I.}~\bibnamefont
            {Hen}}, \bibinfo {author} {\bibfnamefont {J.}~\bibnamefont {Job}}, \bibinfo
        {author} {\bibfnamefont {T.}~\bibnamefont {Albash}}, \bibinfo {author}
        {\bibfnamefont {T.~F.}\ \bibnamefont {Ronnow}}, \bibinfo {author}
        {\bibfnamefont {M.}~\bibnamefont {Troyer}}, \ and\ \bibinfo {author}
        {\bibfnamefont {D.~A.}\ \bibnamefont {Lidar}},\ }\href
    {http://link.aps.org/doi/10.1103/PhysRevA.92.042325} {\bibfield  {journal}
        {\bibinfo  {journal} {{Phys. Rev. A}}\ }\textbf {\bibinfo {volume} {92}},\
        \bibinfo {pages} {042325} (\bibinfo {year} {2015})}\BibitemShut {NoStop}%
    \bibitem [{\citenamefont {Avron}\ \emph {et~al.}(2012)\citenamefont {Avron},
        \citenamefont {Fraas}, \citenamefont {Graf},\ and\ \citenamefont
        {Grech}}]{avronAdiabatic12}%
    \BibitemOpen
    \bibfield  {author} {\bibinfo {author} {\bibfnamefont {J.~E.}\ \bibnamefont
            {Avron}}, \bibinfo {author} {\bibfnamefont {M.}~\bibnamefont {Fraas}},
        \bibinfo {author} {\bibfnamefont {G.~M.}\ \bibnamefont {Graf}}, \ and\
        \bibinfo {author} {\bibfnamefont {P.}~\bibnamefont {Grech}},\ }\href
    {\doibase 10.1007/s00220-012-1504-1} {\bibfield  {journal} {\bibinfo
            {journal} {Communications in Mathematical Physics}\ }\textbf {\bibinfo
            {volume} {314}},\ \bibinfo {pages} {163} (\bibinfo {year}
        {2012})}\BibitemShut {NoStop}%
    \bibitem [{\citenamefont {Hale}(2009)}]{hale_ordinary_2009}%
    \BibitemOpen
    \bibfield  {author} {\bibinfo {author} {\bibfnamefont {J.~K.}\ \bibnamefont
            {Hale}},\ }\href@noop {} {\emph {\bibinfo {title} {Ordinary {Differential}
                {Equations}}}}\ (\bibinfo  {publisher} {Courier Corporation},\ \bibinfo
    {year} {2009})\BibitemShut {NoStop}%
    \bibitem [{\citenamefont {Harris}\ \emph {et~al.}(2009)\citenamefont {Harris},
        \citenamefont {Lanting}, \citenamefont {Berkley}, \citenamefont {Johansson},
        \citenamefont {Johnson}, \citenamefont {Bunyk}, \citenamefont {Ladizinsky},
        \citenamefont {Ladizinsky}, \citenamefont {Oh},\ and\ \citenamefont
        {Han}}]{Harris:2009uw}%
    \BibitemOpen
    \bibfield  {author} {\bibinfo {author} {\bibfnamefont {R.}~\bibnamefont
            {Harris}}, \bibinfo {author} {\bibfnamefont {T.}~\bibnamefont {Lanting}},
        \bibinfo {author} {\bibfnamefont {A.~J.}\ \bibnamefont {Berkley}}, \bibinfo
        {author} {\bibfnamefont {J.}~\bibnamefont {Johansson}}, \bibinfo {author}
        {\bibfnamefont {M.~W.}\ \bibnamefont {Johnson}}, \bibinfo {author}
        {\bibfnamefont {P.}~\bibnamefont {Bunyk}}, \bibinfo {author} {\bibfnamefont
            {E.}~\bibnamefont {Ladizinsky}}, \bibinfo {author} {\bibfnamefont
            {N.}~\bibnamefont {Ladizinsky}}, \bibinfo {author} {\bibfnamefont
            {T.}~\bibnamefont {Oh}}, \ and\ \bibinfo {author} {\bibfnamefont
            {S.}~\bibnamefont {Han}},\ }\href {\doibase 10.1103/PhysRevB.80.052506}
    {\bibfield  {journal} {\bibinfo  {journal} {Physical Review B}\ }\textbf
        {\bibinfo {volume} {80}},\ \bibinfo {pages} {052506} (\bibinfo {year}
        {2009})}\BibitemShut {NoStop}%
    \bibitem [{\citenamefont {Boixo}\ \emph {et~al.}(2013)\citenamefont {Boixo},
        \citenamefont {Albash}, \citenamefont {Spedalieri}, \citenamefont
        {Chancellor},\ and\ \citenamefont {Lidar}}]{q-sig}%
    \BibitemOpen
    \bibfield  {author} {\bibinfo {author} {\bibfnamefont {S.}~\bibnamefont
            {Boixo}}, \bibinfo {author} {\bibfnamefont {T.}~\bibnamefont {Albash}},
        \bibinfo {author} {\bibfnamefont {F.~M.}\ \bibnamefont {Spedalieri}},
        \bibinfo {author} {\bibfnamefont {N.}~\bibnamefont {Chancellor}}, \ and\
        \bibinfo {author} {\bibfnamefont {D.~A.}\ \bibnamefont {Lidar}},\ }\href
    {\doibase 10.1038/ncomms3067} {\bibfield  {journal} {\bibinfo  {journal}
            {Nat. Commun.}\ }\textbf {\bibinfo {volume} {4}},\ \bibinfo {pages} {2067}
        (\bibinfo {year} {2013})}\BibitemShut {NoStop}%
    \bibitem [{\citenamefont {Job}\ and\ \citenamefont {Lidar}(2018)}]{Job:2017aa}%
    \BibitemOpen
    \bibfield  {author} {\bibinfo {author} {\bibfnamefont {J.}~\bibnamefont
            {Job}}\ and\ \bibinfo {author} {\bibfnamefont {D.}~\bibnamefont {Lidar}},\
    }\href {http://stacks.iop.org/2058-9565/3/i=3/a=030501} {\bibfield  {journal}
        {\bibinfo  {journal} {Quantum Science and Technology}\ }\textbf {\bibinfo
            {volume} {3}},\ \bibinfo {pages} {030501} (\bibinfo {year}
        {2018})}\BibitemShut {NoStop}%
\end{thebibliography}
\end{document}